\documentclass[fleqn,usenatbib]{mnras}

\usepackage[T1]{fontenc}

\DeclareRobustCommand{\VAN}[3]{#2}
\let\VANthebibliography\thebibliography
\def\thebibliography{\DeclareRobustCommand{\VAN}[3]{##3}\VANthebibliography}

\usepackage{graphicx}	
\usepackage{amsmath}	
\usepackage{xcolor}
\usepackage{amssymb}	
\usepackage[T1]{fontenc} 
\usepackage{pdflscape}
\usepackage{newtxtext,newtxmath}

\title[Compact groups of dwarfs in TNG50]{Compact groups of dwarf galaxies in TNG50: late hierarchical assembly and delayed stellar build-up in the low-mass regime}

\author[Flores-Freitas et al.]{Rodrigo Flores-Freitas$^{1}$\thanks{E-mail: rodrigoff@ufrgs.br}\thanks{E-mail: rodrigoff.astro@gmail.com},
Marina Trevisan$^{1}$, Mait\^e M\"uckler$^{1}$, Gary A. Mamon$^{2}$, Allan Schnorr-M\"uller$^{1}$, 
\newauthor
Vitor Bootz$^{1}$
\\
$^{1}$Departamento de Astronomia, Universidade Federal do Rio Grande do Sul, Av. Bento Gonçalves 9500, 91501-970 Porto Alegre-RS, Brazil \\
$^2$Institut d'Astrophysique de Paris (UMR 7095: CNRS \& Sorbonne Universit\'e), 98 bis Bd Arago, F-75014 Paris, France\\ 
}

\date{Accepted XXX. Received YYY; in original form ZZZ}

\pubyear{2023}

\begin{document}
\label{firstpage}
\pagerange{\pageref{firstpage}--\pageref{lastpage}}
\maketitle  

\begin{abstract}

Compact groups of dwarf galaxies (CGDs) have been observed at low redshifts ($z<0.1$) and are direct evidence of hierarchical assembly at low masses. 
To understand the formation of CGDs and the galaxy assembly in the low-mass regime, we search for analogues of compact (radius $\leq 100$ kpc) groups of dwarfs ($7 \leq \log[M_{\ast}/{\rm M}_\odot]  \leq 9.5$) in the IllustrisTNG highest-resolution simulation.
Our analysis shows that TNG50-1 can successfully produce CGDs at $z=0$ with realistic total and stellar masses. We also find that the CGD number density decreases towards the present, especially at $z \lesssim 0.26$, reaching $n \approx  10^{-3.5} \ \rm cMpc^{-3}$ at $z = 0$. This prediction can be tested observationally with upcoming surveys targeting the faint end of the galaxy population and is essential to constrain galaxy evolution models in the dwarf regime. The majority of simulated groups at $z \sim 0$ formed recently ($\lesssim 1.5 \ \rm Gyr$), and CGDs identified at $z \leq 0.5$ commonly take more than 1 Gyr to merge completely, giving origin to low- to intermediate-mass  ($8 \leq  \log[M_{\ast}/{\rm M}_\odot] \leq 10$) normally star-forming galaxies at $z=0$. We find that halos hosting CGDs at $z = 0$ formed later when compared to halos of similar mass, having lower stellar masses and higher total gas fractions. The simulations suggest that CGDs observed at $z \sim 0$ arise from a late hierarchical assembly in the last $\sim 3$ Gyr, producing rapid growth in total mass relative to stellar mass and creating dwarf groups with median halo masses of $\sim 10^{11.3} \ \rm M_\odot$ and B-band mass-to-light ratios mostly in the range $10 \lesssim M/L \lesssim 100$, in agreement with previous theoretical and observational studies.

\end{abstract}




\section{Introduction}
Galaxy formation models in a $\Lambda$CDM Universe predict hierarchical galaxy formation \citep{White&Rees1978, Blumenthal+1984, White&Frenk1991, deLucia&Blaizot2007, Frenk&White2012, Primack2012}, where processes like mergers and interactions play a crucial role in the mass assembly of galaxies. Although the relative importance of these processes to the build-up of the galaxy mass varies across different mass regimes \citep{Cattaneo2011}, the hierarchical assembly is thought to occur at all scales, from dwarf to massive galaxies. Dwarf galaxies ($M_\ast \lesssim 10^9\,$M$_\odot$) stand out as the most numerous systems across all cosmic time \citep{Fontana+2006, Grazian+2015} and are often considered the building blocks of more massive galaxies. 
Nonetheless, most studies primarily focus on the assembly of high-mass galaxies \citep{PerezGonzalez2008,Ilbert2010,Rodriguez-Gomez2016}, while the role of interactions and mergers in the evolution and mass assembly of low-mass galaxies remains unclear. Furthermore, comprehending the interplay between baryons and dark matter in this particular low-mass regime holds critical significance in evaluating the standard cosmological model \citep{Sales2022}. Given the inherent challenge of directly measuring dark matter, this knowledge becomes indispensable for thoroughly validating the $\Lambda$CDM model through the observation of galaxies.

With large spectroscopic surveys, the number of dwarf galaxy systems discovered and studied in detail has been increasing \citep{Stierwalt2015, Pearson2016, Stierwalt2017, Pearson2018, Paudel+2018, Besla+2018, Kado-Fong2020, Luber2022, Gao+2023}. 
Most recent studies focus on pairs of dwarfs since they are more frequent than groups of dwarf galaxies. 
Close pairs, which have projected separations $R_{\rm sep} \leq 50\,$kpc, offer valuable insights into the impact of interactions on galaxy evolution in the low-mass regime.
\citet{Stierwalt2015} identified 104 pairs of dwarf galaxies in the Sloan Digital Sky Survey (SDSS) at $0.005 < z < 0.07$ and with pair member masses $7 < \log(M_\ast/{\rm M}_\odot) < 9.7$. Their sample covers a wide range of environments, including isolated dwarf pairs that lie at distances greater than $1.5$ Mpc from their closest massive neighbour ($M_\ast > 5 \times 10^9\,{\rm M}_\odot$). 
They found that the star formation rate (SFR) in close pair dwarfs is enhanced by a factor of $2.3$ compared to a control sample of single dwarfs. The enhancement decreases with increasing $R_{\rm sep}$, suggesting that interactions trigger the starbursts. They also find that the SFR enhancement is independent of the isolation of the pair relative to massive neighbours, and that is $\sim 30\%$ higher than that observed for massive galaxies in close pairs. These results suggest that interactions may play a more important role in the stellar mass assembly of dwarfs compared to that of massive systems. Detailed analysis of a dwarf galaxy pair by \cite{Privon2017} also reveals evidence of SFR enhancements. They found that the specific SFR (sSFR) of the system is more than an order of magnitude higher than that of non-interacting dwarfs in the same mass range. According to their work, in low-mass galaxy mergers, starbursts may be triggered by large-scale interstellar medium (ISM) compression rather than driven by the funnelling of gas to the nucleus and giving rise to a nuclear starburst. Therefore, the SFR enhancements are more distributed throughout the galaxies. These results and the findings from other recent studies \citep{Kimbro2021,Martin2021} indicate that the interaction-driven star formation shows important hydrodynamical differences from what is observed in massive galaxy interactions. 

Theoretical work by \citet{Martin2021} using the NewHorizon simulation revealed that enhancement of SFR in dwarf galaxies with $M_\ast \lesssim 10^8\,$M$_\odot$ due to interactions and mergers can be more than three times compared to single dwarfs at $z = 1$, while the increase is less pronounced in more massive galaxies ($M\ast \gtrsim 10^{9.5} \rm \ M_\odot$). These findings reinforce the differing roles of interactions and mergers in low- and high-mass galaxy evolution. Additionally, they highlight that, although mergers are rare among dwarf galaxies (1 or 2 with a ratio greater than 1:10 between $z = 1$ and $5$), interactions between galaxies are more important than mergers in the mass assembly of dwarf galaxies. Together, these processes are responsible for $\sim 20\%$ of the stellar mass formed in dwarfs at low redshifts ($z \lesssim 0.5$). 

While interactions and mergers might seem to accelerate dwarf galaxy mass build-up and enhance their star-formation \citep{Subramanian2023}, they can also enhance gas turbulence, delaying star formation \citep{Ellison+2018}. Dwarf-dwarf tidal interactions can also remove the gas from the dwarf shallow potential and place it at large distances. This gas needs to be re-accreted for star formation, possibly leading to a delay in the stellar mass build-up of these systems \citep{Pearson2016, Pearson2018}. Although gas in the outskirts can provide a long-lived supply to the system in isolated environments, this pre-processing of baryons in dwarf-only systems can affect the efficiency of gas stripping when dwarfs are accreted by a massive host, preventing gas re-accretion. A cosmological zoom-in N-body simulation predicts that this pre-processing is not negligible, with ~30\% of galaxies with $10^7 - 10^9\,$M$_\odot$ undergoing pre-processing before entering a Milky Way-like halo \citep{Wetzel+2015}. Moreover, recent work by \cite{Kado-Fong2023} suggests that dwarf-dwarf interactions can induce quenching in the low-mass galaxies, although only for a relatively short period of time ($\lesssim 600$~Myr).

In the local Universe, ensembles of dwarfs with three or more members are rare \citep{Besla+2018} and mostly observed as satellites of more massive galaxies, where they are highly susceptible to environmental effects. 
One of the first studies on these type of dwarf-only systems in the local Universe was carried out by \citet{Tully1987, Tully1988}, who identified structures with low luminosity density ($2.5 \times 10^8 < \rho_L < 2.5 \times 10^9 \ \rm L_\odot\,$Mpc$^{-3}$) within the local volume ($\lesssim 5\,$Mpc) containing only faint galaxies. In later work, \citet{Tully2006} confirmed five previously identified associations and argued that these associations are gravitationally bound, which would imply high mass-luminosity ratios ($300 \lesssim M/L \lesssim 1200$). Additionally, these associations were found to have large diameters ($\sim 600\,$kpc), resulting in distances between member galaxies that are considerably large when compared to the typical sizes of dwarf galaxies. Thus, the associations - characterised by large sizes - of dwarfs are not the ideal sites to study the effects of interactions and mergers on the mass build-up in the low-mass regime. For this purpose, the dwarf galaxies must be closer to each other, inhabiting smaller systems like compact groups of dwarfs.

Recently, \citet{Stierwalt2017} spectroscopically confirmed the existence of seven groups of dwarfs with projected group radius $R_{\rm group} \lesssim 100\,$kpc that lie at distances greater than $1.5\,$Mpc from their closest massive neighbour. 
Even though each one of these groups has between three to five members, these groups are more compact and brighter than the associations presented by \citet{Tully2006} and contain more galaxies than systems analysed by \citet{Stierwalt2015}, \citet{Privon2017}, and others. \cite{Stierwalt2017} (hereafter S17) suggests that, given time, merging will turn some of these groups into intermediate-mass galaxies. Therefore, these compact ($R_{\rm group} \leq 100\,$kpc) groups of dwarfs ($7 \leq \log[M_{\ast}/{\rm M}_\odot]  \leq 9.5$) are excellent sites to assess the aforementioned impacts of mergers and interactions in the baryon cycle and mass growth of dwarf galaxies. Furthermore, \textit{isolated} compact groups of dwarf galaxies (CGDs) are excellent cosmic laboratories for studying the effects group environment on the evolution of dwarfs in the absence of massive ($M_{\ast} > 10^{10} \ \rm M_\odot$) galaxies, and to probe the hierarchical assembly in the low-mass regime.
Although rare in the local Universe, these groups are expected to be more common at early epochs \citep{Stierwalt2017}. Therefore, their local analogues provide a unique window into the evolutionary processes that may operate more frequently on dwarf galaxies at higher redshifts \citep{Kimbro2021}. These systems also constitute small-scale structures (sizes~$< 1$~Mpc) within the broader cosmological framework, making their examination pivotal for gaining a deeper comprehension of matter clustering in domains where the standard cosmological model presents tensions with observations \citep{Silk&Mamon2012, Somerville&Dave2015, Bullock2017, Sales2022}. 

Previous studies indicate that $\Lambda$CDM can naturally account for the existence and properties of dwarf-only systems \citep{Besla+2018, Yaryura2020, Yaryura+2023}, such as associations and groups. However,  it is worth noting that these important studies have certain limitations. They do not directly use fully hydrodynamical simulations with mass resolutions high enough to resolve low-mass galaxies ($M_{\ast} < 10^9 \ \rm M_\odot$). Additionally, they do not comprehensively explore the detailed mass assembly of different components within the low-mass regime represented by CGDs, nor conduct a thorough analysis of the formation and merging timescales of dwarf groups. To complement these previous studies, in this work, we use the state-of-the-art TNG50-1 high-resolution cosmological simulation to investigate the properties, formation, and mass assembly of isolated CGDs within a cosmological context. We analyse their dynamical and halo properties and provide quantified predictions of the number-density statistics of these groups. We compare some observable physical quantities of simulated CGDs with observations at $z \sim 0$ to address the consistency between the CGDs in TNG50-1 and the observational constraints. Our goal is to bring new insights into the baryon cycle of dwarf galaxies and the formation of low- to intermediate-mass galaxies through hierarchical merging. 

This paper is organised as follows. In Section \ref{sec:data_methods}, we describe the simulations and observational data used in this work and the methodology adopted to identify and characterise the CGDs in the simulations and the observations. Our results are presented and discussed in Sections~\ref{sec:results} and \ref{sec:discussion}, respectively. Finally, in Section~\ref{sec:conclusions}, we summarise the main findings and conclusions of our study. In Appendix \ref{app:cumul_dists}, we provide extra figures, while in Appendix \ref{app:sup_material}, we provide information about catalogues and supplementary material of this work.


\section{Data and Methods}
\label{sec:data_methods}
\subsection{IllustrisTNG Cosmological Simulations}
\label{methods:TNG}
In this work, we use data from the IllustrisTNG simulation suite \citep{Nelson2019,Pillepich2019}, which is a series of magneto-hydrodynamical simulations developed with the \textsc{arepo} code \citep{Weinberger2020} that incorporates a comprehensive galaxy sub-grid model. These models implement different AGN feedback modes, individual chemical element tracing, stellar evolution and other relevant baryonic processes \citep{Weinberger2017,Pillepich2018}. The simulation self-consistently evolves the dynamics of dark matter, gas, stars and magnetic fields within uniform periodic-boundary cubes (in the case of TNG50-1 run, a cube of side $\sim 51.7$ cMpc). With many different cosmological simulations presented in the literature - with different resolutions, implemented physics and volume - a justification for why we choose one specific simulation is needed. TNG50-1 is one of the few hydrodynamic cosmological simulations able to resolve dwarf galaxies while simulating them in a larger cosmological context. Regarding the simulation resolution, the average initial stellar particle mass is $8.5 \times 10^4 \ \mathrm{M_\odot}$, the dark matter particle mass is $4.5 \times 10^5 \ \mathrm{M_\odot}$, while the Plummer equivalent gravitational softening lengths ($\epsilon$) for both particles is 288 pc at $z = 0$. Thus, we choose TNG50-1 because of its spatial and mass resolution, enabling us to resolve groups of dwarf galaxies. In addition to that, among the modern cosmological simulations presented in recent years, the ones from the IllustrisTNG suite are among the simulations with the easiest access. This is an important feature to consider when carrying on work based on state-of-the-art simulations with limited computational resources. For example, a single snapshot of the TNG50-1 run occupies a disk space greater than 1000 GB; thus, it is not practical to download all the simulation data for local analysis. To facilitate data exploration, the IllustrisTNG team provides access to a server with a \texttt{JupyterLab} interface\footnote{https://www.tng-project.org/data/lab/}, which is an important difference relative to other public simulation suites.

To have a complete picture of the formation of CGDs, ideally, one would create multiple cosmological simulations with varying initial conditions - to avoid spurious results from stochastic variations \citep{Genel2019,Borrow2023} - and varying cosmological parameters - to make predictions robust against changes in cosmology  - however, this would be extremely expensive and is out of the scope of this work. Instead, we use IllustrisTNG suite of cosmological simulations rather than trying to find all the multiple initial conditions that could generate CGDs and doing individual simulations from scratch.
The dwarf groups found in this work are direct by-products of the complex model employed in TNG simulations and, thus, can serve as tests for the model. Moreover, the CGDs found in TNG50-1 have evolved within a larger cosmological environment rather than complete isolation, which is essential if we want to study the large-scale environment of these objects.
However, by studying a cosmological simulation that was not tailored to explore all the possible formation pathways of dwarf galaxy groups, we are subjected to a sampling bias. In other words, the simulated groups studied in this work are generated only from the limited set of initial conditions - e.g. density matter field - in TNG50-1 that can generate CGDs. Thus, the properties of dwarf groups that we present here may not represent all possible CGDs that are allowed to exist in a cosmological simulation within the $\Lambda$CDM framework. Nonetheless, for the purpose of this work, we assume that TNG50-1 is homogeneously sampling the space of initial conditions that create CGDs. 

Throughout this work, when we mention stellar masses, star-formation rates and other properties of individual simulated galaxies, we refer to these quantities computed in the specific aperture of 2 half-mass radius, which is commonly used in the literature. The $r$-band magnitudes of galaxies in the simulation are absolute magnitudes computed from the total luminosity of all stellar particles in the SDSS $r$-band, and dust attenuation is not taken into account. Since the extinction for these galaxies is small, it should not strongly affect our results and conclusions from the comparison with observations. Throughout the text, we use the terms "subhalo" and "galaxy" interchangeably since subhalos in the simulation are supposed to represent galaxies in the real Universe. To be consistent with the TNG50 simulation, in this work we adopt the standard $\Lambda$CDM cosmology with $\Omega_{\rm m,0} = 0.3089$, $\Omega_{\rm \Lambda,0} = 0.6911$, $\Omega_{\rm b,0} = 0.0486$ and $H_0 = 67.74$ km\,s$^{-1}$\, Mpc$^{-1}$, from \cite{Planck2016}.

\subsection{Nearest Neighbours and Environment}
\label{methods:nn_catalog}
To avoid imposing a direct selection bias in the halo mass of the CGDs, we selected the groups based on the distance between galaxies. This approach has the advantage of being more easily reproduced in observations since halo mass is not directly observable. 
For each target galaxy in the simulation box, we computed the distances to the $n$-th nearest neighbours and stored them in catalogues together with neighbours' respective subhalo identifiers. In TNG50-1 we adopted a minimum stellar mass of $M_{\rm \ast, min} = 10^7 \ \rm M_\odot$ for both the target and the neighbour galaxies because we are not interested in galaxies with stellar masses close to the resolution limit of the simulation. Besides, their observational counterparts are very difficult to detect at large distances and restricted to the local volume. Using information about distances, stellar masses and star-formation of nearest neighbours, we created the sample of simulated CGDs analysed in this work (see Section \ref{methods:sim_sample} for details on the selection).

We also use the catalogues described above to compute the densities of the environment around the groups. To analyse the large-scale environment, we define a volumetric density contrast following the approach by \cite{Muldrew2012}, computed within a sphere of radius 1 Mpc centred on the galaxy of interest:
\begin{equation}
    \delta_{\rm 1 \ Mpc} = \frac{N_{g} - \bar{N}_{g}}{\bar{N}_{g}},
\end{equation}

\noindent where $N_{g}$ is the number of galaxies within the sphere around the target galaxy, and $\bar{N}_{g}$ is the average number of galaxies that would be expected in the spherical aperture if galaxies were distributed uniformly throughout the simulation volume. To estimate $\bar{N}_{g}$, we redistribute all the galaxies in new positions and sample the simulation volume with $10^6$ spheres. The rearrangement of the galaxies is done by sampling new positions from uniform distributions - one for each of the three coordinate components - spanning the values in the interval [0, 51.67] Mpc. The positions of the sphere centres are sampled in the same manner.

Alternatively, we also compute the volumetric dark matter density ($\rho_{\rm 1 \ Mpc}$) around the halos of interest in the TNG50-1. We use snapshot data from the full simulation box and sum the masses of all dark matter particles in the simulation within 1 Mpc from the position of interest. Given that the computation of $\rho_{\rm 1 \ Mpc}$ is expensive, we only do it for the halo samples analysed in Section \ref{discussion:assembly}, not all halos in the simulation run.

\subsection{Sample of Simulated CGDs}
\label{methods:sim_sample}
The observational studies on groups of dwarf galaxies are mostly restricted to low redshifts ($z \lesssim 0.1$), but with simulations, we can analyse these objects at higher redshifts. Therefore, we create samples of CGDs for all snapshots over the redshift interval of $0 \leq z \leq 0.5$. 

We are interested in \textit{compact} groups, meaning we must define a group size measurement and use it as a selection criterion. We then define the group radius ($r_{\rm group}$) as the radius of the smallest sphere that encloses all member galaxies. With the coordinates of galaxies, we compute $r_{\rm group}$ and group centres, using the Python package \textsc{miniball}\footnote{\url{https://pypi.org/project/miniball/}}, which has a recursive implementation of Welzl’s algorithm \citep{Welzl91} to solve the smallest bounding sphere problem. In this sense, the group centre is simply the centre of the smallest sphere found to enclose all galaxies of the group, not necessarily the barycenter of the group. To avoid selecting groups suffering the influence of close galaxies, we select only isolated groups, which means that the closest galaxy must be at least 200 kpc away from the group. The distance from the group to neighbour galaxies is obtained using the nearest neighbour catalogue described in Section \ref{methods:nn_catalog} and the coordinates of the group centre. We then generate catalogues of groups by iterative searching for groups of $N$ star-forming dwarf galaxies that are enclosed by a sphere with radius~$< \rm 100 \ kpc$. The search is iterative in the sense that we first look for groups with three galaxies, then four galaxies, and so on, looking for groups up to a maximum of 10 star-forming dwarf galaxies. The criteria described above might lead to the selection of systems that are not exactly representative of the CGDs we want to study; thus, further sample cleaning is needed. For example, the initial catalogue of groups can contain a CGD with $N$ members that are a substructure of other CGD with $N+1$ members, and the first step is to remove groups within groups. We want to study CGDs as isolated structures, so we eliminate CGDs that pass the isolation criteria described above but are still part of a larger halo, such as a more massive group or a cluster.

In summary, to select a CGD in the simulation, we use the following criteria: 
\begin{itemize}
    \item (Compact) The group radius must be smaller than 100 kpc;
    \item (Dwarfs) All members of the group must have $M_{\rm \ast} \leq 10^{9.5} \ \rm M_\odot$;
    \item (Star-forming) All members must have ${\rm sSFR} \geq 10^{-11} \ \rm yr^{-1}$;
    \item (Rich) The group must have 3 to 10 members;
    \item (Isolated) Absence of neighbour galaxies within 200 kpc from the group centre;
    \item (Not substructure) One of the group members must be the primary\footnote{Defined as the subhalo with the largest total number of bound particles/cells, usually the most massive galaxy in its host halo.} galaxy of a halo;
\end{itemize}

The coordinates, stellar masses and star-formation rates used in sample selection were obtained from the TNG subhalo catalogues. Subhalos with $M_{\rm \ast} < 10^7 \ \rm M_\odot$ can be close to or inside CGDs selected in our sample. However, we do not consider these objects in our group selection because they may be at the limit of what we can call a "resolved" galaxy in the simulation (number of particles $N_{\rm particle} \lesssim 200$) and because observations of dwarfs in this mass regime are very limited. 

We want to study the properties of the halos that host CGDs in simulations, and in TNG, the halos are identified using a standard friends-of-friends (FoF) algorithm \citep{Davis1985} with linking length $b=0.2$ which is run over the dark matter particles. For simplicity, we only analyse groups that inhabit a single halo, in opposition to groups composed of more than one halo. This cut removes less than 5\% of the groups initially found at $z=0$, so it does not introduce a strong bias in our final sample of simulated CGDs at $z=0$. In total, 38 CGDs ($\sim 97$~\% triplets and $\sim 3$~\% quartets) are found in the last snapshot of TNG50-1 - which corresponds to $z=0$ - and in Section \ref{results:cgds_at_present}, we show examples of such groups. The total number of CGDs found in other snapshots is presented in Table \ref{tab:group_catalogues} of Section \ref{results:number_density}.

\subsection{Halo Control Sample at $z=0$}
\label{methods:sim_control_sample}
To characterise the halos of CGDs at $z=0$, we can compare their properties with the overall halo population. To control the effect of halo mass in other halo properties, we must build a control sample of halos with a similar distribution of this variable. We adopt as halo mass the total mass of the halo enclosed in a sphere whose mean density is 200 times the mean density of the Universe at the time the halo is considered ($M_{\rm 200,m}$). 

\begin{figure}
    \centering
    \includegraphics[width=0.49\textwidth]{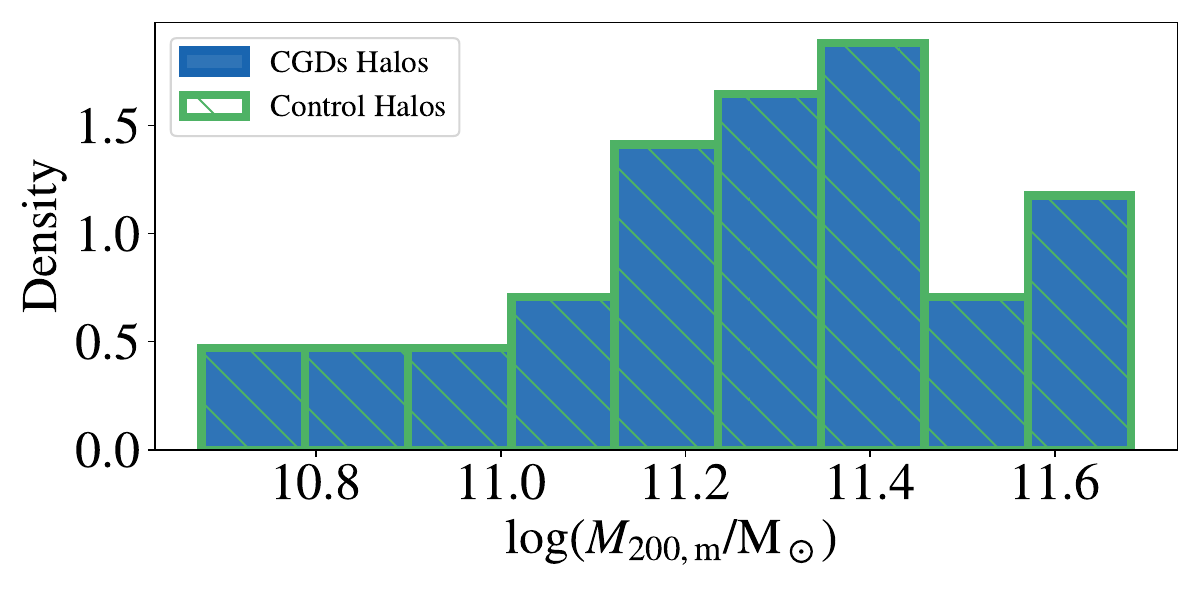}
    \caption{Histograms of halo masses ($M_{\rm 200,m}$) for the sample of TNG50 CGDs at $z=0$ (\emph{blue solid}) and their control sample (\emph{green hashed}). Both samples were matched to have similar distributions of $M_{\rm 200,m}$ following the procedure described in Section~\ref{methods:obs_sample}.}
    \label{fig:halo_match}
\end{figure}

To create the halo control sample at $z=0$, we apply the propensity score matching (PSM) technique \citep{Rosembaum1983}, using the \textsc{MatchIt}\footnote{\url{https://cran.r-project.org/web/packages/MatchIt/}} package \citep{Ho2011}, developed in R language. This method allows us to select from a sample of halos \textit{not} hosting CGDs a control sample in which the distribution of the control variable (i.e. $M_{\rm 200,m}$) is as close as possible to that of those halos hosting CGDs (see Fig.~\ref{fig:halo_match}). For the matching, we adopted the Mahalanobis distance \citep{Mahalanobis1936} approach and the nearest-neighbour method.
The matching process yielded a sample of 190 control halos for the 38 halos hosting CGDs, i.e. five control halos for each CGD. We select more than one control halo for each CGD to have better statistics for the properties of the overall halo population in the same mass interval. We tested different ratios between treat and control samples (1:1, 1:3 and 1:10) and found that our conclusions remain the same.

\subsection{Merger Trees and Cosmic Evolution}
\label{methods:mergertree}
To trace the evolution of the CGDs through different simulation snapshots, we used the subhalo merger trees constructed with the \textsc{sublink} algorithm \citep{Rodriguez-Gomez2015}. The algorithm links progenitor subhaloes to their unique descendants in future snapshots based on the particles that they have in common and a merit function that takes into account the binding energy of the descendants' particles. With the merger trees, we can obtain past properties of the groups identified at $z=0$ and properties of descendants identified in snapshots at $z>0$. We trace progenitors of galaxies using the main progenitor branch of their merger trees, while their descendants are identified using main descendant branches.

The merger trees are used to compute relevant timescales for the evolution of the groups, which are outlined as follows. First, we define $t_{100}$ as the maximum lookback time in the simulation when a CGD has $r_{\rm group} \leq 100$~kpc and all its members have $M_{\ast} \geq 10^7 \ \rm M_\odot$, interpreted as the moment that the group formed and all dwarf galaxies began to interact in a compact region of space. We also use a formation timescale based on halo assignment, $t_{\rm FoF}$, defined as the maximum lookback time in the simulation in which all group members are subhalos of the same FoF group, that is, all dwarfs are hosted by the same halo. Third, we define $t_{\rm 50,halo}$ as the lookback time when the halo hosting a CGD reached half of its current virial mass at present, that is $M_{\rm 200,c} (z_{\rm 50,halo}) = M_{\rm 200,c} (z=0) / 2$, where $z_{\rm 50,halo}$ is the correspondent redshift of $t_{\rm 50,halo}$. The values of $t_{\rm 50,halo}$ are obtained from the values of the scale factor, $a_{\rm form}$, given in the supplementary catalogue of halo structure from TNG50-1 \citep{Anbajagane2022}. We also quantify the coalescence timescale of CGDs, i.e. the amount of time that the galaxies in a group take to merge into one single galaxy. It is defined simply as the difference between the lookback time in which a group was identified ($t_{\rm obs}$) and the maximum lookback time in which all group members have the same descendant subhalo ($t_{\rm coalesc}$).

We also use the merger trees to trace the internal dynamical evolution of the groups until their final state at $z = 0$. Analysing the trees of all group members simultaneously, we determined the fate of their respective groups as one of the following:

\begin{itemize}
    \item "Steady" - all original group members remain the same at $z=0$ and $r_{\rm group}$ did not grow significantly, i.e., $r_{\rm group}(z=0)<200\,$kpc, which corresponds to two times the maximum radius adopted to select the CGDs;
    \item "Coalesced" - all original members of the group merged into a single galaxy;
    \item "Dispersed" - at least one original member of the group is not present anymore or $r_{\rm group}$ grew significantly, i.e., $r_{\rm group}(z=0)>200\,$kpc;
    \item "Altered" - some members of the groups merged, or a new galaxy is part of the group;
\end{itemize}

The four classes described above cover all the possible final states of CGDs in the simulation.

\subsection{Projected Quantities}
\label{methods:proj_quantities}
To make fair comparisons between the physical properties from simulations and observations, it is convenient to project the simulated groups in two dimensions. To do this, we project the positions and velocity vectors of galaxies in the simulations onto 150 randomly oriented two-dimensional planes. For each plane onto which a group is projected, we compute the quantities of interest (e.g., radius, velocity dispersions, etc.) and the 25th, 50th, and 75th percentiles of their respective distributions. In this manner, we do not privilege any specific orientation of the CGDs, and we consider variations on the projected quantities that can arise purely from differences in orientations.

To obtain group masses, we use a virial mass estimator based on projected velocities and positions of galaxies. We use the projected mass estimator described in \cite{Heisler1985}, given as follows:
\begin{equation}
    M_{\rm PME,group} = \frac{f_{\rm pm}}{G(N-\alpha)} \sum_i^N R_{{\rm p},i} \Delta v_i^2 ,
\label{eq:proj_mass_estimator}
\end{equation}
where $R_{{\rm p},i}$ is the projected distance between an $i$-th group member and the geometric centre of the group; $\Delta v_i$ is the difference between the line-of-sight velocity of $i$-th group member and the mean velocity of the group members; $N$ is the number of group members; G is the gravitational constant; $f_{\rm pm} = 20/ \pi$ and $\alpha = 1.5$ are constants as adopted in \cite{Stierwalt2017}.

With the multiple projections on simulations, we are able to access how the group mass estimator adopted is sensible to specific configurations of CGDs. As shown in Fig. \ref{fig:obs_range}, Section \ref{results:obs_properties}, the variation in the values of $M_{\rm PME,group}$ can be large (greater than 0.5 dex). For each simulated CGD, we adopt the 75th (25th) percentiles of the distribution of masses estimated from the 150 projections as the upper (lower) uncertainties in $M_{\rm PME,group}$. However, for the observational data, we have a single projection, so we compute the mean interquartile ranges (IQRs) of $M_{\rm PME,group}$ for all simulated CGDs and assume the uncertainty to be symmetric and equal to half of the mean of IQRs. Here, we consider that the projection effects on the $M_{\rm PME,group}$ estimates in observations are the dominant source of error. These effects are similar to those in simulated systems.

To facilitate the comparison of our results with literature, we also computed velocity dispersions and group radius as defined in other studies addressing dwarf systems. As a measure of group physical size, we computed the projected inertial radius ($R_I$) analogously to \cite{Tully2006} and \cite{Yaryura2020} (hereafter Y20):
\begin{equation}
    R_I = \bigg ( \frac{1}{N} \sum_i^N r_i^2 \bigg )^{1/2} \ ,
\end{equation}
where $r_i$ is the two-dimensional projected distance of a galaxy from the system centroid, and the sum for each group is computed over all $N$ members. We also compute a measure of the groups' velocity dispersion using an analogous expression to that of Y20:
\begin{equation}
    \sigma_{\rm Y} = \bigg [ \frac{1}{N-1} \sum_i^N v_i^2 \bigg]^{1/2}  \ ,
\label{eq:sigma_Y20}
\end{equation}
where $v_i$ is the difference between the line-of-sight velocity of a galaxy and the group mean. As a complementary measure to $\sigma_{\rm Y}$, we also compute the velocity dispersion of groups using the gapper estimator \citep{Wainer1976,Beers1990}. This robust scale estimator is based on the gaps of an order statistics, $x_i$, $xi+1$, ... , $xn$, with gaps defined by:
\begin{equation}
    g_i = x_{i+1} - x_i, \ \ i = 1, . . . , N - 1
\end{equation}
and approximately Gaussian weights given by:
\begin{equation}
    w_i = i(N-1) \ .
\end{equation}
The estimator of scale then is given by:
\begin{equation}
    \sigma_{\rm WT} = \frac{\sqrt{\pi}}{N(N-1)} \sum_i^{N-1} w_i g_i \ ,
\label{eq:sigma_WT76}
\end{equation}
where $x_i$ is the $i$-th element of an ordered set of differences between the line-of-sight velocities of the galaxies and the group mean. We found $\sigma_{\rm Y}$ values that are strongly correlated with $\sigma_{\rm WT}$ values, with both estimators returning similar velocity dispersions, with a relative difference of $\sim 13 \%$ in both simulated and observed CGDs. When computing the velocity dispersions for observations, the scaling factor $1 / (1 + \bar{z})$ multiplies the right-hand side of equations \ref{eq:sigma_Y20} and \ref{eq:sigma_WT76}, where $\bar{z}$ is the mean redshift of the group members.

\subsection{Observational Sample at $z \sim 0$}
\label{methods:obs_sample}
\begin{figure}
    \centering
    \includegraphics[width=0.49\textwidth]{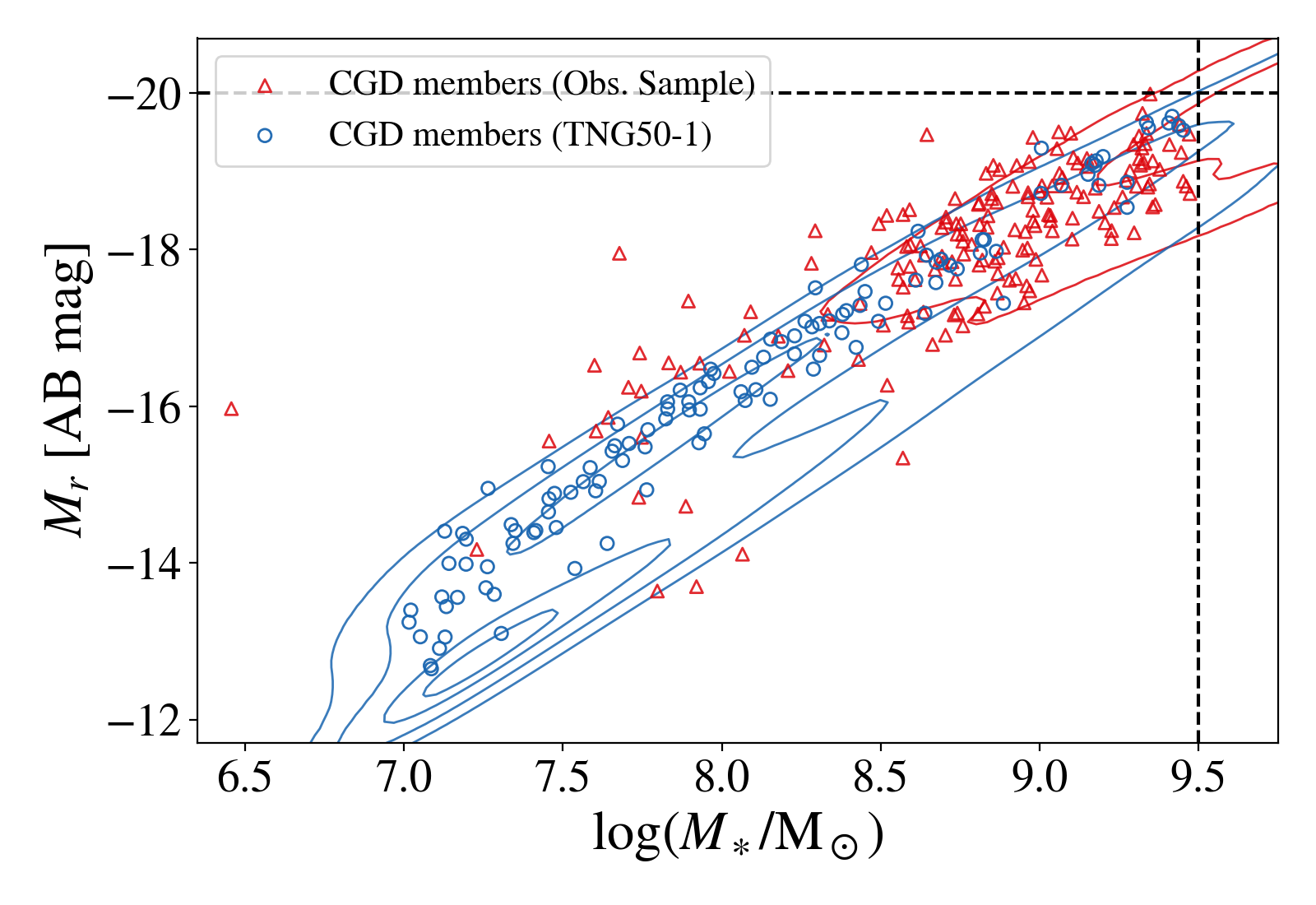}
    \caption{Stellar mass ($M_{\ast}$) versus $r$-band absolute magnitude ($M_{r}$) of CGD members in observations and simulations at $z \sim 0$. Blue circles represent galaxies in simulated CGDs from TNG50-1, while red triangles show galaxies in observed CGDs from SDSS. The distribution of the parent samples of galaxies in which CGD members were selected are shown as blue (TNG50-1, $z=0$) and red (SDSS, $0.01 \leq z \leq 0.1$) contours. These contours indicate Gaussian kernel density estimations, with five levels of density iso-proportions (0.05, 0.25, 0.75 and 0.95). The black dashed lines indicate upper limits of $M_\ast = 10^{9.5} \ \rm M_\odot$ and $M_r = -20$ adopted in the observational sample of CGDs (see Section \ref{methods:obs_sample}).}
    \label{fig:mag_mstar_diagram}
\end{figure}

To compare the simulation results with observations, we built a sample of CGDs using the spectroscopic catalogue from Sloan Digital Sky Survey Data Release 18 \citep[SDSS-DR18]{Almeida2023} in the redshift interval $0.01 \leq z \leq 0.1$. The search for CGDs was made using the CasJobs asynchronous database tool for long, complex queries in the SDSS Catalog Archive Server (CAS)\footnote{The data access is done through the SkyServer (\url{https://skyserver.sdss.org/}), a web interface to the CAS database.}. We selected groups with projected radii $R_{\rm group} < 100$\,kpc and imposed the limits of $M_\ast \leq 9.5 \ \rm M_\odot$ and $M_r \geq -20$ for the stellar masses and $r$-band absolute magnitudes of group members, respectively. The upper limit in stellar mass is compatible with the limit we adopted when searching CGDs in the simulation. On the other hand, the lower limit in $r$-band absolute magnitude guarantees that galaxies in the observational sample will not be brighter than the brightest galaxies in simulated CGDs, and we apply this limit to make comparisons between the simulations and observations more meaningful. As is shown in Figure \ref{fig:mag_mstar_diagram}, all dwarf galaxies inhabiting CGDs in TNG50-1 have absolute magnitudes  $M_r$ fainter than $-20$, justifying the adopted magnitude limit mentioned above. The additional limit in magnitude also accounts for the fact that not all galaxies in the SDSS have estimates of stellar mass, so we can select them using $M_r$.

The criteria described above to select CGDs in observations are similar to those employed in TNG50-1, but some differences are important to mention. The projected radius cannot exceed the real group radius in three dimensions. However, due to projection effects, observed CGDs can have a projected radius that is below the upper limit used in the selection ($R_{\rm group} < 100$ kpc) but a real radius that is actually larger. Additionally, chance alignments can significantly contaminate observational samples of compact groups \citep{Taverna2022}. Thus, to search for real compact groups, we also impose a maximum difference between line-of-sight velocity ($\Delta v$) of group members of 300 km/s.

In the spectroscopic catalogue of SDSS, some sources are not individual galaxies but clumps of larger galaxies. Thus, after we built the group catalogues using CasJobs queries, we visually inspected all CGD candidates and removed false group members from the sample, in which a clump of a larger galaxy was interpreted as an individual galaxy. At the end of the sample selection process, the observational sample of CGDs was composed of 58 groups ($\sim 93$~\% triplets and $\sim 7$~\% quartets). A table with their properties is provided as supplementary material for this work (see Appendix \ref{app:sup_material}).

Once we had a clean sample of CGDs from SDSS data, we cross-matched the SDSS CGDs with data from the Arecibo Legacy Fast ALFA (ALFALFA) survey \citep{Haynes.etal:2011, Haynes.etal:2018} to obtain information about the atomic neutral gas content of observed groups. The ALFALFA extragalactic neutral hydrogen (HI) source catalogue provides estimated masses for individual detections based on the 21-cm line flux density. Thus, we obtain the HI gas masses ($M_{\rm HI}$) of CGDs by summing the individual masses of all detections within $1.5 R_{\rm group}$ and $\Delta v < 500\,{\rm km\, s}^{-1}$ from the centre of each group. Since the SDSS spectroscopic catalogue has only a partial overlap with the ALFALFA extragalactic catalogue, we do not have information about $M_{\rm HI}$ for all groups in our observational sample. More precisely, we only present results about the neutral hydrogen content of 20 CGDs in SDSS (see Section \ref{results:obs_properties}).

We consider as CGD candidates all groups having at least three members with spectroscopic redshifts, which introduces a limit in the magnitude of the faintest members. Thus, throughout the work, we always use the magnitude of the third brightest group member ($M_{r,\rm3rd}$) as a reference when comparing the faintest members in simulated and observed groups. It is possible that fainter objects are part of the CGDs in the observational sample, but we are only considering group members with spectroscopic redshifts because we want to avoid contamination from chance alignments and analyse only groups of galaxies that are truly close in real space.

To compare some properties of simulations and observations more directly, we created sub-samples of simulated and observed CGDs. We match groups in simulations with groups in observations with the PSM technique described in Section~\ref{methods:sim_control_sample} using as control variables their total absolute magnitudes ($M_{r,\rm tot}$) and third brightest member magnitudes, both in $r$-band. We use both magnitudes to constrain, at the same time, the luminosity of the faintest group members and of the whole group. Due to the different distributions of $M_{r,\rm tot}$ and $M_{r,\rm 3rd}$ in both simulations and observations, only 9 CGDs are paired for each sample.

The magnitudes of galaxies selected in SDSS are Petrosian magnitudes corrected by extinction. The luminosity distance used to compute the absolute magnitudes is obtained using the \texttt{FlatLambdaCDM} class of \textsc{astropy} cosmology package\footnote{\url{https://docs.astropy.org/en/stable/cosmology/}} and the parameters mentioned at the end of Section \ref{methods:TNG}. We also apply K-correction to the magnitudes using the code \textsc{kcorrect} \citep{Blanton2007}. Magnitudes in the B-band are computed using the following Lupton (2005) transformation\footnote{\url{https://www.sdss3.org/dr10/algorithms/sdssUBVRITransform.php}}: 
\begin{equation}
    M_{\mathrm{B}} = M_g + 0.3130(M_g - M_r) + 0.2271 \ ,
\end{equation}
where $M_g$ and $M_r$ are absolute magnitudes in the $g$- and $r$-band, respectively. To convert the magnitudes to luminosities, we consider $M_\mathrm{B} = 5.44$ for the Sun \citep{Willmer2018}.

\subsection{General Caveats}
\label{methods:caveats}

It is important to note that the simulated systems analysed in this work are taken as \textit{analogues} of objects found in observations, not perfect representations. Cosmological simulations assume recipes to model physical phenomena that occur in scales smaller than their particle resolution; these coarse-grained models may not be appropriate for all situations \citep{Crain2023}, resulting in galaxies with spurious properties. If we assume that the recipes work well as effective models, then the coarse-grained physics may not be a problem. In other words, we can expect the cosmological simulations to fail in reproducing some properties of galaxies or galaxy groups. However, we can also expect that the simulations will give us useful insights about the formation and evolution of galaxy \textit{populations}, even when the objects simulated are not \textit{perfect} representations of observed galaxies. In addition, here we are focusing on the analysis of IllustrisTNG simulations, which adopt a specific code, cosmology and sub-grid physics. Other simulations may present different results about the nature and properties of CGDs. Thus, further analysis is needed in order to address this dependence on simulation parameters.

Regarding the number counts statistic of CGDs, we expect the dominant effect in our observational sample to be the limiting magnitude that severely limits the observed volume for dwarf galaxies. Due to our choice for high-resolution simulations, we also have a limited simulated volume, which may influence our results. Simulations with similar resolution to TNG50-1 but large volumes (volume $> 100^3 \ \rm Mpc^3$) are not available at the time, thus, the dependence of our results with simulation volume may be tested only in the future. Additionally, cosmic variance may also play a role in our results since the volume we are sampling - in simulations and observations - is relatively small.


\section{Results}
\label{sec:results}

\subsection{Space Number Density}
\label{results:number_density}
After applying the selection criteria described in Section \ref{methods:sim_sample}, we obtained catalogues of simulated groups for different snapshots corresponding to different redshifts. In Table \ref{tab:group_catalogues}, we show basic information about each catalogue of CGDs. Catalogues with properties of simulated CGDs will be provided as supplementary material (see Appendix \ref{app:sup_material}).

\begin{table}
\centering
\begin{tabular}{ccccc} \hline
Snapshot & $z$ & Lookback Time [Gyr] & $N_{\rm groups}$ & $\log(n)$ [${\rm cMpc^{-3}}$] \\ \hline
99 & 0.00 & 0.00 & 38 & $-3.56$ \\ 
96 & 0.03 & 0.42 & 42 & $-3.52$ \\ 
92 & 0.08 & 1.09 & 51 & $-3.43$ \\ 
89 & 0.13 & 1.71 & 51 & $-3.43$ \\ 
86 & 0.17 & 2.18 & 61 & $-3.35$ \\ 
83 & 0.21 & 2.62 & 74 & $-3.27$ \\ 
80 & 0.26 & 3.14 & 81 & $-3.23$ \\ 
77 & 0.31 & 3.62 & 78 & $-3.25$ \\ 
74 & 0.38 & 4.25 & 85 & $-3.21$ \\ 
70 & 0.44 & 4.74 & 98 & $-3.15$ \\ 
67 & 0.50 & 5.20 & 87 & $-3.20$ \\ \hline
\end{tabular}%
\caption{Metadata about the catalogues of CGDs in TNG50-1 at different snapshots. From left to right, the columns present the snapshot of the simulation run, correspondent redshift, lookback time, total number of CGDs in the snapshot, and the comoving number density of CGDs.}
\label{tab:group_catalogues}
\end{table}

\begin{figure}
    \centering
    \includegraphics[width=0.48\textwidth]{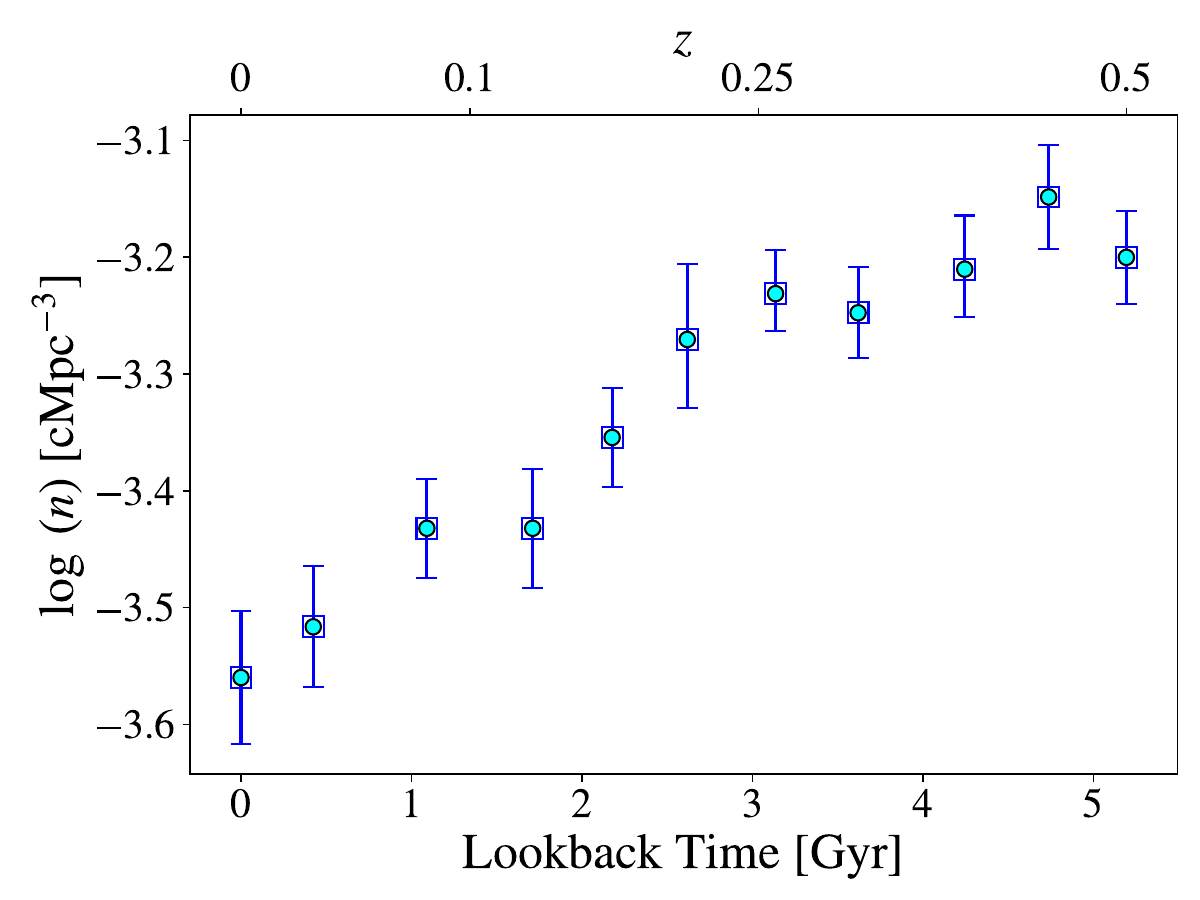}
    \includegraphics[width=0.49\textwidth]{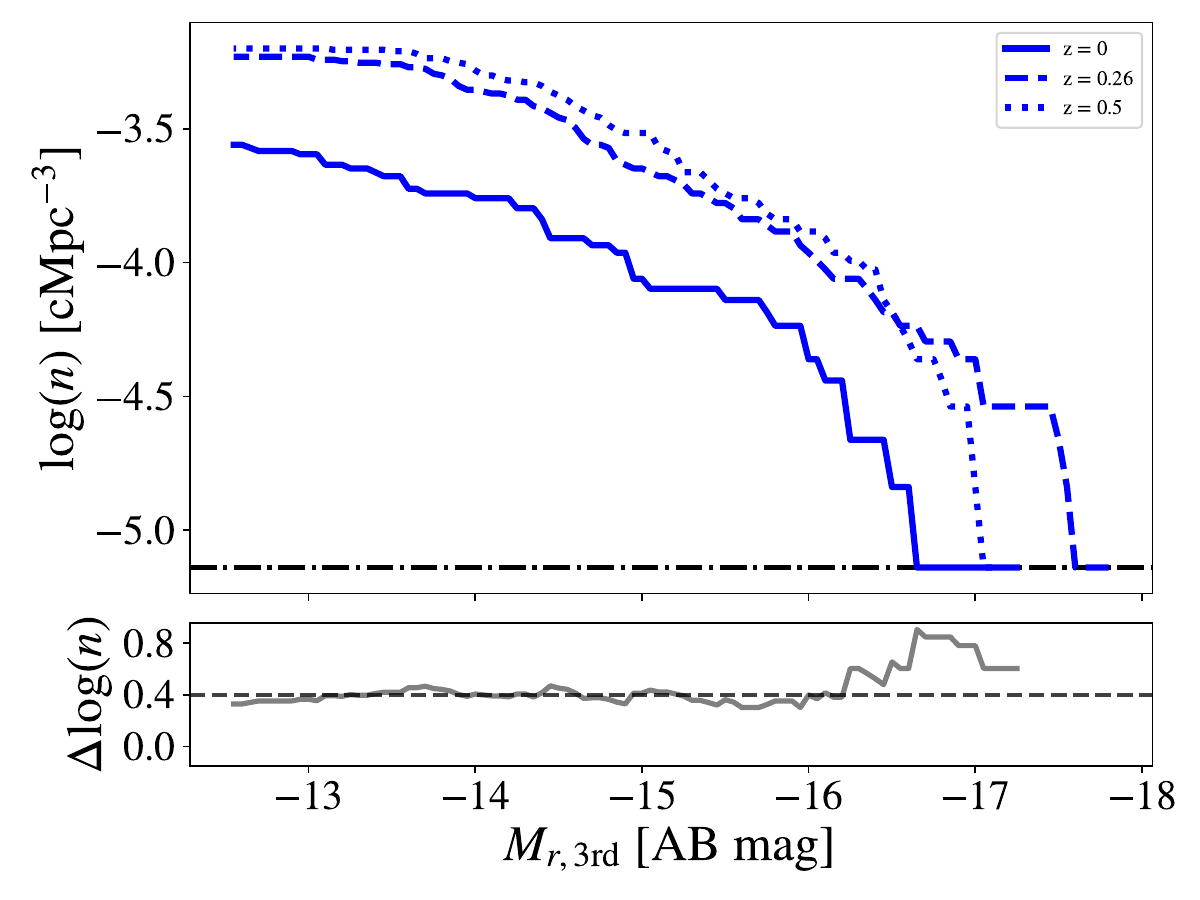}
    \caption{Comoving number density of CGDs in TNG50-1 cosmological simulation. \textbf{Top:} Evolution of CGD comoving number density with time. Blue squares are median number densities estimated by counting the number of CGDs in the simulation using randomly located boxes. Error bars are conservative uncertainties obtained from the 16th and 84th percentiles (see Section \ref{results:number_density}). Cyan circles are number densities estimated from the number of groups shown in Table \ref{tab:group_catalogues}. \textbf{Bottom:} CGDs comoving number density versus the $r$-band absolute magnitude of the third most brightest group member ($M_{r \rm ,3rd}$). Different shades of blue illustrate CGD samples at different redshifts in the simulation. The grey line in the sub-panel below represents the difference between $\log(n)$ at $z=0.26$ and $z=0$, and the black dashed line indicates the median difference over the whole magnitude range covered.}
    \label{fig:nd_evolution}
\end{figure}

The space number density is an important quantity to be evaluated when studying populations of extragalactic sources. Thus, we estimate the comoving number density of CGDs in TNG50-1 and interpret it as a prediction of the cosmological simulation. 
In the top panel of Fig. \ref{fig:nd_evolution}, we show the evolution of the comoving number density of CGDs. To estimate the errors for the number densities found in TNG50, we create a "super-box" of size $\sim 310$ cMpc by randomly rotating $6^3$ identical TNG50 boxes. Then, we assemble them in a cube to create a larger box, 216 times the size of TNG50. We sample the super-box with $10^5$ cubes of the same size as TNG50, computing the number density of CGDs. The errors shown in the top panel of Fig.~\ref{fig:nd_evolution} are the 16th and 84th percentiles of the $\log(n)$ distribution obtained with this sampling. The number density estimated by counting the number of CGDs in the TNG50-1 standard box (shown as cyan circles in the figure) coincides with the median value of our sampling in all snapshots, indicating that the method is working as expected.

Our results for the number density at $z=0$ are in agreement with previous theoretical work from Y20, who found $n \approx 10^{-3.4} \ \rm Mpc^{-3}$ for the systems they define as "Groups", with three members and adopting similar mass limits in the stellar mass of group members as those used in this work. As can be seen in Fig. \ref{fig:nd_evolution}, the density of groups decreases towards the present, with a more pronounced decline at $z \leq 0.26$. From the values of the number densities in Fig. \ref{fig:nd_evolution}, we can also see that CGDs are scarce at redshifts $z < 0.1$, which is expected from $\Lambda$CDM cosmologies and is in agreement with previous studies which suggest that groups and associations of dwarf galaxies are rare \citep{Stierwalt2017,Yaryura2020}. 

In the bottom panel of Fig.~\ref{fig:nd_evolution}, we show the number density as a function of a minimum absolute magnitude, in this case, the $r$-band magnitude of the third brightest group member ($M_{r \rm ,3rd}$). For most simulated CGDs, $M_{r \rm,3rd}$ is the magnitude of the faintest galaxy in the group. We see that the number of groups decreases monotonically with $M_{r \rm ,3rd}$, with a more pronounced decline for $M_{r \rm, 3rd} \lesssim -15$, meaning that groups with fainter $M_{r \rm ,3rd}$ are relatively more common. An interpretation for this result is that we should observe more groups that contain fainter galaxies ($M_{r \rm, 3rd} \gtrsim -15$), implying that some observed dwarf galaxy pairs may have fainter companions that are not detected in current spectroscopic surveys. However, the size of the simulation box of TNG50-1 is relatively small ($L_{\rm box} = 35 \ {\rm cMpc}/h$), meaning there may be an insufficient sampling of the bright end of luminosity functions estimated from this simulation. Ideally, our result for TNG50-1 should be compared with other simulations with the same resolution - to resolve dwarf galaxies - but with a much larger volume to check for variations in the number density of CGDs. Nevertheless, it is relevant to study if TNG50 correctly estimates the number of CGDs that we should find in a flux-limited survey, as we discuss in Section~\ref{discussion:CGDs_in_TNG}.

In the bottom panel of Fig. \ref{fig:nd_evolution}, we show how the number density of groups as a function of $M_{r, \rm 3rd}$ varies with redshift. The largest difference is between the number densities at $z=0$ and $z=0.26$, as is expected from the steeper $\log (n)$ decline in this redshift interval on the top panel of the same figure. Except for the bright end ($M_{r \rm ,3rd} \lesssim -16$), the difference in number density does not seem to depend strongly on the magnitude of the third brightest member, with the median of differences being $\Delta \log (n) \approx 0.4$. This shows that the number of groups decreases by more than half from $z=0.26$ until the present, almost uniformly for all groups with $M_{r \rm ,3rd} \gtrsim -16$. Thus, at least in the TNG50-1 simulation, the formation and destruction rates of CGDs do not seem to be strongly affected by the luminosity of the faintest group members.

\subsection{CGDs at $z = 0$}
\label{results:cgds_at_present}
\begin{figure*}
    \centering
    \includegraphics[width=0.48\textwidth]{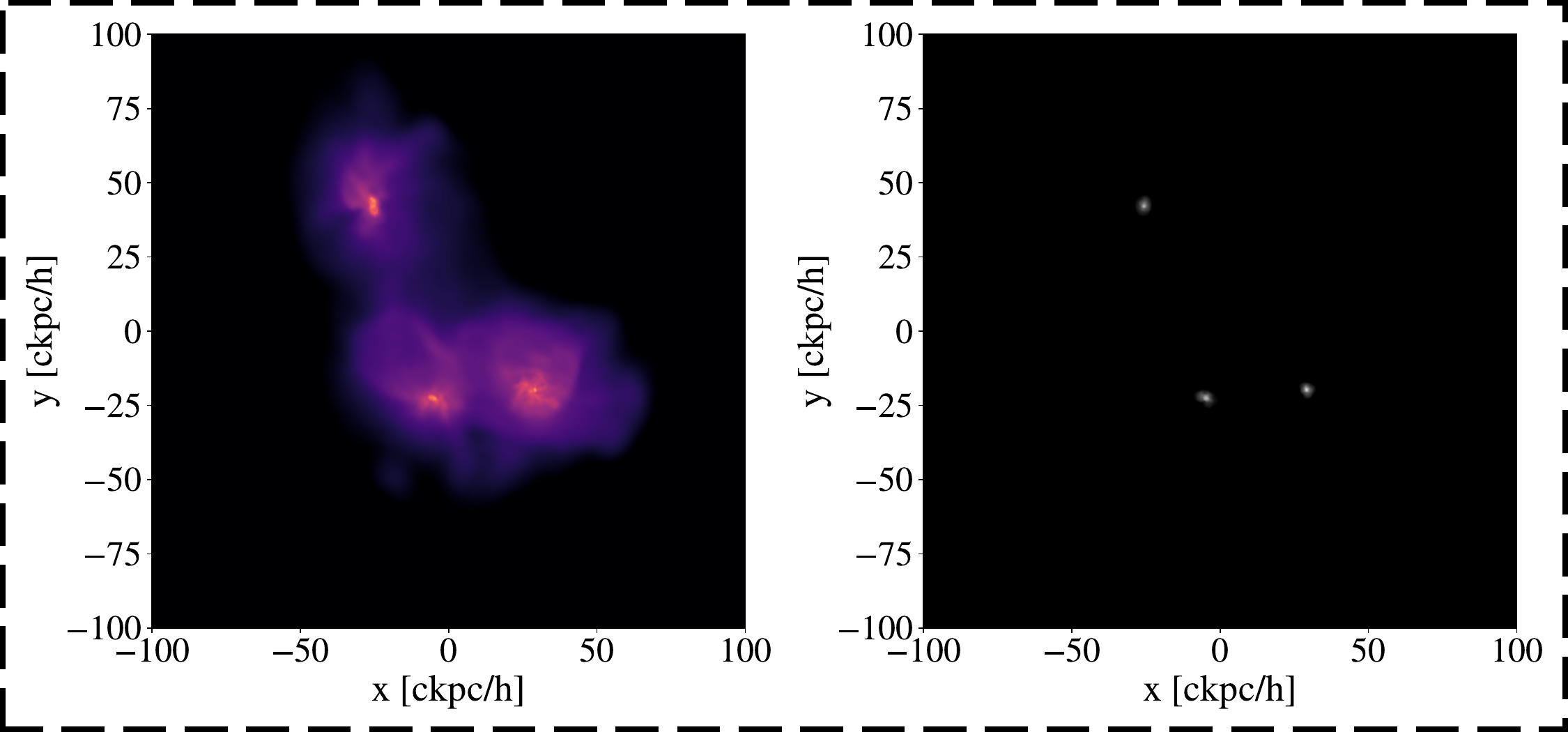}
    \includegraphics[width=0.48\textwidth]{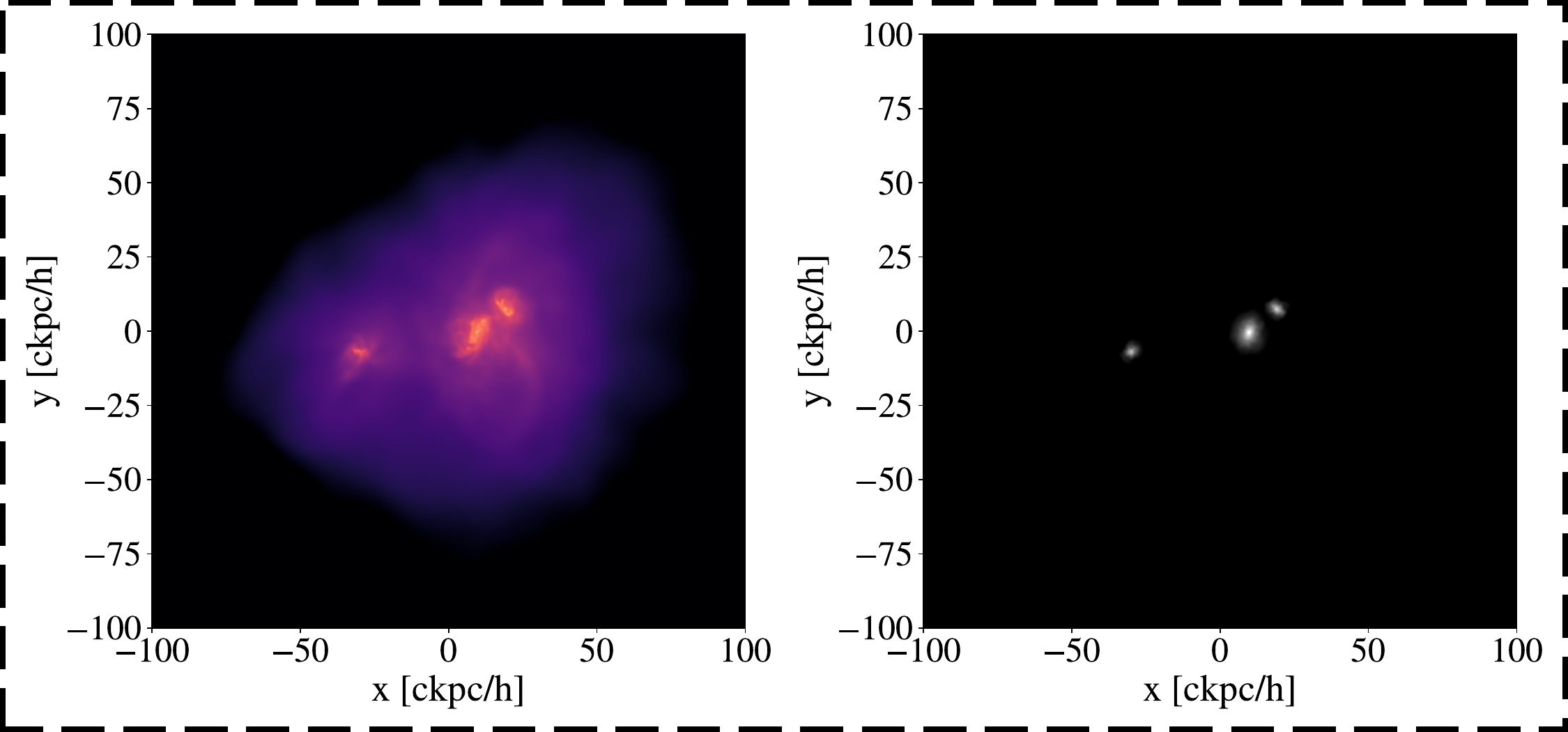}
    \includegraphics[width=0.48\textwidth]{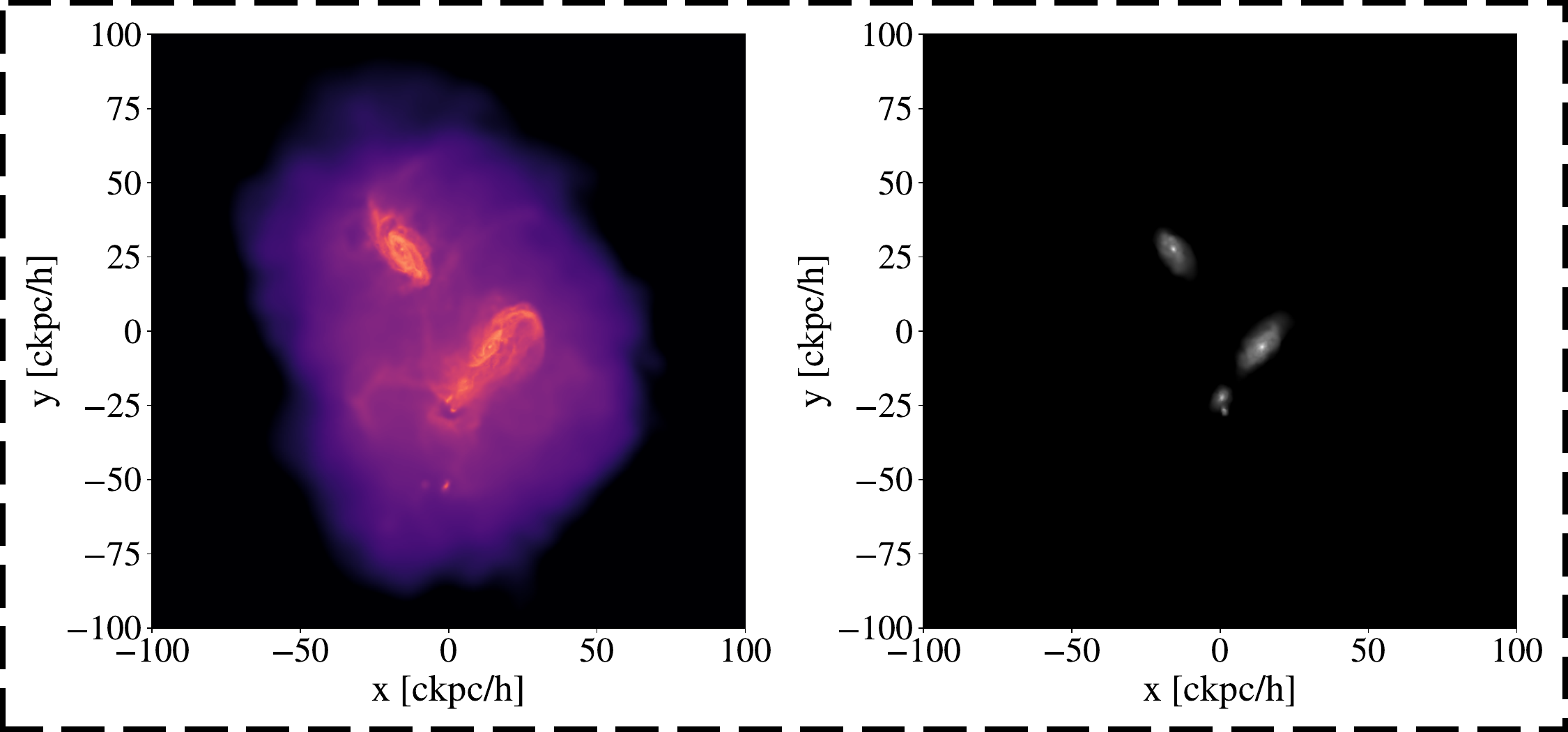}
    \includegraphics[width=0.48\textwidth]{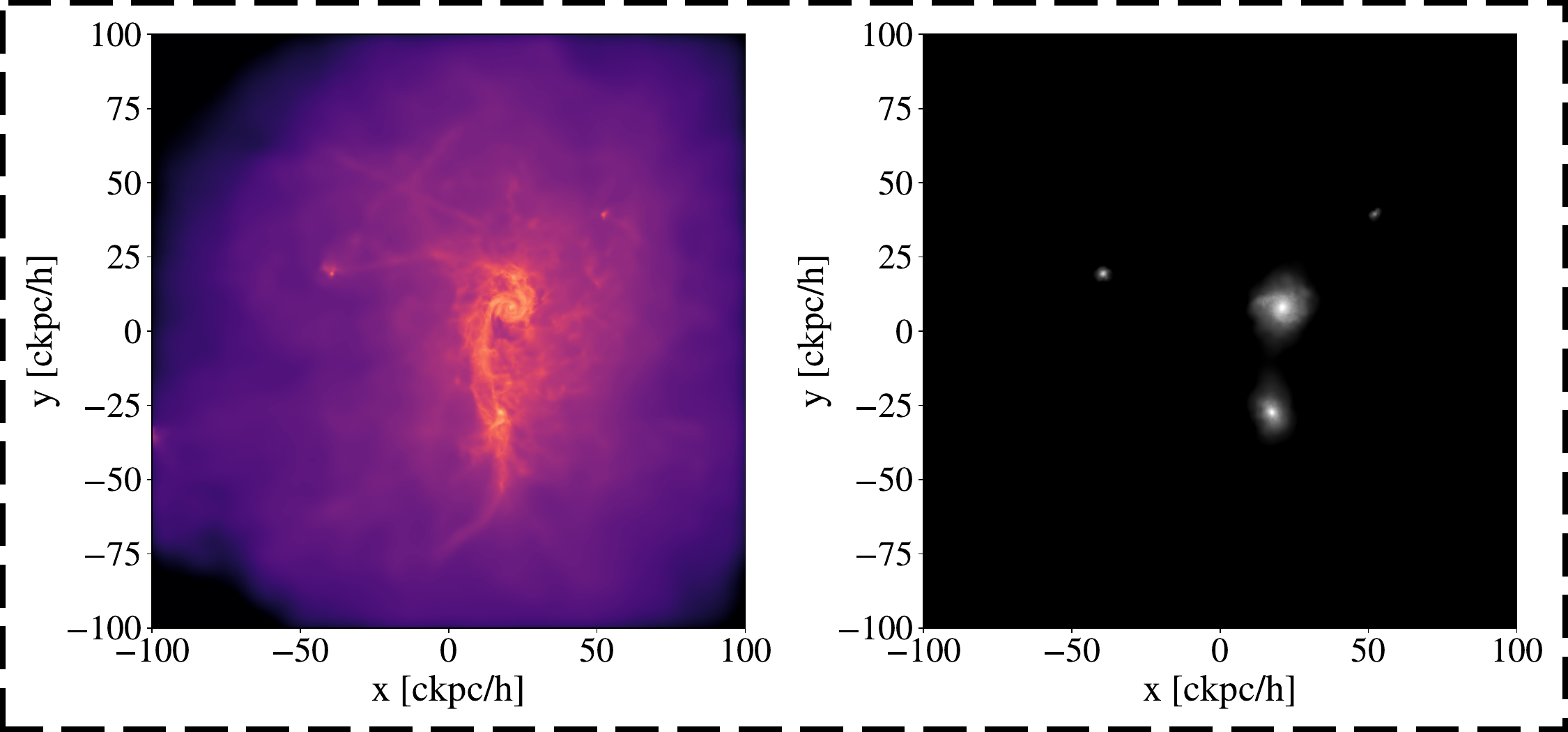}
    \caption{Examples of CGDs found at $z=0$ in TNG50-1. Each pair of density maps shows the distribution of gas (\emph{left}) and stars (\emph{right}) of the groups. We show four groups with different halo masses: CGD-26 with $M_{\rm 200,m} = 10^{10.68} \ \rm M_\odot$ (\emph{upper left}), CGD-6 with $M_{\rm 200,m} = 10^{11.1} \ \rm M_\odot$ (\emph{upper right}), CGD-41 with $M_{\rm 200,m} = 10^{11.31} \ \rm M_\odot$ (\emph{lower left}), CGD-18 with $M_{\rm 200,m} = 10^{11.68} \ \rm M_\odot$ (\emph{lower right}).}
    \label{fig:example_groups}
\end{figure*}

In what follows, we focus on the properties of the 38 CGDs found in the latest snapshot of TNG50-1, corresponding to $z = 0$. In Figure \ref{fig:example_groups}, we show examples of such CGDs. As one can see in the different panels, the stellar component of the groups resembles images of observations of CGDs in the optical \citep[e.g.][]{Stierwalt2017}. As for the gas distribution in these groups, it is clear that it can be very scattered and complex, in agreement with HI observations of dwarf pairs by \citet{Pearson2016, Pearson2018, Luber2022}.

Given that dwarf galaxies are challenging to detect at large distances, the observational samples of CGDs are mostly restricted to the local Universe. Therefore, predictions for the properties of CGDs found at earlier times in the simulation cannot be properly constrained by current observations. Nonetheless, we explore the predictions of TNG50-1 about CGDs at higher redshifts ($0< z \leq 0.5$) in Sections \ref{results:descendants} and \ref{discussion:coalescence}. 

\subsubsection{Observable Properties}
\label{results:obs_properties}
One goal of this work is to use simulations to understand the formation of CGDs found in observations. But first, we investigate whether the general properties of CGDs in simulations are in accordance with what is expected from observations.
To achieve this, we use two observational samples as reference for the properties of CGDs. The first observational sample is built in this work (see Section~\ref{methods:obs_sample}), and the second observational sample is from S17. Neither are supposed to be complete in volume/magnitude; thus, the observational constraints that they impose are also restricted to a small luminosity/mass range. Nonetheless, we can use the ranges of the properties of the groups in the observations as weak constraints for what we can expect cosmological simulations to reproduce. 

\begin{figure}
    \centering
    \includegraphics[width=0.49\textwidth]{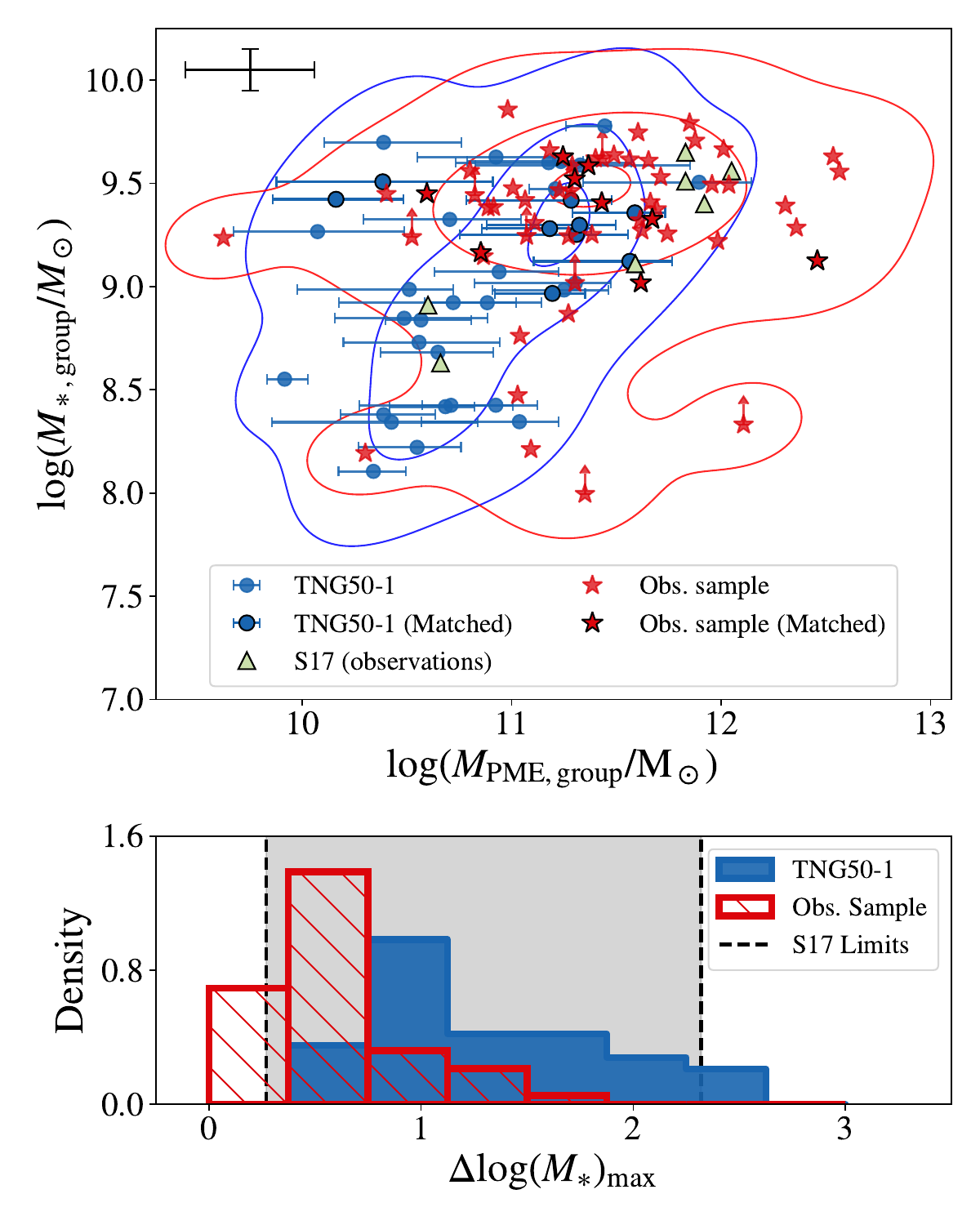}
    \caption{Total and stellar mass scale of CGDs in TNG50-1 and observations ($z \sim 0$). \textbf{Top:} Diagram of total mass (estimated using Equation \ref{eq:proj_mass_estimator}) versus the total stellar mass (sum of $M_\ast$ of all group members). Blue circles are simulated CGDs in TNG50, red stars are CGDs in the observational sample of this work and green triangles are CGDs in the sample of S17. Red stars and blue circles with black edges belong to the magnitude-matched samples described in Section \ref{methods:obs_sample}. Error bars of $M_{\rm PME,group}$ for TNG50 groups are purely from the variation on the estimated masses (25th and 75th percentiles) according to the projection plane. Errors for the observational samples are shown in the top left, CGDs with members lacking measurements of $M_\ast$ have upward arrows to indicate that $M_{\ast \rm , group}$ is a lower limit to the real value. For details on the error bars, see Section \ref{methods:proj_quantities}. Blue and red contours indicate Gaussian kernel density estimations, with three levels of density iso-proportions (0.05, 0.5 and 0.95). \textbf{Bottom:} Maximum difference in $\log (M_{\ast})$ between members of a group. CGDs in TNG50 are presented in the blue histogram, and limits of $\log (M_{\ast})$ in the S17 sample are shown as vertical black dashed lines.}
    \label{fig:obs_range}
\end{figure}

As shown in the top panel of Fig. \ref{fig:obs_range}, we have a diagram region where both simulated and observed CGDs overlap within the error bars. Specially groups from S17, lie within the 95\% contour of the simulated groups. Regarding the sub-samples in observations and simulations, which were matched by magnitude (see Section \ref{methods:obs_sample}), we see that they are almost all within the limits of the 95\% contours of each other parent samples. However, TNG50-1 produces many low-mass groups with no counterparts in the observational sample.
The incompleteness of our observational sample partially causes this disagreement, but the possibility that the simulation is producing spurious low-mass CGDs also cannot be excluded. Unfortunately, the current data is not enough to confidently discard any of the hypotheses. Additionally, the simulation does not produce CGDs as massive as the most massive CGDs in observations, which is more challenging to reconcile with observations but may be due to the limited simulation volume of TNG50-1. Since more massive groups will be relatively less numerous, they may be under-sampled in the cosmological volume of TNG50-1 ($13.8 \times 10^4 \ \mathrm{cMpc^3}$). On the other hand, in our observational sample, the more massive CGDs ($M_{\rm PME,group} > 10^{12} \ \rm M_\odot$) have a median absolute magnitude $M_{r,\rm 3rd}=-17.8$, which, with a magnitude limit of 17.77 in the $r$-band, corresponds to a maximum detection redshift of $z \approx 0.0287$. Considering approximately 7500 $\rm deg^2$ as the observed sky area of the SDSS Legacy spectroscopic sample \citep{Abazajian2009}, the cosmological volume observed up to $z=0.0287$ would be more than 12 times the volume of TNG50-1. Thus, the bright end of our observational sample is selected within a larger volume than the simulation. 

Regarding the maximum difference in stellar mass of group members ($\Delta \log (M_\ast)_{\rm max}$), we see that the most simulated CGDs lie within the limits imposed by S17. Still, the agreement between simulated CGDs and observed CGDs is limited. This means that the analogue systems only partially capture the stellar mass ranges within observed groups. However, we stress that the values of $\Delta \log (M_\ast)_{\rm max}$ may be directly affected by the stellar mass or magnitude selection criteria. 

As we can see in Fig. \ref{fig:obs_range}, for almost the entire range of total masses, the observations lie in the same regions as the simulations. This is an important result because group total masses are never used as a selection criterion to find CGDs in the simulation. We only look for associations of dwarf galaxies by looking at their distances to each other. Thus, the similarity of the group masses in simulations and observations indicates that, at least in terms of mass, the analogue CGDs in the simulation reproduce the scale of these systems in observations.

\begin{figure}
    \centering
    \includegraphics[width=0.47\textwidth]{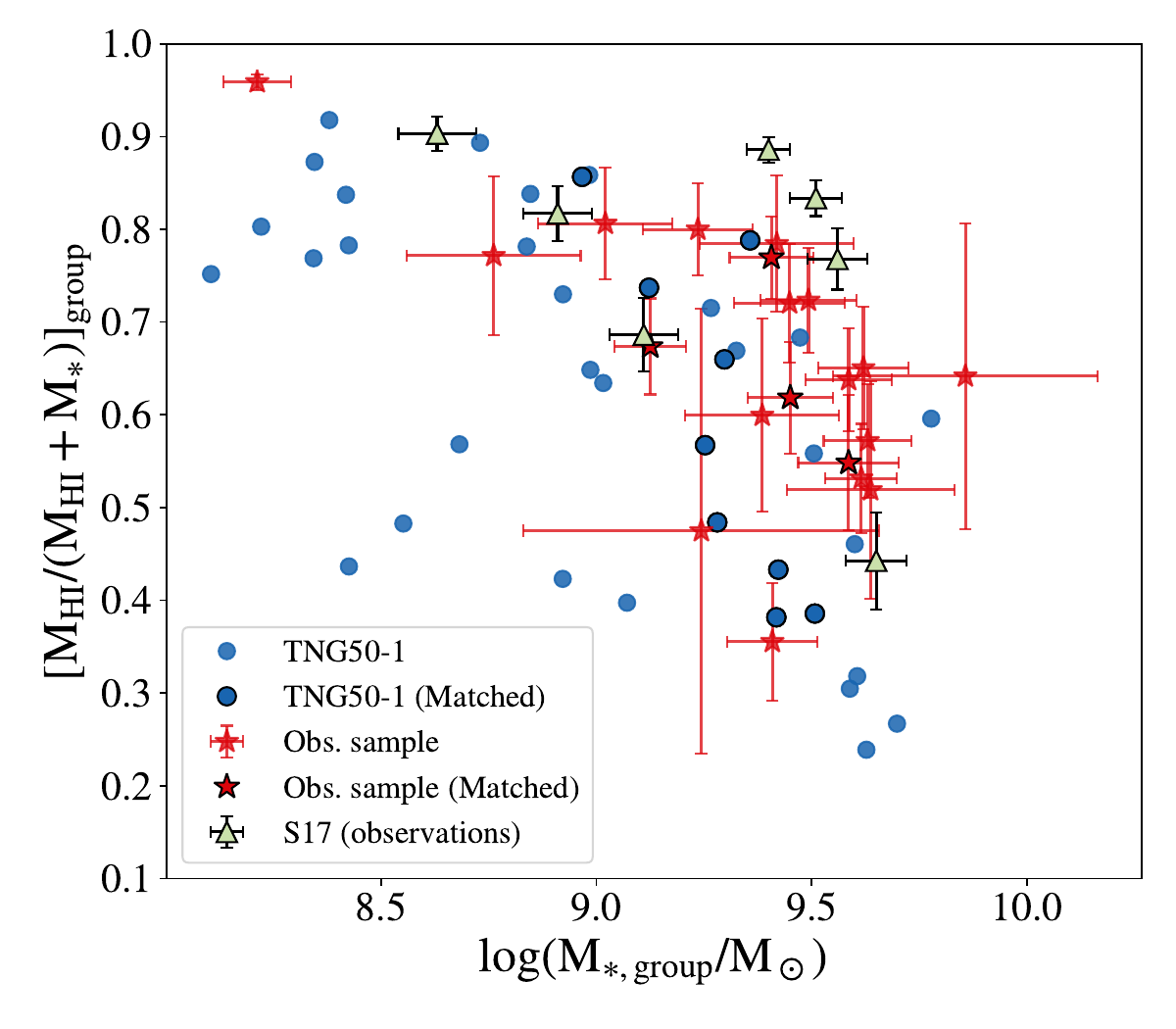}
    \caption{Fraction of HI gas versus total stellar mass in groups. Blue circles are simulated CGDs in TNG50, red stars are CGDs in the observational sample with data in the ALFALFA survey and green triangles are CGDs from the sample of S17. Red stars and blue circles with black edges belong to the magnitude-matched samples described in Section \ref{methods:obs_sample}.}
    \label{fig:HI_frac}
\end{figure}

Regarding the gas content of the groups, if we consider HI mass fractions (Fig. \ref{fig:HI_frac}), we see hints that the simulated CGDs may have lower HI fractions at fixed stellar mass. However, this is only a first approximation of the total HI mass fraction in groups, and the observational samples are even smaller than those shown in Fig. \ref{fig:obs_range}, thus, further work and larger samples are needed to draw more robust conclusions. 

\subsubsection{Host halo properties}
\label{results:halo_properties}
\begin{figure*}
    \centering
    \includegraphics[width=0.495\textwidth]{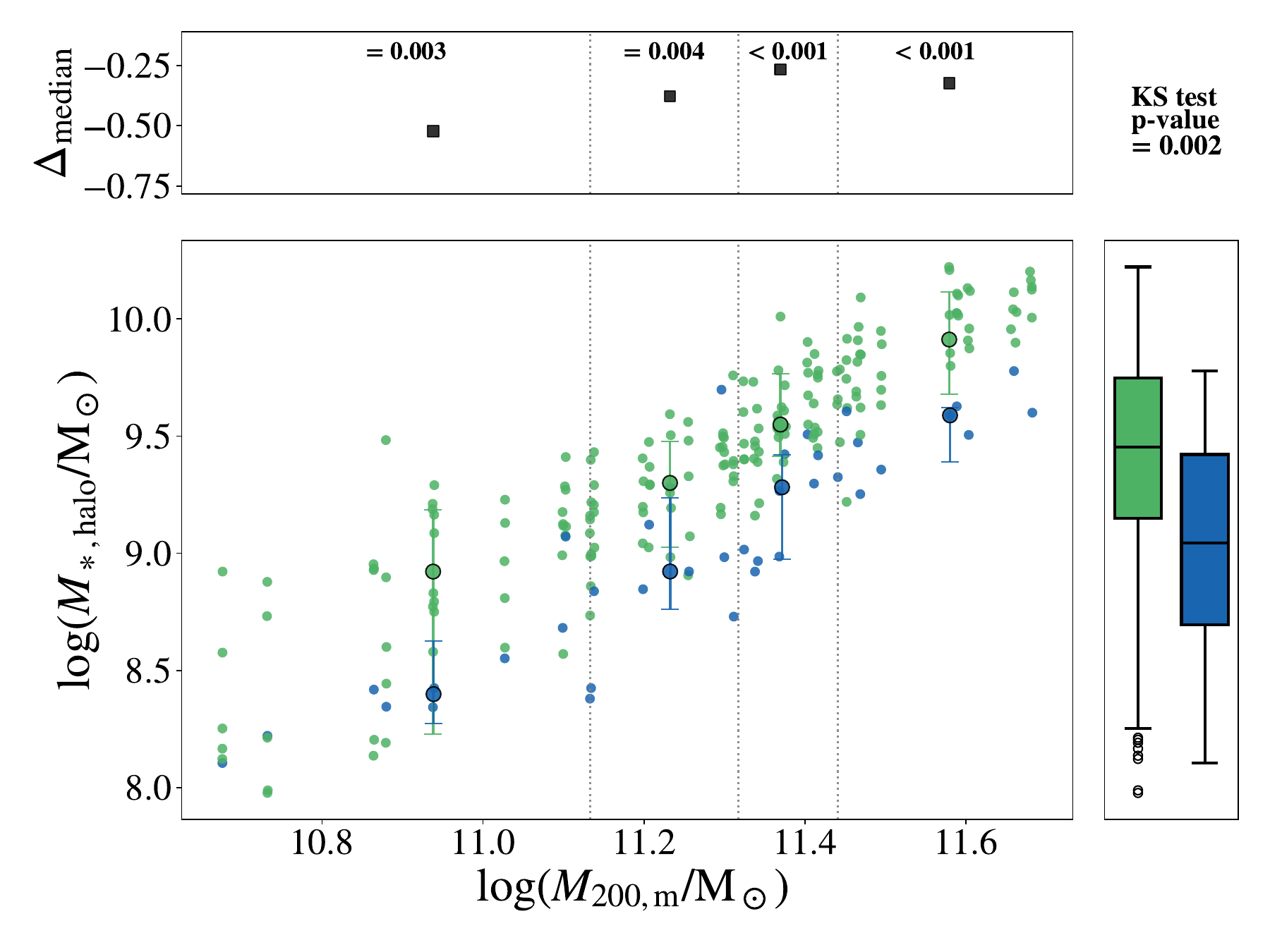}
    \includegraphics[width=0.495\textwidth]{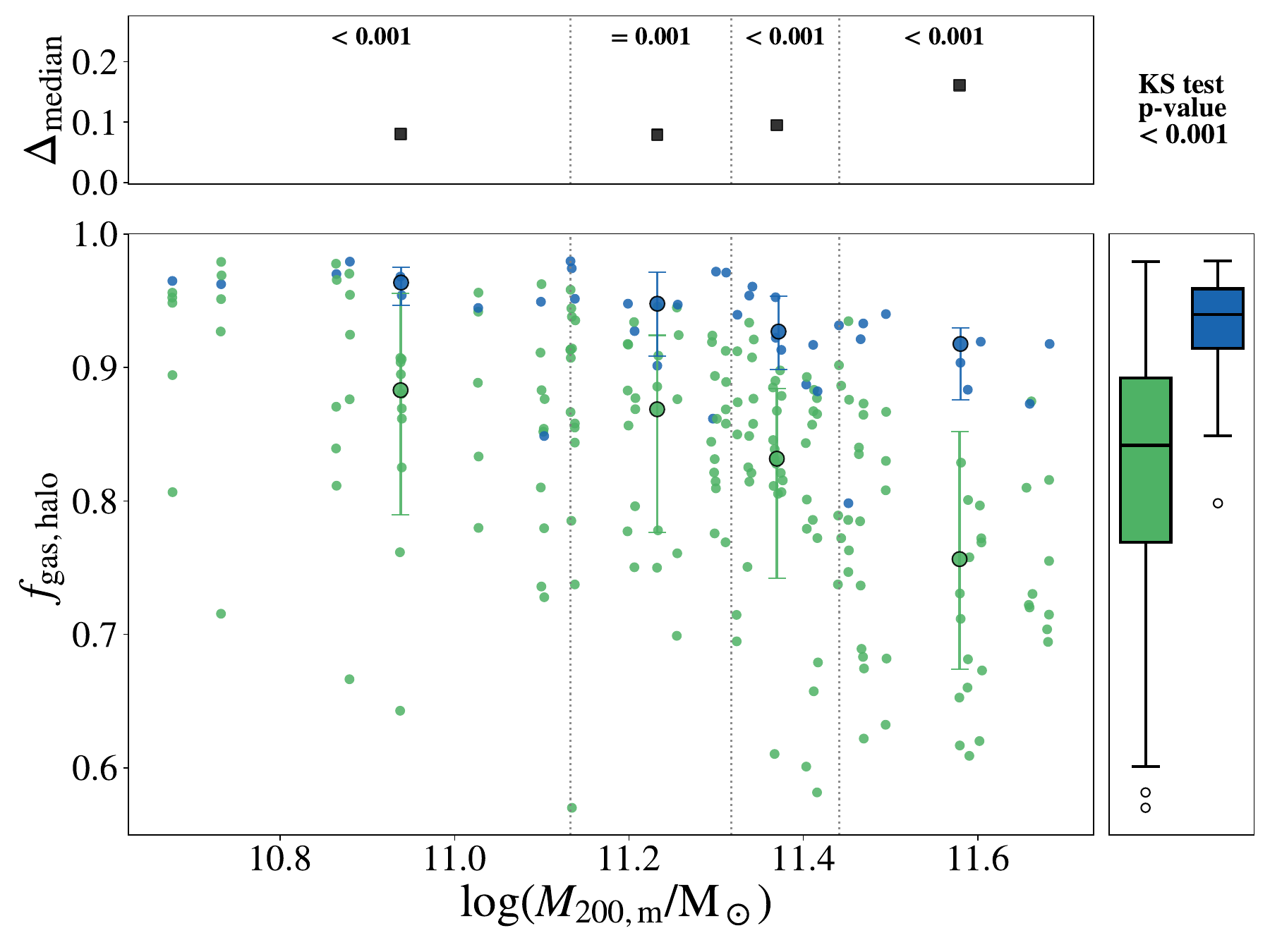}
    \includegraphics[width=0.495\textwidth]{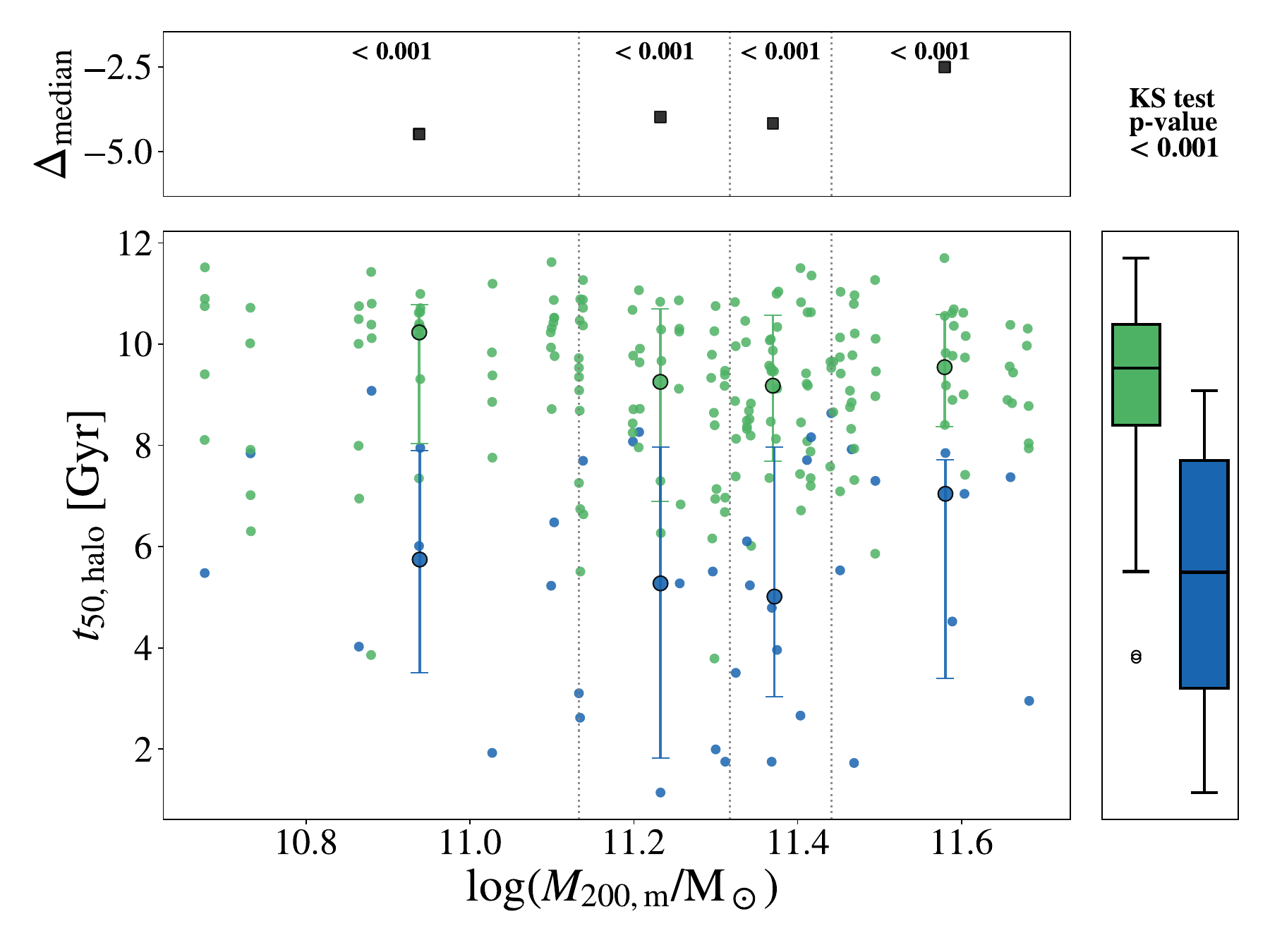}
    \includegraphics[width=0.495\textwidth]{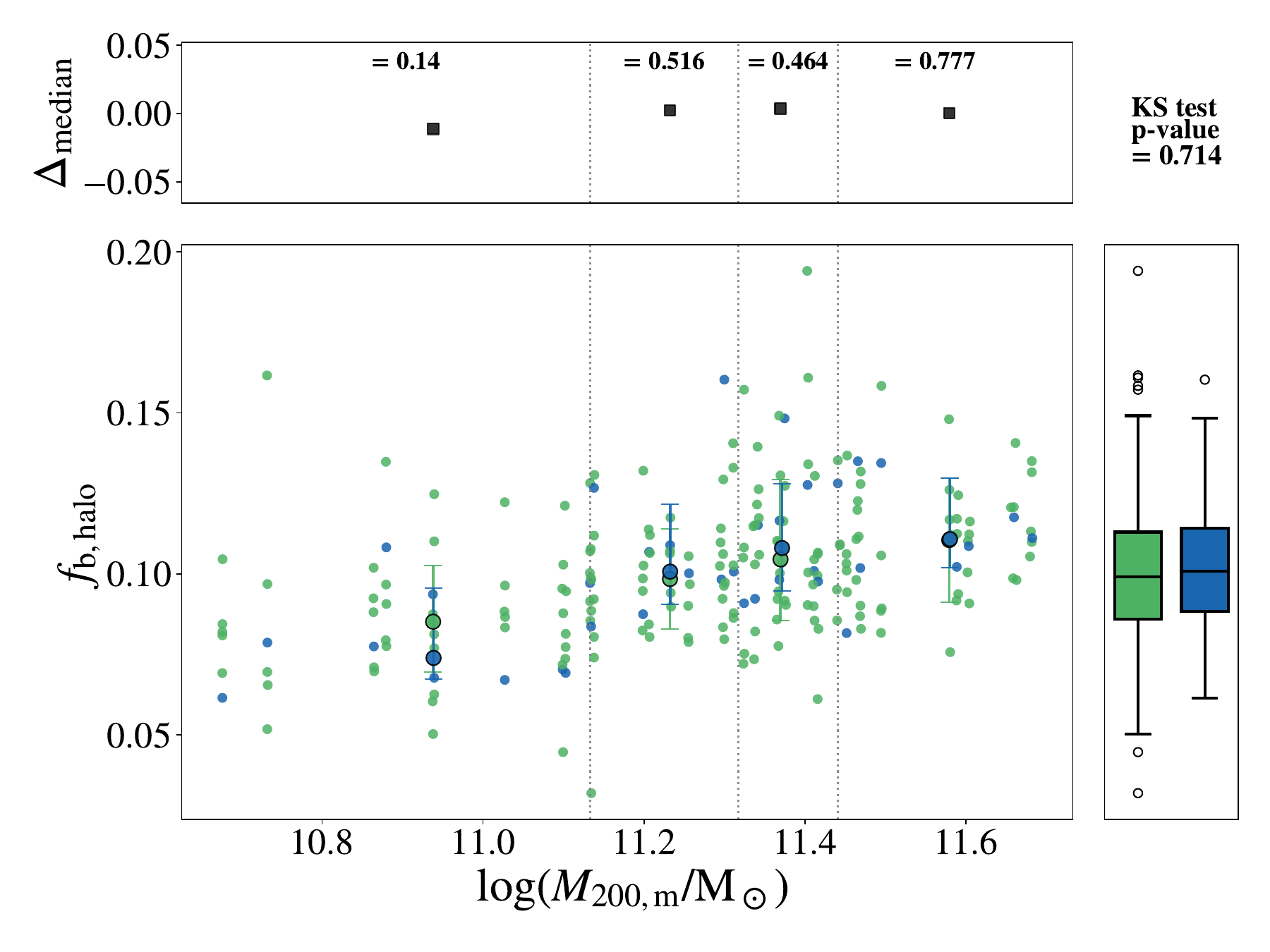}
    \caption{Properties of CGDs host halos. Halos hosting CGDs are shown in blue, and halos in the control sample are shown in green. In all panels, we show the adopted halo mass on the horizontal axes. The properties shown on the vertical axes are total stellar mass (top left), total gas mass fraction (top right), 50\%-mass age (bottom left), and baryonic mass fraction (bottom right). In each panel, we plot grey vertical dotted lines to indicate halo-mass bin edges - defined by the quartiles of $M_{\rm 200,m}$ distribution of halos in control and CGD samples - used to compute median values for the properties. The median values in each mass bin are shown as green/blue circles, with the error bars corresponding to the 16th and 84th percentiles. For each bin, we perform Kolmogorov-Smirnov (KS) tests to check if the distribution of properties of control and CGDs halos are significantly different. We show the test's p-values at the top of the sub-panels above each scatter plot. In the vertical axes of the sub-panels, we show the difference between the medians ($\Delta_{\rm median}$) in each mass bin as black squares. Box plots are shown to illustrate the distribution of the samples over the whole halo mass range (without binning). KS test p-values for the whole samples are shown above the box plots. For completeness, in Appendix \ref{app:cumul_dists}, we present Anderson-Darling tests and cumulative distributions of the halo properties used in this figure.}
    \label{fig:host_halo_properties}
\end{figure*}

In this work, we consider the simulated groups as possible models of the real CGDs. Thus, it is relevant to analyse the basic properties of the halos that host CGDs in simulations to access quantities that can be difficult to measure with observations. By physically characterising the simulated CGDs, we can obtain insights about possible formation scenarios for their observed counterparts and how they fit in the larger context of galaxy formation and evolution. It is expected that properties of halos scale with their mass \citep{Yang2008}, so it is essential to consider this scaling during analysis. Thus, we compare the properties of halos hosting CGDs with control halos having similar masses and not hosting a CGD. In Fig. \ref{fig:host_halo_properties}, we show scatter plots of halo mass ($M_{\rm 200,m}$) versus four different halo properties for the control sample and the CGD hosts sample. The properties are defined as total stellar mass ($M_{\rm \ast,halo}$), which is all stellar mass in the halo; total gas fraction ($f_{\rm gas,halo}$), which is the fraction $M_{\rm gas,halo} / (M_{\rm gas,halo} + M_{\rm \ast,halo})$; 50\%-mass age ($t_{\rm 50,halo}$), which is the lookback time when the halo reached half of its current virial mass, that is $M_{\rm 200,c}(z = z_{\rm 50,halo}) = M_{\rm 200,c}(z = 0)/2$; and a baryonic mass fraction ($f_{\rm b,halo}$), which is the fraction $(M_{\rm gas,halo} + M_{\rm \ast,halo}) / M_{\rm 200,m}$.

The median differences in total stellar mass are of the order of $-0.4$ dex, with CGDs having less than half of the stellar mass present in the control halos. Regarding how gas-rich the CGD halos are, we see that they tend to have $\sim 10 \%$ more gas than other halos of similar mass, with this difference being slightly higher for CGDs in the most massive bin ($11.45 \lesssim M_{\rm 200,c} \lesssim 11.8$). Additionally, analysing the difference in the halo ages, we see that, on average, CGD host halos assembled $\sim 4 $~Gyr later than other halos of similar mass, with a smaller difference ($\sim 2.5$~Gyr) for the most massive bin. By analysing the fraction of baryons in both halo samples, we see that CGD hosts have similar amounts of baryons to control halos, indicating that that the differences in halo properties mentioned so far are not a by-product of differences in the baryonic mass fraction.

In summary, we found that CGDs tend to inhabit halos with less stellar mass and higher gas fractions than the control sample. The halos that host the groups also assembled half of their virial mass later when compared to halos of similar mass. The tendencies on halo properties described so far are significant over all the halo mass range probed and suggest that CGDs inhabit halos with a different assembly history from the overall population of halos within the same mass regime at $z = 0$. We explored this possibility further in the results presented in Section \ref{results:assembly} and discuss them in Section \ref{discussion:assembly}.

From the results presented so far, we see that the CGD host halos are a special population because they have specific properties that deviate from the overall halo population at the same redshift and mass interval. However, the sample of CGDs in TNG50-1 is small ($N < 50$) and may be biased. \cite{Borrow2023} draws attention to the stochastic effects on the properties of individual galaxies, and they show that re-simulations of a galaxy can result in differences up to 25\% in stellar mass. However, median values for scaling relations should not be significantly affected. Thus, the predictions for populations of galaxies are more robust against stochastic variations than predictions for individual galaxies. We can apply this logic to the population of CGDs found in this work and argue that the overall median values of halo properties can possibly be robust against the stochastic effects of a single simulation run (TNG50-1).

\subsubsection{Assembly History}
\label{results:assembly}
\begin{figure}
    \centering
    \includegraphics[width=0.47\textwidth]{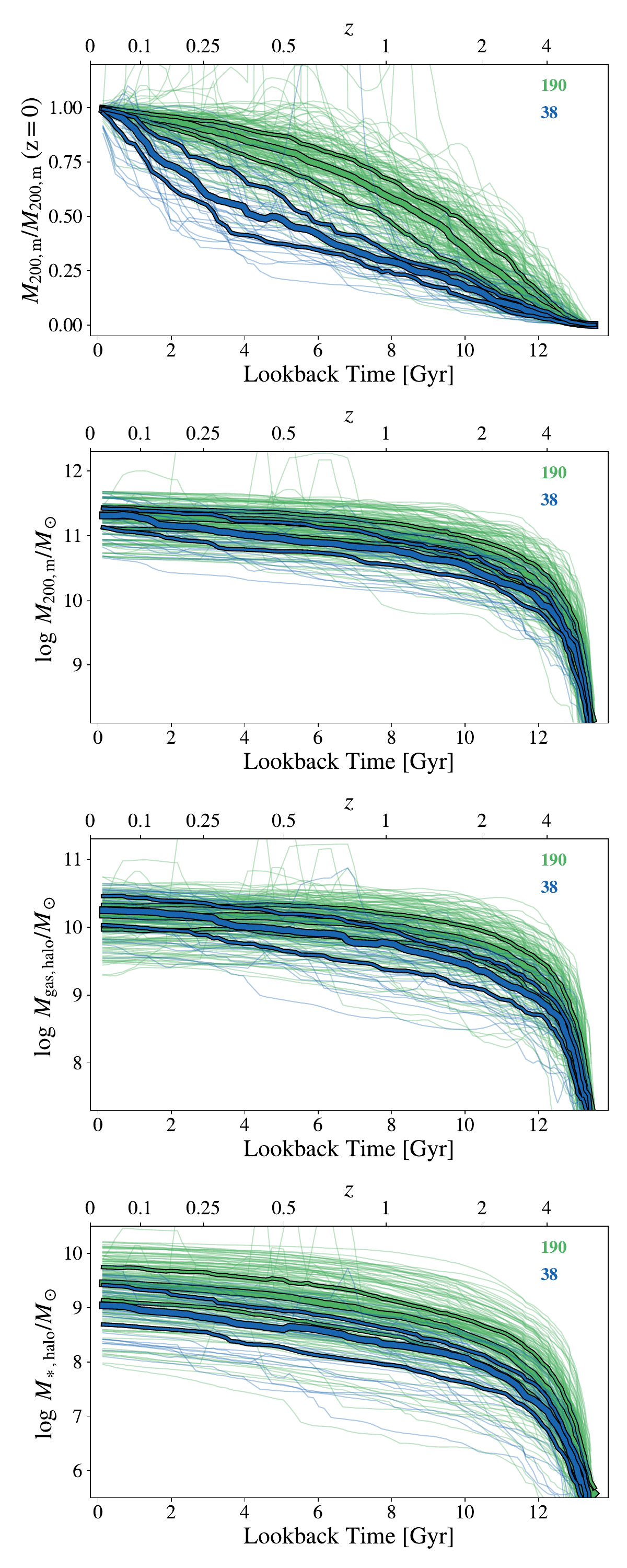}
    \caption{Mass assembly histories (MAH) of halos hosting CGDs (blue) and halos in the control sample (green). From top to bottom: normalised total MAH, total MAH, gas MAH and stellar MAH. Median values for each sample are shown as thick solid lines, while thick dashed lines show the 25th and 75th percentiles. MAHs of individual halos in each sample are shown as thin semitransparent solid lines.}
    \label{fig:assembly_halos}
\end{figure}

To fully characterise the CGDs found at $z=0$ in TNG50-1, we can go beyond their current properties and analyse their mass assembly history (MAH), readily available in a cosmological simulation. To understand the mass assembly of CGDs, it is useful to compare them with the control sample of halos with similar mass. Thus, in Figure \ref{fig:assembly_halos}, we present the mass growth of control and CGD halos over the whole time span of the cosmological simulation.

As we can see in the normalised total MAH, the halos hosting CGDs got their masses later when compared to the halos in the control sample. In other words, CGDs, on average, assembled relatively more of their mass in the last few Gyr. As is shown in the gas MAH, the CGD and control halos also only reached similar values of $M_{\rm gas,halo}$ recently. The median gas accretion rate of CGDs was higher for several Gyr, and this is illustrated by the steeper slope of the CGDs gas MAH since $z \sim 0.5$. By analysing the evolution of gas distribution in simulated CGDs (see videos in Supplementary Material), it becomes clear that the accretion of gas in these systems happens hierarchically, that is, from the merger of similar low-mass halos that only recently assembled into a single structure. This would be contrasted with the accretion of a single, more massive halo continuously receiving gas from its surroundings and the cosmic web. This crucial difference in the accretion mode could explain why the total and gas MAHs look steeper than in the control sample for $z \lesssim 0.5$. Although the total and gas masses were accreted at higher rates and reached similar values for control and CGD halos at $z=0$, the picture differs for the stellar masses. As we can see in the stellar MAH, on average, CGDs have always had fewer stars than other halos of similar mass at $z = 0$. The results presented in this subsection indicate that the stellar and gas fractions of CGD halos differs from that of other halos with similar mass because of the particular assembly history of these compact groups. More specifically, the halo properties of CGDs found at $z=0$ seem to be the result of a late hierarchical assembly of low-mass ($M_{\rm 200,m} \lesssim 10^{10} \ \rm M_\odot$) halos.

\begin{figure}
    \centering
    \includegraphics[width=0.46\textwidth]{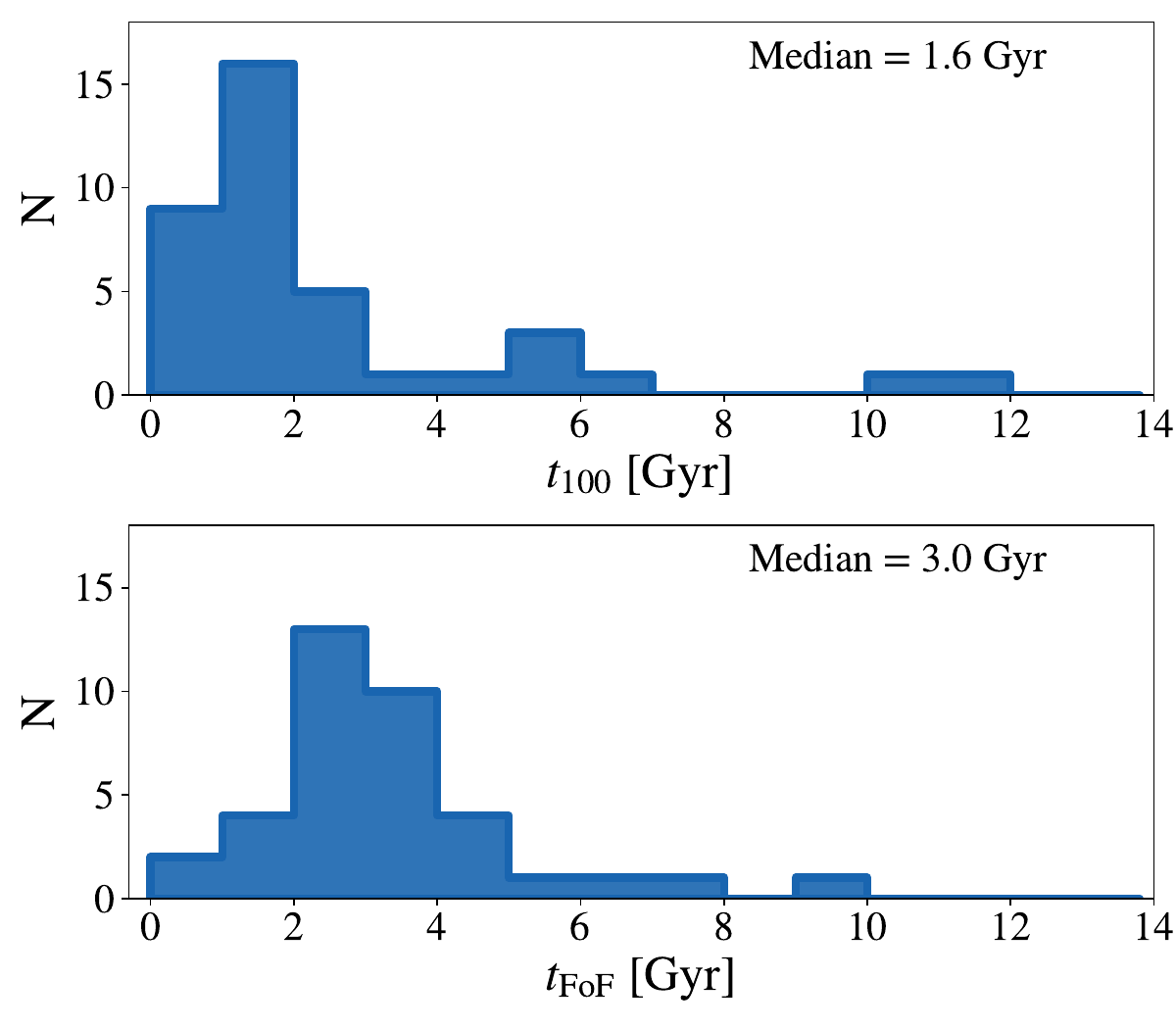}
    \caption{Histograms of different group formation time indicators. \textbf{Top:} maximum lookback time which $r_{\rm group} < 100 \ \rm kpc$. \textbf{Bottom:} maximum lookback time in which all group members belong to the same halo (FoF group). The median values of the indicators are shown on the top right of each panel.}
    \label{fig:timescales}
\end{figure}

We can also study the history of the CGDs to quantify how long these groups can exist in the simulation. In Fig. \ref{fig:timescales}, we show a histogram of group formation time indicators derived from the analysis of merger trees (see Section \ref{methods:mergertree}), which can be adopted as criteria to determine the moment a group can be considered to be "formed". In practice, both timescale indicators presented in Fig. \ref{fig:timescales} are challenging or even impossible to infer in observations, but through this work, we are going to consider the $t_{100}$ estimator as the closest to what we can call an \textit{age} of the CGDs. We choose it because it is defined more straightforwardly than $t_{\rm FoF}$ and, in principle, can be more easily inferred from observations of the kinematics of galaxies within the groups. As is shown in the histogram of $t_{100}$, the majority of CGDs formed compact associations of dwarf galaxies in the last 2 Gyrs, but they also can be much older, with two of the groups being older than 10~Gyr. The results for $t_{\rm FoF}$ are quite different, with most groups forming between 2 and 4 Gyr ago, where the median for all the CGDs is $\sim 3$~Gyr. These lookback times coincide with the epochs in which the CGDs hosts present steeper slopes in their normalised total MAH (see Figure \ref{fig:assembly_halos}). Thus, $t_{\rm FoF}$ may capture the epoch of the hierarchical assembly that builds up CGDs halos in the simulation. While $t_{100}$ may better trace the time since which galaxies start to interact more strongly inside the groups.

\subsubsection{Position in the Large-Scale Structure}
Groups of galaxies in the real Universe and simulations are not formed in random positions of space, they are usually found close to matter clustering in the large-scale structure (LSS). Thus, it is relevant to explore whether low-mass groups - such as CGDs - follow this trend and if they are situated in specific regions of the LSS. The advantage of doing this kind of analysis in simulations is that we can access all positions in real space, precisely computing local densities and distances between galaxies in the simulation. For this brief exploration of the location of CGDs in the LSS, we use a proxy of the underlying matter density, the local galaxy density contrast ($\delta_{\rm 1 Mpc}$), described in Section \ref{methods:nn_catalog}. This proxy of total matter density has the advantage of being accessible also in observations, especially with the advent of modern facilities, which will map the large-scale structure of the Universe through spectroscopic surveys, e.g. DESI \citep{DESI2023} and Euclid \citep{Euclid2022}. 

\begin{figure*}
    \centering
    \includegraphics[width=0.99\textwidth]{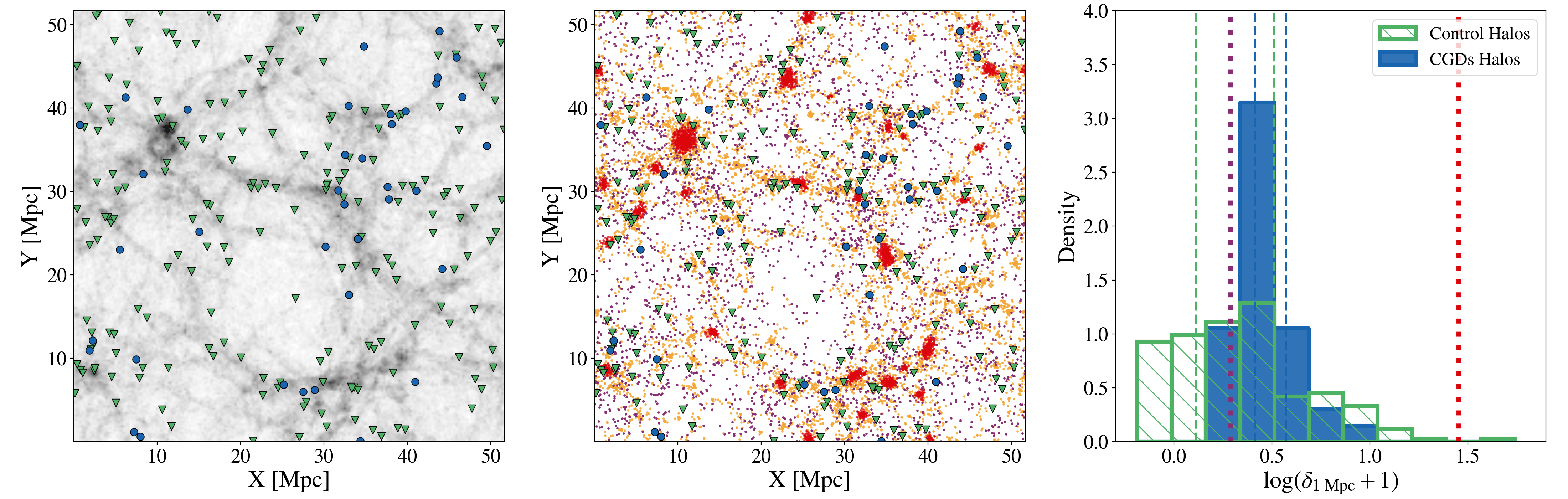}
    \caption{Locations of CGDs in the large-scale structure of TNG50-1. \textbf{Left:} Projected halo density along the z-axis of the simulation. High-density regions are shown as darker shades of grey, while low-density regions are shown as lighter shades. Halos hosting CGDs and control halos are shown as blue circles and green triangles, respectively. \textbf{Center:} Projected position of galaxies along the z-axis of the simulation. Each point represents a galaxy, colour-coded by its local galaxy density contrast ($\delta_{\rm 1 \ Mpc}$). Purple points are galaxies in the sparsest regions (1st $\delta_{\rm 1 \ Mpc}$ quartile), red points are galaxies in the densest regions (4th $\delta_{\rm 1 \ Mpc}$ quartile), and orange points are galaxies in regions with densities in between (2nd and 3rd $\delta_{\rm 1 \ Mpc}$ quartiles). The quartiles are defined from the $\delta_{\rm 1 \ Mpc}$ distribution of all galaxies in TNG50-1 with $M_{\ast} \geq 10^7 \ \rm M_\odot$. The colours of the triangles and circles are as in the left panel. \textbf{Right:} Histogram of local galaxy density around galaxies. Host halos of CGDs are represented by blue, and control halos by green. Blue and green vertical dashed lines are the 25th and 75th $\delta_{\rm 1 \ Mpc}$ percentiles within each of the halo samples. Thick vertical dotted lines represent the 25th (violet) and 75th (red) percentiles of the whole galaxy sample in TNG50-1. The most massive galaxies of CGD and control halos represent respective parent halos in all panels of this figure.}
    \label{fig:pos_in_LSS}
\end{figure*}

As is qualitatively shown in the left and centre panels of Fig. \ref{fig:pos_in_LSS}, the CGDs tend to be situated away from the centres of voids and knots in the cosmic web, differently from the control halo sample. When we look at the distribution of the adopted large-scale environment proxy - right panel of Fig. \ref{fig:pos_in_LSS} - we see that CGDs are concentrated in a narrow range of local density, while the halos in the control sample are distributed over a wider range of $\delta_{\rm 1 \ Mpc}$ values. Although both CGDs and control halos avoid the highest densities, only the CGD hosts avoid the lowest density extremes in the simulation volume, being primarily found in regions with densities slightly above the 25th percentile. The avoidance of high-density environments of the cosmic web (e.g. knots) can be expected because low-mass halos ($M_{\rm 200,m} < 10^{12} \ \rm M_\odot$) may be easily accreted into the larger halos that inhabit these environments, becoming subhalos of them or being disrupted. However, the fact that CGDs also avoid low-density environments (e.g. voids) is less straightforward. That CGDs do not inhabit the lowest density environments at present implies the possibility that these groups can only form above a certain threshold of local matter density, while control halos do not have this restriction. Given that the control halos are paired by mass only at $z=0$, the differences in the large-scale environment at $z=0$ may be a consequence of past local over-density and/or different halo assembly histories (see Sections \ref{results:assembly} and \ref{discussion:assembly}).

\subsection{Coalescence Fractions and Timescales}
\label{results:descendants}
\begin{figure}
    \centering
    \includegraphics[width=0.49\textwidth]{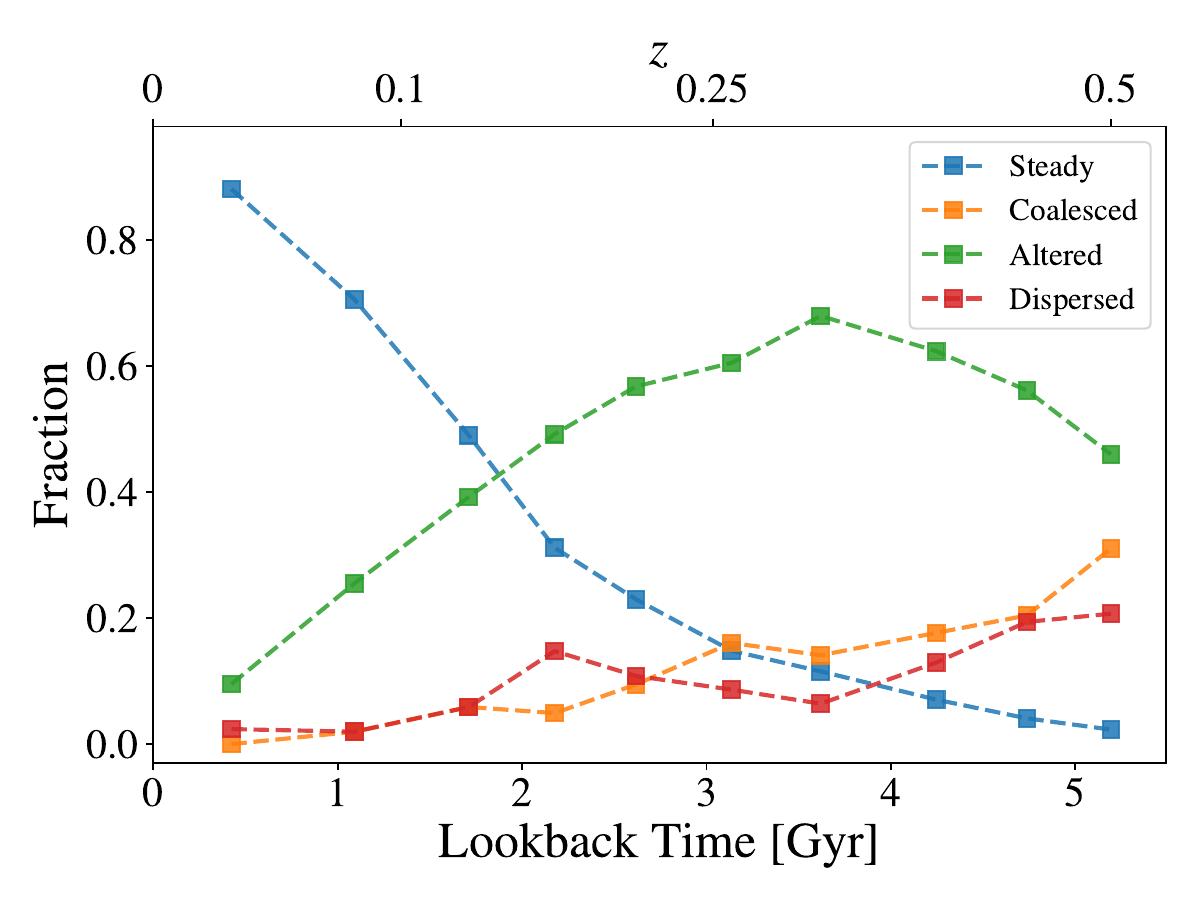}
    \caption{Fraction of CGDs final states at $z = 0$ as a function of time in which they were identified. Each line represents one of the states defined in Section \ref{methods:mergertree}.}
    \label{fig:group_final_states}
\end{figure}

One way to understand how the population of CGDs evolves with redshift is to look at their fate in the last snapshot of the simulation. We categorise the final states of CGDs found at $0 < z \leq 0.5$ using the merger trees of their group members and divide them into four classes: Steady, Altered, Coalesced and Dispersed (see Section \ref{methods:mergertree}). In Fig. \ref{fig:group_final_states}, we show the fraction of each class as a function of redshift. The fractions are computed for each group catalogue in a given redshift, that is, for each $z$, the sum of the fractions of the four classes is 1. In this way, we can see the fractions as probabilities of a CGD at $z$ having a certain final state at $z=0$. 

The most obvious trend that we see in Figure \ref{fig:group_final_states} is that the number of "Steady" groups increases rapidly as we approach $z=0$, and the number of "Coalesced" groups increases slowly with redshift. Both trends are related to the coalescence timescale of CGDs. For example, $\sim 30$ \% of the CGDs found in $z=0.5$ coalesced into a single galaxy, while only $\sim 2$ \% remained unaltered ("Steady") up to $z=0$. Meanwhile, the fraction of dispersed groups remains below 15\% for most of the redshift interval presented. From this, we can state some main results: first, even though it is a minority, the simulation predicts that a tiny fraction of CGDs can survive for more than 5 Gyr with all its original group members; second, the most common final state for CGDs identified at $z > 0.2$ is to be altered, that is, accrete new galaxies (massive or not) in the group and/or have mergers between original members of the group; third, the dismantling of groups is not common, meaning that a small fraction of them are transient unbound structures. These fractions presented in Fig. \ref{fig:group_final_states} provide important information to infer how much time we can expect the CGDs in observations to serve as isolated laboratories to study environmental effects in dwarf-only contexts. 

\begin{figure}
    \centering
    \includegraphics[width=0.49\textwidth]{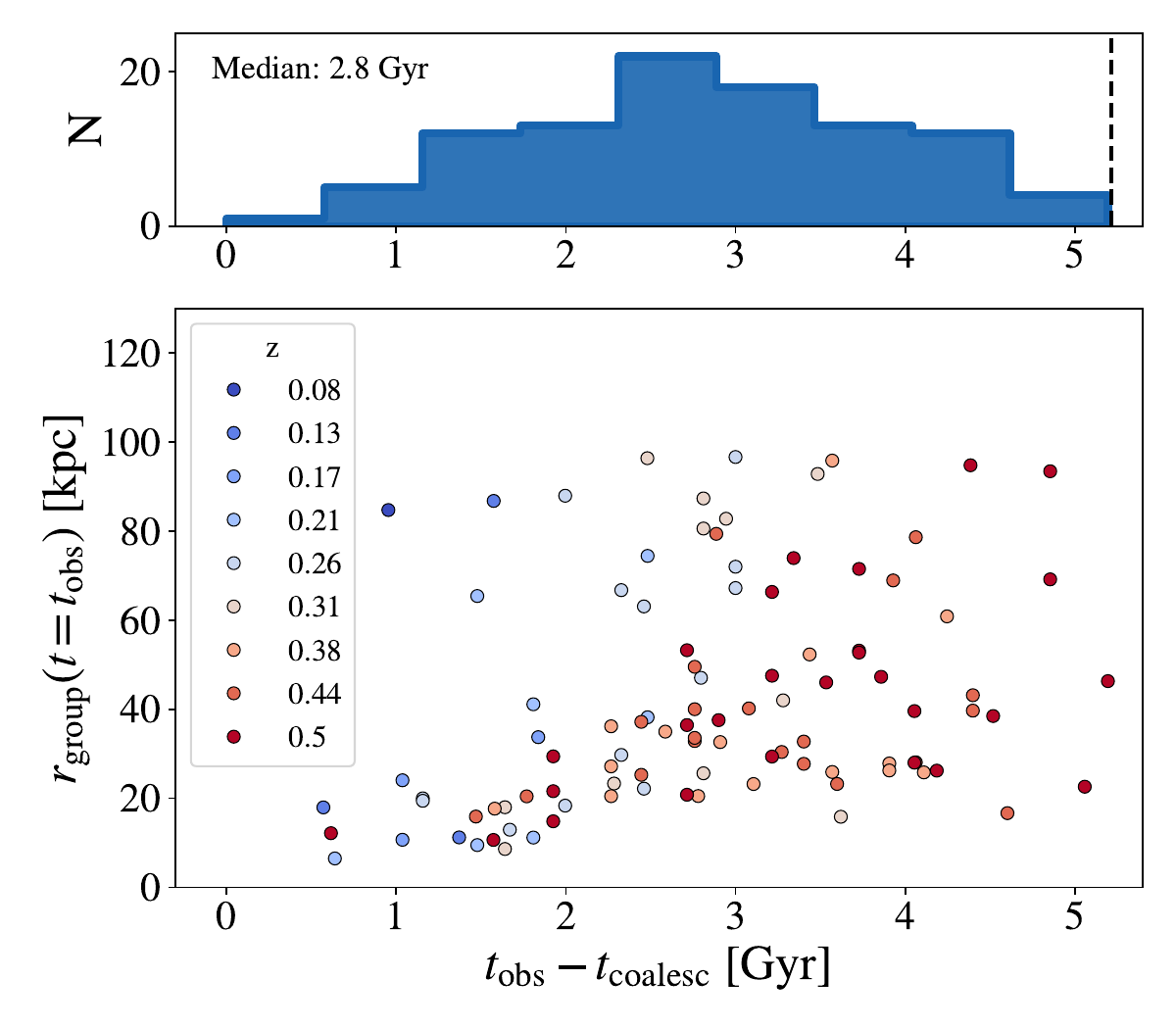}
    \caption{\textbf{Top:} Histogram of coalescence timescales for CGDs found at $0 < z \leq 0.5$. The timescale shown is the difference between the lookback time in which a group was identified ($t_{\rm obs}$) and the earliest lookback time in which the same group was fully coalesced ($t_{\rm coalesc}$). The black dashed line indicates an upper limit (lookback time at $z=0.5$), imposed by the maximum redshift in which the CGDs were selected. \textbf{Bottom:} Coalescence timescale versus group radius. The colour of the circles indicates the redshift of the snapshot in which the CGD was identified, corresponding to $t_{\rm obs}$ in which $r_{\rm group}$ is determined.}
    \label{fig:tcoal_and_Rgroup}
\end{figure}

Looking more specifically at the class of coalesced groups, we can determine the coalescence timescales of CGDs and provide typical timescales for the maximum interaction time of the dwarf galaxies in these environments. As shown in Fig. \ref{fig:tcoal_and_Rgroup}, most of the CGDs found at $z \leq 0.5$ take more than 1 Gyr to completely merge into a single galaxy, with a typical timescale of $\sim 3$ Gyr. This is a reasonable amount of time in terms of galaxy evolution, allowing for the dwarf galaxies in CGDs to have multiple episodes of star formation that may be triggered due to close encounters or fly-bys \citep{Martin2021}. 

We also investigate if the time it takes for a CGD to merge fully has any dependence on the radius of the group at the time it is identified. As shown in Fig. \ref{fig:tcoal_and_Rgroup}, there is no clear trend between the coalescence time and $r_{\rm group}$ over the whole range of group radius. Yet, we can see a tentative trend for $r_{\rm group} \lesssim 40 \ \rm kpc$, where the larger the radius, the longer it takes for the group to coalesce.


\section{Discussion}
\label{sec:discussion}

\subsection{CGDs in TNG50-1}
\label{discussion:CGDs_in_TNG}
 
\begin{figure}
    \centering
    \includegraphics[width=0.49\textwidth]{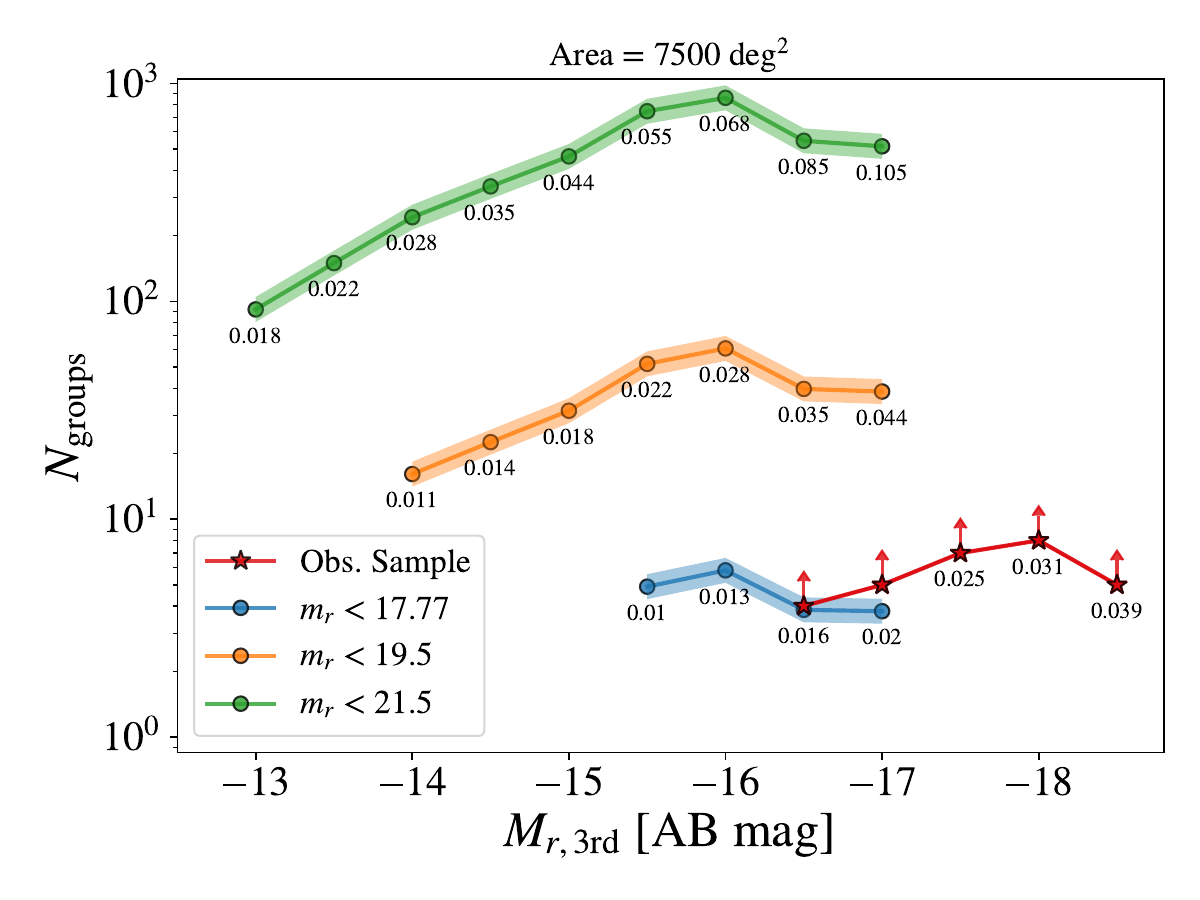}
    \caption{Predicted number of CGDs found in magnitude-limited surveys. The circles are predictions of the number of CGDs given an absolute magnitude of the third brightest group member ($M_{r,\rm 3rd}$) and an apparent magnitude limit on SDSS $r$-band ($m_r$). In the top panel, we show constraints from our observational sample as red stars, representing the number of CGDs in the observational sample with $M_{r,\rm 3rd}$ smaller than a given value on the horizontal axis. The constraints are only lower limits for the number of CGDs due to our selection criteria. Below each point is annotated the redshift, $z_{\rm max}$, corresponding to a maximum luminosity distance at which a group with a specific $M_{r,\rm 3rd}$ is detected; in other words, the redshift corresponding to the comoving volume in which the detected CGDs are enclosed.}
    \label{fig:nd_prediction}
\end{figure}

In Section \ref{results:number_density}, we presented the number density of simulated CGDs and showed how it evolved within the redshift interval of $0 \leq z \leq 0.5$. This is an important result for the statistics of CGDs, being relevant to predicting the number of observed CGDs, and possibly bringing useful insights for improvement in the numerical models. Considering this, we use the z=0 CGD number density estimated in the TNG50-1 to predict the number of CGDs that may be found, given different $r$-band magnitude limits and a fixed sky coverage\footnote{We choose this to be approximately the same area from SDSS Legacy Survey because our observation sample is constructed on it.} of 7500 $\rm deg^2$. In Fig. \ref{fig:nd_prediction}, we show the predicted number of CGDs ($N_{\rm groups}$) that should be found in observations according to the results we describe in Section \ref{results:number_density}. For the comparison with our observational sample, we considered only absolute magnitudes $M_{r \rm , 3rd}$ where the maximum redshift of detection ($z_{\rm max}$) is greater than 0.01, to have a reasonable sampling volume. Since our observational sample is constructed from the spectroscopic catalogue of the SDSS Legacy Survey, our reference magnitude limit for comparison in Fig. \ref{fig:nd_prediction} is $m_r =17.77$ (shown in blue). This causes the confrontation with the observations in Fig. \ref{fig:nd_prediction} to be limited to a short magnitude range, however, it is still helpful to check whether, even in this range, we see some disagreement. By construction, the observational sample of CGDs in this work gives only a lower limit to the estimated number of CGDs in the local Universe, and we see that the predictions from TNG50-1 are roughly in agreement with such estimates, at least in the same order of magnitude. Still, it is essential to have larger samples of dwarf galaxy groups complete to fainter magnitudes ($M_r > -16$), to more strongly constrain current models of galaxy formation and evolution in the low-mass regime. Current \citep{Luo2023,Darragh-Ford2023} and future deeper surveys - such as those conducted by Vera Rubin Observatory \citep{Ivezic2019}, DESI \citep{DESI2023} and Euclid \citep{Euclid2022} - will be essential for this goal. For example, the DESI Bright Galaxy Survey \citep{Hahn2023} will have a magnitude limit of $M_r < 19.5$ for its "Bright" target sample, allowing for the spectroscopic confirmation of fainter and further CGDs not present in our observational sample (see orange line on Fig. \ref{fig:nd_prediction}). Even without spectra, deep photometric surveys will show low surface brightness tidal features resulting from interactions, which are important signals to infer the proximity of dwarfs in pairs or groups. 

The number density of CGDs can be viewed as a prediction from the TNG50-1 cosmological simulation, which may be helpful to test it as a model for galaxy formation and evolution. If the number density of CGDs is above the values expected from observations, it may indicate that this kind of group is created too easily in the cosmological simulation. On the other hand, if the number density is below the values expected from observations, then the models employed in IllustrisTNG may suppress the formation of CGDs. Nonetheless, if simulation and observations agree within the error bars, it does not necessarily mean that the physical model of the simulation is completely correct; that is, it reproduces observations for the right reasons. Still, it is a good indicator in favour of it, as long as the number densities of other populations of galaxies and groups are also correctly reproduced. The analysis we make in this work is only part of a bigger series of ongoing comparisons that the community may carry in the literature (\citealt{Jackson2020,Oppenheimer2020,Donnari2021,Goddy2023}; to cite a few), and we hope the predictions presented here will be useful in future observational studies exploring the faint end of galaxy groups.  

Regarding the properties of CGDs compared to observations at $z \sim 0$, we see that even without directly selecting the CGDs by group mass, we could obtain reasonable agreement for the mass scales of the simulated groups (see Fig. \ref{fig:obs_range}). This means that, at least in terms of stellar and total masses, TNG50-1 can produce acceptable analogues for a restricted interval of these parameters. The fact that the simulation can predict the existence of a population of CGDs without them being explicitly put in the simulation is a positive point for the model. Thus, it is reasonable to try to use the simulation to understand the formation and nature of this kind of group. 

The masses that we find for the CGDs in TNG50-1 are comparable to the theoretical results from Y20, where they find virial masses in the range $\sim 10^{10} \ \rm M_\odot$ to $\sim 10^{11.7} \ \rm M_\odot$ for groups of dwarfs\footnote{The maximum and minimum stellar mass of the dwarfs in their sample being $\log(M_{\rm \ast,min}/ \rm M_\odot) = 6.969$ and $\log(M_{\rm \ast,max}/ \rm M_\odot) = 9.169$ respectively, which are close to the limits adopted in this work.} in which all members inhabit the same halo. These groups described by Y20 also have similar mass-to-light ratios in the B-band (see their Figures 4 and 7) to the majority of CGDs we found in TNG50-1 ($10 \lesssim M/L \lesssim 100$, see Figure \ref{fig:m_l_ratio}). Additionally, the typical inertial radius ($R_I$) that \cite{Yaryura+2023} finds for dwarf groups are of the same order of magnitude of $R_I$ values that we find for the CGDs in TNG50-1 (see their Figure 2 and our Appendix \ref{app:sup_material}). 

Our work independently demonstrates that compact groups of dwarf galaxies can emerge naturally in a $\Lambda$CDM framework, which evolves not only dark matter but also takes into account the baryonic processes and self-consistent hydrodynamical evolution of gas.

\subsection{Formation of CGDs at $z = 0$}
\label{discussion:assembly}

\begin{figure}
    \centering
    \includegraphics[width=0.49\textwidth]{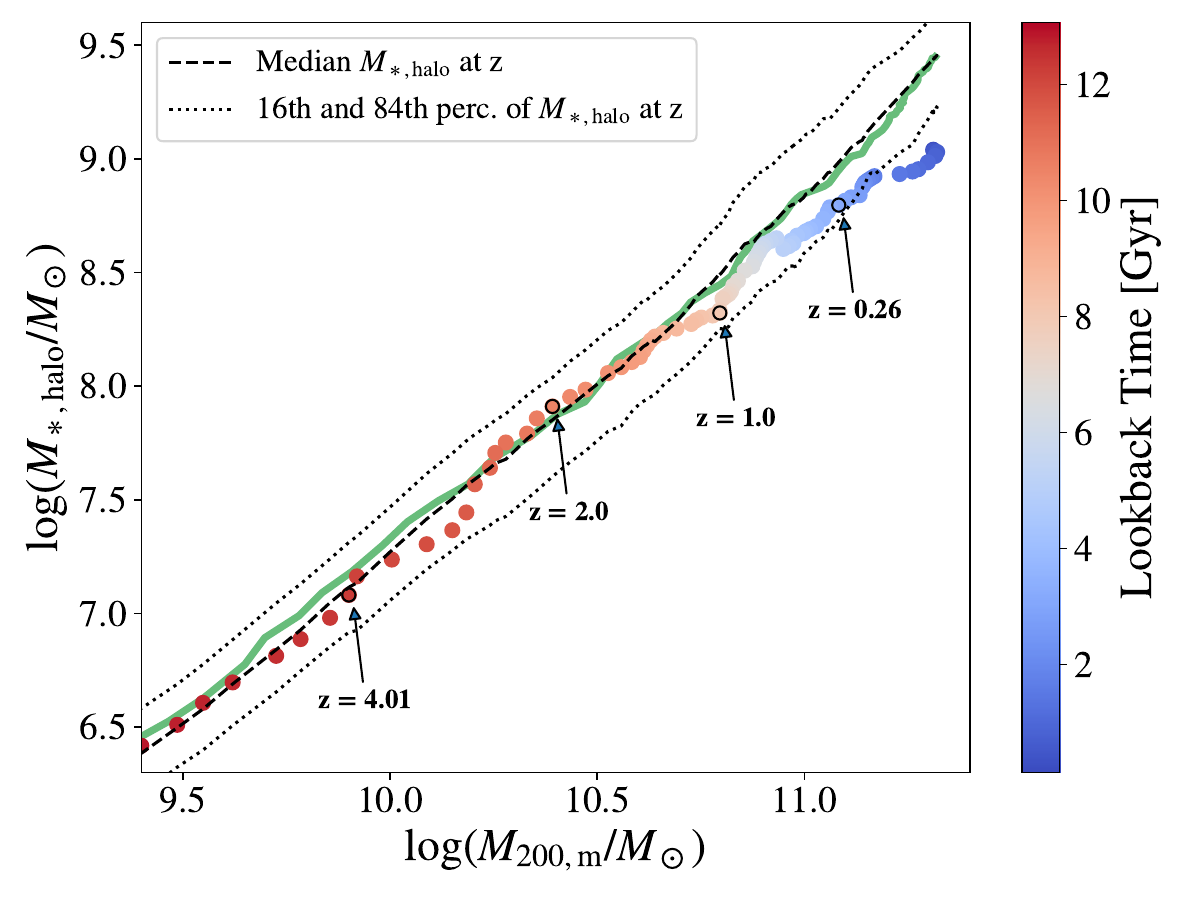}
    \caption{Evolution of $z=0$ CGDs in the $M_{\rm 200,m}$ - $M_{\rm \ast,halo}$ diagram. The median evolution of the CGD halos is shown as circles coloured by the lookback time. The black dashed line shows the median stellar mass expected for halos of similar $M_{\rm 200,m}$ at each snapshot. Specific redshifts in the median evolution of CGDs are indicated by the open black circles.
    }
    \label{fig:evo_M200_Mstar}
\end{figure}

By comparing the host halos of CGDs with control halos of similar mass, we are trying to obtain insights into specific properties that CGD host halos must have. In other words, we can explore what differentiates the halos of CGDs at $z=0$ from the general population of halos at the same mass scale and redshift. As we present in Sections \ref{results:halo_properties} and \ref{results:assembly}, the host halos of CGDs have lower amounts of stellar mass, have higher total gas fractions and assemble all their mass later when compared with halos of similar mass. Since we systematically find lower stellar masses in the host halos of CGDs at present, it is important to understand if CGD hosts always deviate from the overall halo population through cosmic time. In Figure \ref{fig:evo_M200_Mstar}, we see that, on average, the main progenitor halos of CGDs deviate noticeably from the expected evolution for their masses at $z \gtrsim 0.26$. As expected, this redshift interval is compatible with the epochs at which the CGD were first hosted by a single halo, the lookback time traced by $t_{\rm FoF}$ (see Section \ref{results:assembly}). 

\begin{figure}
    \centering
    \includegraphics[width=0.49\textwidth]{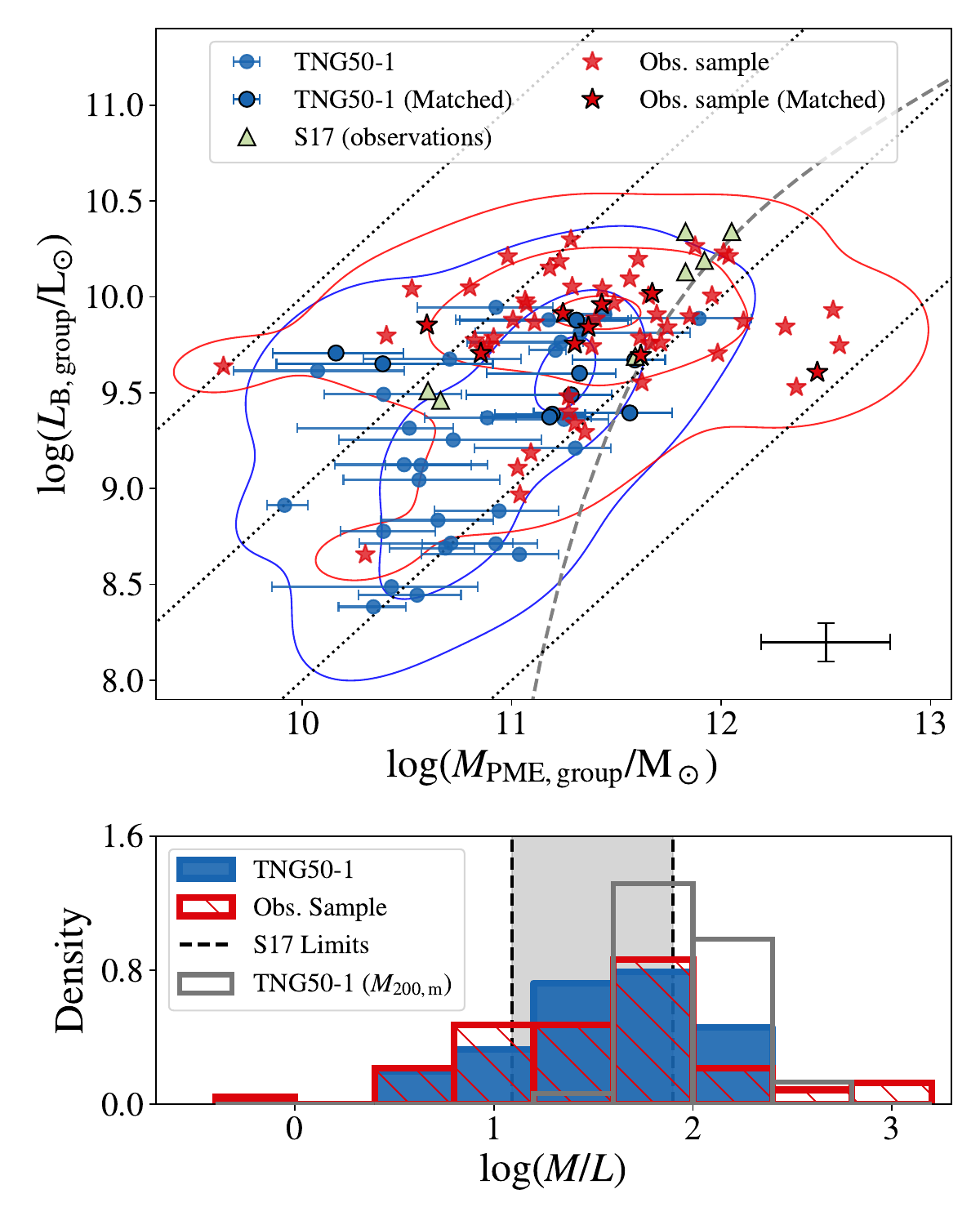}
    \caption{ \textbf{Top:} Total B-band luminosity versus total group mass of CGDs. Symbols and contours as in Figure \ref{fig:obs_range}. The grey dashed line is the relation between luminosity and virial mass from \citealt{Tully2015} (their Equation 16), and is used here as a reference line to compare with other studies. Black dotted lines indicate fixed values of mass-to-light ($M/L$) ratio, from left to right, they indicate $M/L$ equal to 1, 10, 100 and 1000. \textbf{Bottom:} Histogram of CGDs' mass-to-light ratios in the B-band. Histogram styles and shaded area are the same as in Figure \ref{fig:obs_range}. The histogram shown in grey represents the M/L of simulated groups computed using halo mass ($M_{\rm 200,m}$) instead of $M_{\rm PME,group}$.}
    \label{fig:m_l_ratio}
\end{figure}

From the results presented so far, we can interpret that the host halos of CGDs at $z=0$, on average, only reached their halo masses at present-day due to recent - last $\sim 3$ Gyr - hierarchical merging, which increased their masses quickly. However, the stellar component did not grow at the same rate as the total mass, making CGD halos more deficient in stellar mass. If the simulated CGDs are good representations of their observational counterparts, then we should expect both samples to have similar mass-to-light ratios. In Figure \ref{fig:m_l_ratio}, we present a diagram of the group masses versus luminosities in the B-band ($L_{\rm B,group}$) and a histogram of mass-to-light ratios in the same band. It is evident from this diagram that most CGDs in the simulation have mass-to-light ratios spanning 10 to 100, with a median value of $M/L \sim 38$ (25th and 75th $M/L$ percentiles being $\sim 24$ and $\sim 89$, respectively), comparable to the median of groups from S17 ($M/L \sim 50$). The observational sample of CGDs presents similar values of mass-to-light ratio to the simulated groups, with a median of $M/L \sim 42$ (25th and 75th $M/L$ percentiles being $\sim 14$ and $\sim 83$, respectively). Thus, the mass-to-light ratios estimated for observed groups could possibly be explained by a recent hierarchical formation, driving rapid growth in total mass with a delayed and relatively inefficient build-up in stellar mass. Additionally, by visually comparing the gas distributions of the CGDs and control halos, we see that the gas can be greatly scattered in the groups by the interaction between the members. This may create an even larger delay in the stellar assembly of halos hosting CGDs as compared to the control halos, which typically have a single dominant galaxy accreting gas and smaller satellites, thus concentrating the gas into a single region.

\begin{figure}
    \centering
    \includegraphics[width=0.49\textwidth]{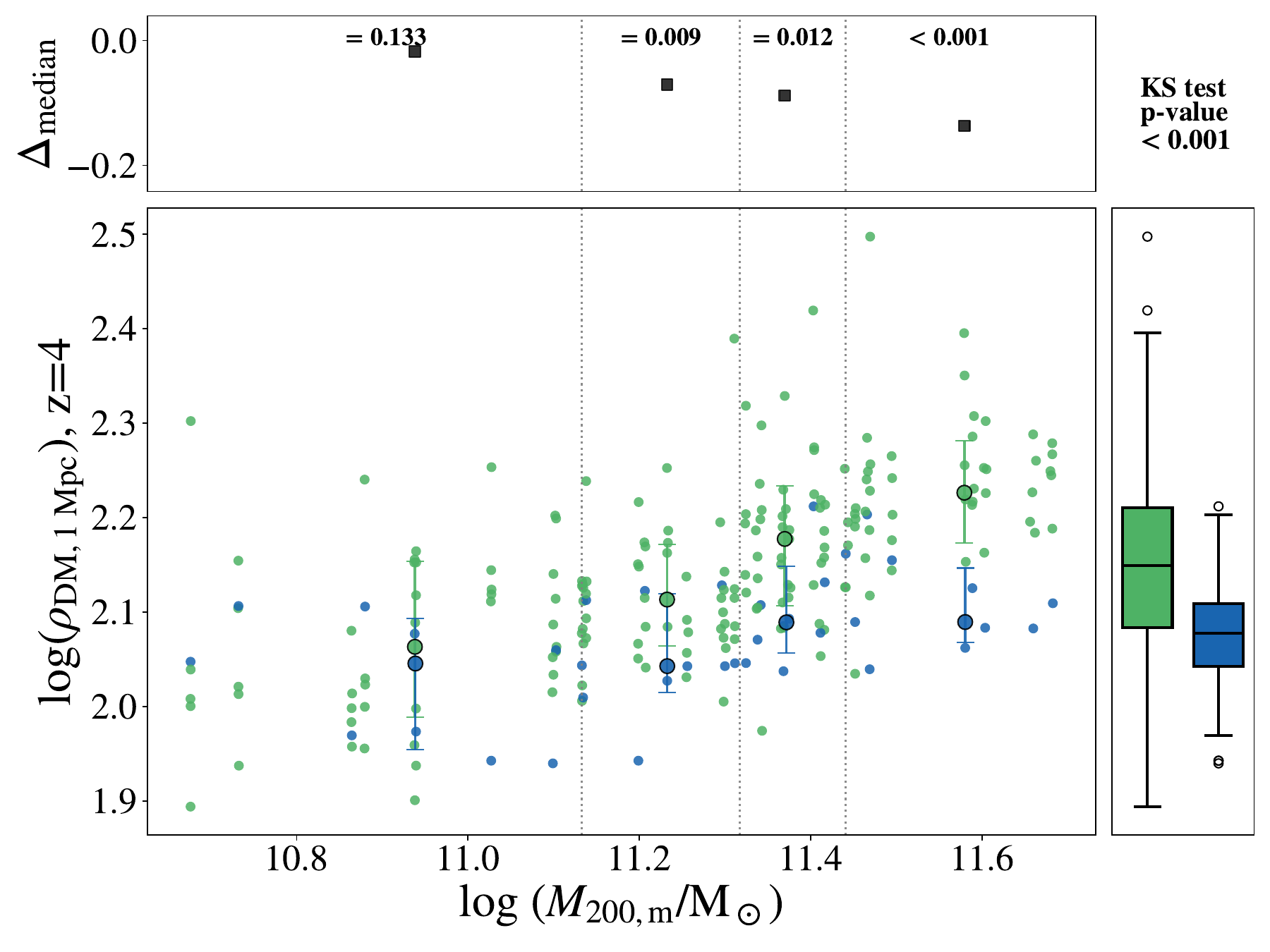}
    \caption{Same as in Fig. \ref{fig:host_halo_properties}. But in the vertical axes, instead of a halo property, it is plotted the volumetric dark matter density at the position of the progenitor halos at $z=4$; for details, see Section \ref{methods:mergertree}.}
    \label{fig:high_z_dm_density}
\end{figure}

In the simulation large-scale context, the CGDs at $z=0$ can be viewed as structures that formed recently, and their formation can be interpreted as the direct product of a late hierarchical mass assembly of halos inhabiting regions of the cosmic web with low-to-intermediate density - away from the centres of either knots or voids. A recent study by \cite{Martizzi2020} analyses the correlation between galaxy stellar masses, halo masses and large-scale environment in the TNG100 simulation. The authors show that for a fixed halo mass, galaxies with stellar masses lower than the median are more likely to be found in voids and sheets, while galaxies with stellar masses higher than the median are more likely to be found in filaments and knots. Naturally, it is tempting to ask if CGD halos had a different assembly history due to differences in the primordial large-scale environment of these halos and other halos of similar mass. To briefly explore this hypothesis, we show in Fig. \ref{fig:high_z_dm_density} the difference in the dark matter density at $z=4$ around the progenitor halos of CGDs and control sample.

It is possible to see that at least for the present-day halos with $M_{\rm 200,m} \gtrsim 10^{11.1} \ \rm M_\odot$, there was a difference in the local dark matter density at $z=4$ between CGD and control progenitor halos, with the groups progenitors inhabiting regions slightly less dense ($\sim 15 \%$ lower than control). If, at a fixed cosmic time, less massive halos have a higher probability of forming in less dense regions of the cosmic web, then the difference in the assembly of CGDs when compared to control halos could be partially explained by a difference in the past large-scale environment.

\subsection{Group Coalescence and Descendants}
\label{discussion:coalescence}
Dwarf galaxies inside the CGDs may suffer from cumulative environmental effects, so it is important to quantify the lifespan of these groups.
The ages of most CGDs found at $z=0$ in TNG50-1 indicate that these systems can survive for more than one billion years without the merging of their members (see Fig. \ref{fig:timescales}). This means dwarf galaxies inhabiting CGDs may suffer dwarf-only environmental effects for a non-negligible amount of time. The typical coalescence timescale of simulated CGDs that we found at $z > 0$ is even larger ($> 1$ Gyr), implying that the dwarf galaxies may suffer even longer with tidal forces and gas shocks inside the groups. In their work, S17 extrapolates the merger timescale of groups with massive galaxies (\citealt{Barnes89}) to predict the time until the observed CGDs would merge and suggests a timescale of $\lesssim 1$ Gyr. On the other hand, the CGDs that we found at $z=0$ in the simulation exist as compact groups for timescales longer than 1 Gyr. As is shown in Fig. \ref{fig:tcoal_and_Rgroup}, the typical timescale for the coalescence of CGDs in the simulation is even longer, being $\sim 3 \ \rm Gyr$ for groups found at $z\leq 0.5$. This suggests that the extrapolation of results found for massive groups may not be adequate to predict properties of groups - such as their merger timescales - in the low-mass regime.

More than half of the groups found at $z=0$ already existed more than 1 Gyr ago, meaning that the galaxies in these groups may have been interacting for long enough to suffer the effects of the group environment. In the dwarf regime, interactions which do not end in a merger can be more important than mergers for the amount of stellar mass formed \citep{Martin2021}, so to understand how these interactions affect the star formation history (SFH) of dwarf galaxies that inhabit compact groups, we are also conducting a detailed study of the CGDs members SFH's.

\begin{figure}
    \centering
    \includegraphics[width=0.49\textwidth]{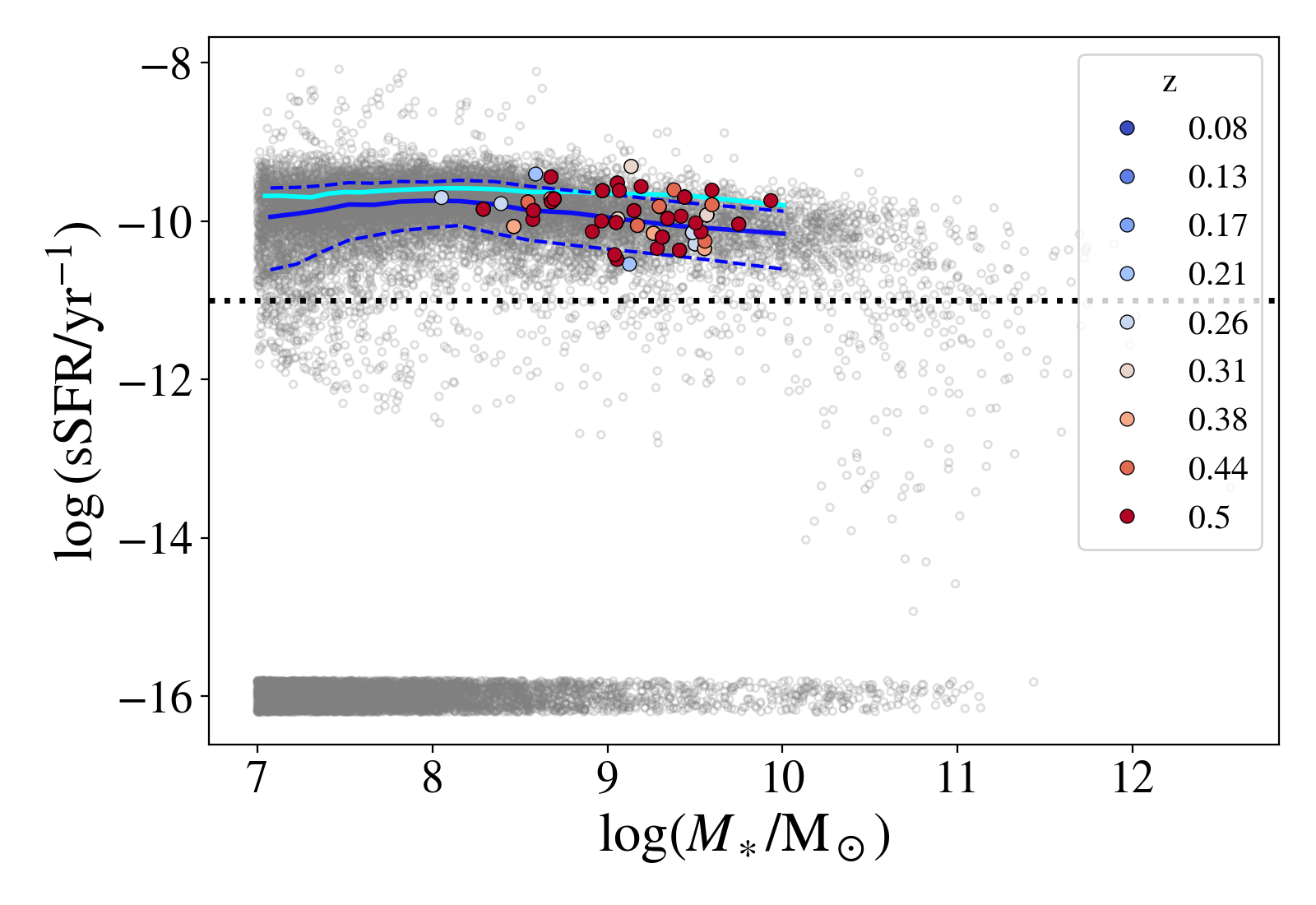}
    \caption{Position of coalesced CGDs descendants (coloured circles) in the stellar mass versus sSFR diagram at $z=0$. The colour of the circles indicates the redshift in which the progenitor CGDs were identified. Grey open circles are all other galaxies in the TNG50-1 run. Galaxies with zero star-formation rate are assigned random values of sSFR between $10^{-15.8} \ \rm yr^{-1}$ and $10^{-16.2} \ \rm yr^{-1}$. The blue solid line shows the median sSFR for different stellar masses at $z=0$, representing the main sequence (MS) of star formation. Blue dashed lines indicate the 16th and 84th percentiles of sSFR at $z=0$ and represent the scatter around the MS. We also show the MS at $z=0.5$ as a cyan solid line, which is, on average, only 0.2 dex above the MS at $z=0$. The black dotted line indicates the limit $\rm sSFR = 10^{-11} \ yr^{-1}$.}
    \label{fig:coalesc_galaxies_MS}
\end{figure}

Considering the CGDs as an intermediate stage for the formation of more massive systems, it is relevant to characterise the properties of the galaxy that form from the complete coalescence of these groups. In Fig. \ref{fig:coalesc_galaxies_MS}, we present the position of the descendants of all coalesced groups at $z  \leq 0.5$ in the $M_\ast$ - sSFR diagram. We also plot reference lines to represent the main sequence (MS) of star formation at $z=0$ and $z=0.5$. The MS at each redshift was obtained from the linear interpolation between the median sSFR in different stellar mass bins. The bins are defined within $7 \leq \log(M_\ast / \rm M_\odot) \leq 10$, with $\sim 750$ data points each. As we can see in the figure, the coalescence of CGDs gives origin to normally star-forming galaxies (SFGs) in the mass range $8 \lesssim \log(M_\ast / \rm M_\odot) \lesssim 10$. These objects are a minority in the population of SFGs at $z=0$, however, CGDs are expected to be more common at high-redshift, so they can possibly be an important evolutionary stage in the formation of low- to intermediate-mass galaxies at earlier epochs. 

Furthermore, galaxies that recently formed from the coalescence of CGDs may differ from other galaxies of similar mass in terms of their gas reservoir. The interaction of the dwarf galaxies during group coalescence may have similar effects on the diffuse gas as in dwarf pairs, where stripping and scattering of the gas creates a large gas reservoir that can fuel future star formation in the absence of a more massive galaxy \citep{Pearson2016}. The fact that all merger remnants of CGDs in the simulation lie on the star-forming main sequence means that star-formation can be maintained for a long time after the group coalescence. This is in line with results found for dwarf-dwarf mergers and their remnants in other studies \citep{Amorisco2012,Pearson2018,Kado-Fong2020}, which point to a scenario where the merger remnants have a gas supply to sustain star formation continuously.

Regarding the structure of the galaxies formed from group coalescence, we found that disks of gas with active star formation are easily formed short (less than 2 Gyr) after the complete merging process. This finding is valid for a subset of galaxies with $M_{\ast} < 10^{10} \ \rm M_\odot$ living in gas-rich halos, but according to results from \cite{Sotillo-Ramos2022}, this easiness to form discs may also be valid for higher masses since they find that Milky Way/Andromeda analogues in TNG50 can easily reform their stellar discs after a major merger given that there is sufficient gas available.

\section{Conclusions}
\label{sec:conclusions}
In this work, we search for analogues of CGDs in the IllustrisTNG cosmological simulations and compare their properties with some expectations from observations. We analyse the properties of analogue CGDs, investigating how these groups can form in a full cosmological context and what type of galaxies will result from their coalescence. We can summarise our conclusions as follows:
\begin{itemize}
    \item The TNG50-1 simulation can reproduce isolated compact groups of dwarf galaxies in the Local Universe ($z<0.1$), with the majority of these groups being triplets (Figure \ref{fig:example_groups}) and having a median halo mass of $10^{11.3} \ \rm M_\odot$ (Figure \ref{fig:halo_match}). 
    \item The comoving number density of CGDs in TNG50-1 at $z=0$ is $n \approx 10^{-3.56} \ \rm cMpc^{-3}$, and increases with redshift (Figure \ref{fig:nd_evolution}), effectively doubling the number of CGDs by $z = 0.26$.
    \item The predicted number of CGDs observed in a survey like the SDSS Legacy Survey partially agrees with the lower limits imposed by the observational sample used in this work. However, the confrontation between simulation and observations is limited to a small range of magnitudes ($-17.5 \lesssim M_{r \rm ,3rd} \lesssim -16.5$) due to the SDSS sample detection limit and the TNG50-1 volume (Figure \ref{fig:nd_prediction}).  
    \item The total group masses (Figure \ref{fig:obs_range}) and mass-to-light ratios ($10 \leq M/L \leq 100$) (Figure \ref{fig:m_l_ratio}) of the simulated CGD are mostly compatible with observations, but the most massive observed groups are missing in the simulation, probably due to the limited volume of TNG50-1.
    \item Host halos of CGDs at $z=0$ have lower stellar masses, higher total gas fractions and were assembled later compared to other halos of similar mass at the same epoch (Figure \ref{fig:host_halo_properties}).
    \item Simulated CGDs found at $z = 0$ have median formation times of $\sim 1.6$ Gyr (Figure \ref{fig:timescales}), with $\sim 20$\% of the groups being already formed for more than 3 Gyr.
    \item In the large-scale structure at $z=0$, CGDs avoid regions of high matter density, such as the meeting point of filaments (knots), and regions with the lowest matter density, such as voids (Figure \ref{fig:pos_in_LSS}).
    \item About 14\% of all unique CGDs identified since $z \leq 0.5$ merged into a single galaxy, and their median coalescence timescale is $\sim 3$ Gyr (Figure \ref{fig:timescales}).  
    \item All the CGDs that coalesced at $z \leq 0.5$ formed star-forming galaxies with $8 < \log(M_\ast/M_\odot) < 10$ at $z=0$ (Figure \ref{fig:coalesc_galaxies_MS}). Part of these galaxies have intermediate stellar mass ($9.5 \leq \log(M_\ast/M_\odot) \leq 10$) and represent 2.7 \% of the galaxies within this mass regime in the latest snapshot of TNG50-1.
\end{itemize}

In summary, we showed that the IllustrisTNG highest resolution simulation can create CGDs with total masses comparable to observations. We also found that the simulated CGDs at $z=0$ are formed from a late hierarchical assembly of halos in the low-mass regime and that CGDs at $z < 0.5$ can commonly take more than 1 Gyr to fully merge in the simulations. Investigating the formation of galaxy groups in the low-mass regime is important to understand the environmental effect of dwarf groups on galaxies even before they infall in denser environments and also to obtain insights into the hierarchical assembly of low-mass halos that may be common at the first stages of galaxy formation. In this manner, we hope that our work can help connect the models of galaxy formation with present and future observations in the still vastly unexplored low-luminosity regime.

\section*{Acknowledgements}
The authors thank the referee for comments that led to an improved version of the manuscript.
RFF thanks the support of Conselho Nacional de Desenvolvimento Científico e Tecnológico (CNPq). MT thanks the support of CNPq (process \#312541/2021-0).
We acknowledge the use of SDSS data (\url{http://www.sdss.org/collaboration/credits.html}), 
{\tt TOPCAT} Table/VOTable Processing Software 
\citep[][\url{http://www.star.bris.ac.uk/mbt/topcat/}]{Taylor:2005}, and R language and environment for statistical computing \citep{R:2015}.
The IllustrisTNG simulations were undertaken with compute time awarded by the Gauss Centre for Supercomputing (GCS) under GCS Large-Scale Projects GCS-ILLU and GCS-DWAR on the GCS share of the supercomputer Hazel Hen at the High Performance Computing Center Stuttgart (HLRS), as well as on the machines of the Max Planck Computing and Data Facility (MPCDF) in Garching, Germany.
\section*{Data Availability}
The IllustrisTNG simulations are publicly available at \url{www.tng-project.org/data} \citep{Nelson2019}. Nearest neighbour catalogues for IllustrisTNG are available at \url{http://www.tng-project.org/floresfreitas24}. Catalogues of simulated and observed CGDs analysed in this work are provided as supplementary material. Additional data directly related to this publication are available on request from the corresponding author.



\bibliographystyle{mnras}
\bibliography{main} 

\begin{thebibliography}{}
\makeatletter
\relax
\def\mn@urlcharsother{\let\do\@makeother \do\$\do\&\do\#\do\^\do\_\do\%\do\~}
\def\mn@doi{\begingroup\mn@urlcharsother \@ifnextchar [ {\mn@doi@}
  {\mn@doi@[]}}
\def\mn@doi@[#1]#2{\def\@tempa{#1}\ifx\@tempa\@empty \href
  {http://dx.doi.org/#2} {doi:#2}\else \href {http://dx.doi.org/#2} {#1}\fi
  \endgroup}
\def\mn@eprint#1#2{\mn@eprint@#1:#2::\@nil}
\def\mn@eprint@arXiv#1{\href {http://arxiv.org/abs/#1} {{\tt arXiv:#1}}}
\def\mn@eprint@dblp#1{\href {http://dblp.uni-trier.de/rec/bibtex/#1.xml}
  {dblp:#1}}
\def\mn@eprint@#1:#2:#3:#4\@nil{\def\@tempa {#1}\def\@tempb {#2}\def\@tempc
  {#3}\ifx \@tempc \@empty \let \@tempc \@tempb \let \@tempb \@tempa \fi \ifx
  \@tempb \@empty \def\@tempb {arXiv}\fi \@ifundefined
  {mn@eprint@\@tempb}{\@tempb:\@tempc}{\expandafter \expandafter \csname
  mn@eprint@\@tempb\endcsname \expandafter{\@tempc}}}

\bibitem[\protect\citeauthoryear{{Abazajian} et~al.,}{{Abazajian}
  et~al.}{2009}]{Abazajian2009}
{Abazajian} K.~N.,  et~al., 2009, \mn@doi [\apjs]
  {10.1088/0067-0049/182/2/543}, \href
  {https://ui.adsabs.harvard.edu/abs/2009ApJS..182..543A} {182, 543}

\bibitem[\protect\citeauthoryear{{Almeida} et~al.,}{{Almeida}
  et~al.}{2023}]{Almeida2023}
{Almeida} A.,  et~al., 2023, \mn@doi [\apjs] {10.3847/1538-4365/acda98}, \href
  {https://ui.adsabs.harvard.edu/abs/2023ApJS..267...44A} {267, 44}

\bibitem[\protect\citeauthoryear{{Amorisco} \& {Evans}}{{Amorisco} \&
  {Evans}}{2012}]{Amorisco2012}
{Amorisco} N.~C.,  {Evans} N.~W.,  2012, \mn@doi [\apjl]
  {10.1088/2041-8205/756/1/L2}, \href
  {https://ui.adsabs.harvard.edu/abs/2012ApJ...756L...2A} {756, L2}

\bibitem[\protect\citeauthoryear{{Anbajagane}, {Evrard}  \&
  {Farahi}}{{Anbajagane} et~al.}{2022}]{Anbajagane2022}
{Anbajagane} D.,  {Evrard} A.~E.,   {Farahi} A.,  2022, \mn@doi [\mnras]
  {10.1093/mnras/stab3177}, \href
  {https://ui.adsabs.harvard.edu/abs/2022MNRAS.509.3441A} {509, 3441}

\bibitem[\protect\citeauthoryear{{Barnes}}{{Barnes}}{1989}]{Barnes89}
{Barnes} J.~E.,  1989, \mn@doi [\nat] {10.1038/338123a0}, \href
  {https://ui.adsabs.harvard.edu/abs/1989Natur.338..123B} {338, 123}

\bibitem[\protect\citeauthoryear{{Beers}, {Flynn}  \& {Gebhardt}}{{Beers}
  et~al.}{1990}]{Beers1990}
{Beers} T.~C.,  {Flynn} K.,   {Gebhardt} K.,  1990, \mn@doi [\aj]
  {10.1086/115487}, \href
  {https://ui.adsabs.harvard.edu/abs/1990AJ....100...32B} {100, 32}

\bibitem[\protect\citeauthoryear{{Besla} et~al.,}{{Besla}
  et~al.}{2018}]{Besla+2018}
{Besla} G.,  et~al., 2018, \mn@doi [\mnras] {10.1093/mnras/sty2041}, \href
  {https://ui.adsabs.harvard.edu/abs/2018MNRAS.480.3376B} {480, 3376}

\bibitem[\protect\citeauthoryear{{Blanton} \& {Roweis}}{{Blanton} \&
  {Roweis}}{2007}]{Blanton2007}
{Blanton} M.~R.,  {Roweis} S.,  2007, \mn@doi [\aj] {10.1086/510127}, \href
  {https://ui.adsabs.harvard.edu/abs/2007AJ....133..734B} {133, 734}

\bibitem[\protect\citeauthoryear{{Blumenthal}, {Faber}, {Primack}  \&
  {Rees}}{{Blumenthal} et~al.}{1984}]{Blumenthal+1984}
{Blumenthal} G.~R.,  {Faber} S.~M.,  {Primack} J.~R.,   {Rees} M.~J.,  1984,
  \mn@doi [\nat] {10.1038/311517a0}, \href
  {https://ui.adsabs.harvard.edu/abs/1984Natur.311..517B} {311, 517}

\bibitem[\protect\citeauthoryear{{Borrow}, {Schaller}, {Bah{\'e}}, {Schaye},
  {Ludlow}, {Ploeckinger}, {Nobels}  \& {Altamura}}{{Borrow}
  et~al.}{2023}]{Borrow2023}
{Borrow} J.,  {Schaller} M.,  {Bah{\'e}} Y.~M.,  {Schaye} J.,  {Ludlow} A.~D.,
  {Ploeckinger} S.,  {Nobels} F. S.~J.,   {Altamura} E.,  2023, \mn@doi
  [\mnras] {10.1093/mnras/stad2928}, \href
  {https://ui.adsabs.harvard.edu/abs/2023MNRAS.526.2441B} {526, 2441}

\bibitem[\protect\citeauthoryear{{Bullock} \& {Boylan-Kolchin}}{{Bullock} \&
  {Boylan-Kolchin}}{2017}]{Bullock2017}
{Bullock} J.~S.,  {Boylan-Kolchin} M.,  2017, \mn@doi [\araa]
  {10.1146/annurev-astro-091916-055313}, \href
  {https://ui.adsabs.harvard.edu/abs/2017ARA&A..55..343B} {55, 343}

\bibitem[\protect\citeauthoryear{{Cattaneo}, {Mamon}, {Warnick}  \&
  {Knebe}}{{Cattaneo} et~al.}{2011}]{Cattaneo2011}
{Cattaneo} A.,  {Mamon} G.~A.,  {Warnick} K.,   {Knebe} A.,  2011, \mn@doi
  [\aap] {10.1051/0004-6361/201015780}, \href
  {https://ui.adsabs.harvard.edu/abs/2011A&A...533A...5C} {533, A5}

\bibitem[\protect\citeauthoryear{{Crain} \& {van de Voort}}{{Crain} \& {van de
  Voort}}{2023}]{Crain2023}
{Crain} R.~A.,  {van de Voort} F.,  2023, \mn@doi [\araa]
  {10.1146/annurev-astro-041923-043618}, \href
  {https://ui.adsabs.harvard.edu/abs/2023ARA&A..61..473C} {61, 473}

\bibitem[\protect\citeauthoryear{{DESI Collaboration} et~al.,}{{DESI
  Collaboration} et~al.}{2023}]{DESI2023}
{DESI Collaboration} et~al., 2023, \mn@doi [arXiv e-prints]
  {10.48550/arXiv.2306.06307}, \href
  {https://ui.adsabs.harvard.edu/abs/2023arXiv230606307D} {p. arXiv:2306.06307}

\bibitem[\protect\citeauthoryear{{Darragh-Ford} et~al.,}{{Darragh-Ford}
  et~al.}{2023}]{Darragh-Ford2023}
{Darragh-Ford} E.,  et~al., 2023, \mn@doi [\apj] {10.3847/1538-4357/ace902},
  \href {https://ui.adsabs.harvard.edu/abs/2023ApJ...954..149D} {954, 149}

\bibitem[\protect\citeauthoryear{{Davis}, {Efstathiou}, {Frenk}  \&
  {White}}{{Davis} et~al.}{1985}]{Davis1985}
{Davis} M.,  {Efstathiou} G.,  {Frenk} C.~S.,   {White} S.~D.~M.,  1985,
  \mn@doi [\apj] {10.1086/163168}, \href
  {https://ui.adsabs.harvard.edu/abs/1985ApJ...292..371D} {292, 371}

\bibitem[\protect\citeauthoryear{{De Lucia} \& {Blaizot}}{{De Lucia} \&
  {Blaizot}}{2007}]{deLucia&Blaizot2007}
{De Lucia} G.,  {Blaizot} J.,  2007, \mn@doi [\mnras]
  {10.1111/j.1365-2966.2006.11287.x}, \href
  {https://ui.adsabs.harvard.edu/abs/2007MNRAS.375....2D} {375, 2}

\bibitem[\protect\citeauthoryear{{Donnari}, {Pillepich}, {Nelson}, {Marinacci},
  {Vogelsberger}  \& {Hernquist}}{{Donnari} et~al.}{2021}]{Donnari2021}
{Donnari} M.,  {Pillepich} A.,  {Nelson} D.,  {Marinacci} F.,  {Vogelsberger}
  M.,   {Hernquist} L.,  2021, \mn@doi [\mnras] {10.1093/mnras/stab1950}, \href
  {https://ui.adsabs.harvard.edu/abs/2021MNRAS.506.4760D} {506, 4760}

\bibitem[\protect\citeauthoryear{{Ellison}, {Catinella}  \&
  {Cortese}}{{Ellison} et~al.}{2018}]{Ellison+2018}
{Ellison} S.~L.,  {Catinella} B.,   {Cortese} L.,  2018, \mn@doi [\mnras]
  {10.1093/mnras/sty1247}, \href
  {https://ui.adsabs.harvard.edu/abs/2018MNRAS.478.3447E} {478, 3447}

\bibitem[\protect\citeauthoryear{{Euclid Collaboration} et~al.,}{{Euclid
  Collaboration} et~al.}{2022}]{Euclid2022}
{Euclid Collaboration} et~al., 2022, \mn@doi [\aap]
  {10.1051/0004-6361/202141938}, \href
  {https://ui.adsabs.harvard.edu/abs/2022A&A...662A.112E} {662, A112}

\bibitem[\protect\citeauthoryear{{Fontana} et~al.,}{{Fontana}
  et~al.}{2006}]{Fontana+2006}
{Fontana} A.,  et~al., 2006, \mn@doi [\aap] {10.1051/0004-6361:20065475}, \href
  {https://ui.adsabs.harvard.edu/abs/2006A&A...459..745F} {459, 745}

\bibitem[\protect\citeauthoryear{{Frenk} \& {White}}{{Frenk} \&
  {White}}{2012}]{Frenk&White2012}
{Frenk} C.~S.,  {White} S.~D.~M.,  2012, \mn@doi [Annalen der Physik]
  {10.1002/andp.201200212}, \href
  {https://ui.adsabs.harvard.edu/abs/2012AnP...524..507F} {524, 507}

\bibitem[\protect\citeauthoryear{{Gao}, {Gu}, {Liu}, {Zhang}, {Shi}, {Dou},
  {Li}  \& {Kong}}{{Gao} et~al.}{2023}]{Gao+2023}
{Gao} Y.,  {Gu} Q.,  {Liu} G.,  {Zhang} H.,  {Shi} Y.,  {Dou} J.,  {Li} X.,
  {Kong} X.,  2023, \mn@doi [\aap] {10.1051/0004-6361/202346753}, \href
  {https://ui.adsabs.harvard.edu/abs/2023A&A...677A.179G} {677, A179}

\bibitem[\protect\citeauthoryear{{Genel} et~al.,}{{Genel}
  et~al.}{2019}]{Genel2019}
{Genel} S.,  et~al., 2019, \mn@doi [\apj] {10.3847/1538-4357/aaf4bb}, \href
  {https://ui.adsabs.harvard.edu/abs/2019ApJ...871...21G} {871, 21}

\bibitem[\protect\citeauthoryear{{Goddy}, {Stark}, {Masters}, {Bundy}, {Drory}
  \& {Law}}{{Goddy} et~al.}{2023}]{Goddy2023}
{Goddy} J.~S.,  {Stark} D.~V.,  {Masters} K.~L.,  {Bundy} K.,  {Drory} N.,
  {Law} D.~R.,  2023, \mn@doi [\mnras] {10.1093/mnras/stad298}, \href
  {https://ui.adsabs.harvard.edu/abs/2023MNRAS.520.3895G} {520, 3895}

\bibitem[\protect\citeauthoryear{{Grazian} et~al.,}{{Grazian}
  et~al.}{2015}]{Grazian+2015}
{Grazian} A.,  et~al., 2015, \mn@doi [\aap] {10.1051/0004-6361/201424750},
  \href {https://ui.adsabs.harvard.edu/abs/2015A&A...575A..96G} {575, A96}

\bibitem[\protect\citeauthoryear{{Hahn} et~al.,}{{Hahn}
  et~al.}{2023}]{Hahn2023}
{Hahn} C.,  et~al., 2023, \mn@doi [\aj] {10.3847/1538-3881/accff8}, \href
  {https://ui.adsabs.harvard.edu/abs/2023AJ....165..253H} {165, 253}

\bibitem[\protect\citeauthoryear{{Haynes} et~al.,}{{Haynes}
  et~al.}{2011}]{Haynes.etal:2011}
{Haynes} M.~P.,  et~al., 2011, \mn@doi [\aj] {10.1088/0004-6256/142/5/170},
  \href {https://ui.adsabs.harvard.edu/abs/2011AJ....142..170H} {142, 170}

\bibitem[\protect\citeauthoryear{{Haynes} et~al.,}{{Haynes}
  et~al.}{2018}]{Haynes.etal:2018}
{Haynes} M.~P.,  et~al., 2018, \mn@doi [\apj] {10.3847/1538-4357/aac956}, \href
  {https://ui.adsabs.harvard.edu/abs/2018ApJ...861...49H} {861, 49}

\bibitem[\protect\citeauthoryear{{Heisler}, {Tremaine}  \& {Bahcall}}{{Heisler}
  et~al.}{1985}]{Heisler1985}
{Heisler} J.,  {Tremaine} S.,   {Bahcall} J.~N.,  1985, \mn@doi [\apj]
  {10.1086/163584}, \href
  {https://ui.adsabs.harvard.edu/abs/1985ApJ...298....8H} {298, 8}

\bibitem[\protect\citeauthoryear{Ho, Imai, King  \& Stuart}{Ho
  et~al.}{2011}]{Ho2011}
Ho D.,  Imai K.,  King G.,   Stuart E.~A.,  2011, \mn@doi [Journal of
  Statistical Software] {10.18637/jss.v042.i08}, 42, 1–28

\bibitem[\protect\citeauthoryear{{Ilbert} et~al.,}{{Ilbert}
  et~al.}{2010}]{Ilbert2010}
{Ilbert} O.,  et~al., 2010, \mn@doi [\apj] {10.1088/0004-637X/709/2/644}, \href
  {https://ui.adsabs.harvard.edu/abs/2010ApJ...709..644I} {709, 644}

\bibitem[\protect\citeauthoryear{{Ivezi{\'c}} et~al.,}{{Ivezi{\'c}}
  et~al.}{2019}]{Ivezic2019}
{Ivezi{\'c}} {\v{Z}}.,  et~al., 2019, \mn@doi [\apj]
  {10.3847/1538-4357/ab042c}, \href
  {https://ui.adsabs.harvard.edu/abs/2019ApJ...873..111I} {873, 111}

\bibitem[\protect\citeauthoryear{{Jackson}, {Pasquali}, {Pacifici}, {Engler},
  {Pillepich}  \& {Grebel}}{{Jackson} et~al.}{2020}]{Jackson2020}
{Jackson} T.~M.,  {Pasquali} A.,  {Pacifici} C.,  {Engler} C.,  {Pillepich} A.,
    {Grebel} E.~K.,  2020, \mn@doi [\mnras] {10.1093/mnras/staa2306}, \href
  {https://ui.adsabs.harvard.edu/abs/2020MNRAS.497.4262J} {497, 4262}

\bibitem[\protect\citeauthoryear{{Kado-Fong}, {Greene}, {Greco}, {Beaton},
  {Goulding}, {Johnson}  \& {Komiyama}}{{Kado-Fong}
  et~al.}{2020}]{Kado-Fong2020}
{Kado-Fong} E.,  {Greene} J.~E.,  {Greco} J.~P.,  {Beaton} R.,  {Goulding}
  A.~D.,  {Johnson} S.~D.,   {Komiyama} Y.,  2020, \mn@doi [\aj]
  {10.3847/1538-3881/ab6ef3}, \href
  {https://ui.adsabs.harvard.edu/abs/2020AJ....159..103K} {159, 103}

\bibitem[\protect\citeauthoryear{{Kado-Fong}, {Robinson}, {Nyland}, {Greene},
  {Suess}, {Stierwalt}  \& {Beaton}}{{Kado-Fong} et~al.}{2023}]{Kado-Fong2023}
{Kado-Fong} E.,  {Robinson} A.,  {Nyland} K.,  {Greene} J.~E.,  {Suess} K.~A.,
  {Stierwalt} S.,   {Beaton} R.,  2023, \mn@doi [arXiv e-prints]
  {10.48550/arXiv.2311.09280}, \href
  {https://ui.adsabs.harvard.edu/abs/2023arXiv231109280K} {p. arXiv:2311.09280}

\bibitem[\protect\citeauthoryear{{Kimbro}, {Reines}, {Molina}, {Deller}  \&
  {Stern}}{{Kimbro} et~al.}{2021}]{Kimbro2021}
{Kimbro} E.,  {Reines} A.~E.,  {Molina} M.,  {Deller} A.~T.,   {Stern} D.,
  2021, \mn@doi [\apj] {10.3847/1538-4357/abec6a}, \href
  {https://ui.adsabs.harvard.edu/abs/2021ApJ...912...89K} {912, 89}

\bibitem[\protect\citeauthoryear{{Luber}, {Pearson}, {Putman}, {Besla},
  {Stierwalt}  \& {Meyers}}{{Luber} et~al.}{2022}]{Luber2022}
{Luber} N.,  {Pearson} S.,  {Putman} M.~E.,  {Besla} G.,  {Stierwalt} S.,
  {Meyers} J.~P.,  2022, \mn@doi [\aj] {10.3847/1538-3881/ac3750}, \href
  {https://ui.adsabs.harvard.edu/abs/2022AJ....163...49L} {163, 49}

\bibitem[\protect\citeauthoryear{{Luo} et~al.,}{{Luo} et~al.}{2023}]{Luo2023}
{Luo} Y.,  et~al., 2023, \mn@doi [arXiv e-prints] {10.48550/arXiv.2305.19310},
  \href {https://ui.adsabs.harvard.edu/abs/2023arXiv230519310L} {p.
  arXiv:2305.19310}

\bibitem[\protect\citeauthoryear{{Mahalanobis}}{{Mahalanobis}}{2018}]{Mahalanobis1936}
{Mahalanobis} P.,  2018, \mn@doi [Sankhya A] {10.1007/s13171-019-00164-5}, 80,
  1

\bibitem[\protect\citeauthoryear{{Martin} et~al.,}{{Martin}
  et~al.}{2021}]{Martin2021}
{Martin} G.,  et~al., 2021, \mn@doi [\mnras] {10.1093/mnras/staa3443}, \href
  {https://ui.adsabs.harvard.edu/abs/2021MNRAS.500.4937M} {500, 4937}

\bibitem[\protect\citeauthoryear{{Martizzi}, {Vogelsberger}, {Torrey},
  {Pillepich}, {Hansen}, {Marinacci}  \& {Hernquist}}{{Martizzi}
  et~al.}{2020}]{Martizzi2020}
{Martizzi} D.,  {Vogelsberger} M.,  {Torrey} P.,  {Pillepich} A.,  {Hansen}
  S.~H.,  {Marinacci} F.,   {Hernquist} L.,  2020, \mn@doi [\mnras]
  {10.1093/mnras/stz3418}, \href
  {https://ui.adsabs.harvard.edu/abs/2020MNRAS.491.5747M} {491, 5747}

\bibitem[\protect\citeauthoryear{{Muldrew} et~al.,}{{Muldrew}
  et~al.}{2012}]{Muldrew2012}
{Muldrew} S.~I.,  et~al., 2012, \mn@doi [\mnras]
  {10.1111/j.1365-2966.2011.19922.x}, \href
  {https://ui.adsabs.harvard.edu/abs/2012MNRAS.419.2670M} {419, 2670}

\bibitem[\protect\citeauthoryear{{Nelson} et~al.,}{{Nelson}
  et~al.}{2019}]{Nelson2019}
{Nelson} D.,  et~al., 2019, \mn@doi [Computational Astrophysics and Cosmology]
  {10.1186/s40668-019-0028-x}, \href
  {https://ui.adsabs.harvard.edu/abs/2019ComAC...6....2N} {6, 2}

\bibitem[\protect\citeauthoryear{{Oppenheimer} et~al.,}{{Oppenheimer}
  et~al.}{2020}]{Oppenheimer2020}
{Oppenheimer} B.~D.,  et~al., 2020, \mn@doi [\apjl] {10.3847/2041-8213/ab846f},
  \href {https://ui.adsabs.harvard.edu/abs/2020ApJ...893L..24O} {893, L24}

\bibitem[\protect\citeauthoryear{{Paudel}, {Smith}, {Yoon},
  {Calder{\'o}n-Castillo}  \& {Duc}}{{Paudel} et~al.}{2018}]{Paudel+2018}
{Paudel} S.,  {Smith} R.,  {Yoon} S.~J.,  {Calder{\'o}n-Castillo} P.,   {Duc}
  P.-A.,  2018, \mn@doi [\apjs] {10.3847/1538-4365/aad555}, \href
  {https://ui.adsabs.harvard.edu/abs/2018ApJS..237...36P} {237, 36}

\bibitem[\protect\citeauthoryear{{Pearson} et~al.,}{{Pearson}
  et~al.}{2016}]{Pearson2016}
{Pearson} S.,  et~al., 2016, \mn@doi [\mnras] {10.1093/mnras/stw757}, \href
  {https://ui.adsabs.harvard.edu/abs/2016MNRAS.459.1827P} {459, 1827}

\bibitem[\protect\citeauthoryear{{Pearson} et~al.,}{{Pearson}
  et~al.}{2018}]{Pearson2018}
{Pearson} S.,  et~al., 2018, \mn@doi [\mnras] {10.1093/mnras/sty2052}, \href
  {https://ui.adsabs.harvard.edu/abs/2018MNRAS.480.3069P} {480, 3069}

\bibitem[\protect\citeauthoryear{{P{\'e}rez-Gonz{\'a}lez}
  et~al.,}{{P{\'e}rez-Gonz{\'a}lez} et~al.}{2008}]{PerezGonzalez2008}
{P{\'e}rez-Gonz{\'a}lez} P.~G.,  et~al., 2008, \mn@doi [\apj] {10.1086/523690},
  \href {https://ui.adsabs.harvard.edu/abs/2008ApJ...675..234P} {675, 234}

\bibitem[\protect\citeauthoryear{{Pillepich} et~al.,}{{Pillepich}
  et~al.}{2018}]{Pillepich2018}
{Pillepich} A.,  et~al., 2018, \mn@doi [\mnras] {10.1093/mnras/stx2656}, \href
  {https://ui.adsabs.harvard.edu/abs/2018MNRAS.473.4077P} {473, 4077}

\bibitem[\protect\citeauthoryear{{Pillepich} et~al.,}{{Pillepich}
  et~al.}{2019}]{Pillepich2019}
{Pillepich} A.,  et~al., 2019, \mn@doi [\mnras] {10.1093/mnras/stz2338}, \href
  {https://ui.adsabs.harvard.edu/abs/2019MNRAS.490.3196P} {490, 3196}

\bibitem[\protect\citeauthoryear{{Planck Collaboration} et~al.,}{{Planck
  Collaboration} et~al.}{2016}]{Planck2016}
{Planck Collaboration} et~al., 2016, \mn@doi [\aap]
  {10.1051/0004-6361/201525830}, \href
  {https://ui.adsabs.harvard.edu/abs/2016A&A...594A..13P} {594, A13}

\bibitem[\protect\citeauthoryear{{Primack}}{{Primack}}{2012}]{Primack2012}
{Primack} J.~R.,  2012, \mn@doi [Annalen der Physik] {10.1002/andp.201200077},
  \href {https://ui.adsabs.harvard.edu/abs/2012AnP...524..535P} {524, 535}

\bibitem[\protect\citeauthoryear{{Privon} et~al.,}{{Privon}
  et~al.}{2017}]{Privon2017}
{Privon} G.~C.,  et~al., 2017, \mn@doi [\apj] {10.3847/1538-4357/aa8560}, \href
  {https://ui.adsabs.harvard.edu/abs/2017ApJ...846...74P} {846, 74}

\bibitem[\protect\citeauthoryear{{R Core Team}}{{R Core Team}}{2015}]{R:2015}
{R Core Team} 2015, R: A Language and Environment for Statistical Computing.
R Foundation for Statistical Computing, Vienna, Austria, \url
  {https://www.R-project.org}

\bibitem[\protect\citeauthoryear{{Rodriguez-Gomez} et~al.,}{{Rodriguez-Gomez}
  et~al.}{2015}]{Rodriguez-Gomez2015}
{Rodriguez-Gomez} V.,  et~al., 2015, \mn@doi [\mnras] {10.1093/mnras/stv264},
  \href {https://ui.adsabs.harvard.edu/abs/2015MNRAS.449...49R} {449, 49}

\bibitem[\protect\citeauthoryear{{Rodriguez-Gomez} et~al.,}{{Rodriguez-Gomez}
  et~al.}{2016}]{Rodriguez-Gomez2016}
{Rodriguez-Gomez} V.,  et~al., 2016, \mn@doi [\mnras] {10.1093/mnras/stw456},
  \href {https://ui.adsabs.harvard.edu/abs/2016MNRAS.458.2371R} {458, 2371}

\bibitem[\protect\citeauthoryear{{Rosembaum} \& {Rubin}}{{Rosembaum} \&
  {Rubin}}{1983}]{Rosembaum1983}
{Rosembaum} P.~R.,  {Rubin} D.~B.,  1983, \mn@doi [Biometrika]
  {10.1093/biomet/70.1.41}, 70, 41

\bibitem[\protect\citeauthoryear{{Sales}, {Wetzel}  \& {Fattahi}}{{Sales}
  et~al.}{2022}]{Sales2022}
{Sales} L.~V.,  {Wetzel} A.,   {Fattahi} A.,  2022, \mn@doi [Nature Astronomy]
  {10.1038/s41550-022-01689-w}, \href
  {https://ui.adsabs.harvard.edu/abs/2022NatAs...6..897S} {6, 897}

\bibitem[\protect\citeauthoryear{{Silk} \& {Mamon}}{{Silk} \&
  {Mamon}}{2012}]{Silk&Mamon2012}
{Silk} J.,  {Mamon} G.~A.,  2012, \mn@doi [Research in Astronomy and
  Astrophysics] {10.1088/1674-4527/12/8/004}, \href
  {https://ui.adsabs.harvard.edu/abs/2012RAA....12..917S} {12, 917}

\bibitem[\protect\citeauthoryear{{Somerville} \& {Dav{\'e}}}{{Somerville} \&
  {Dav{\'e}}}{2015}]{Somerville&Dave2015}
{Somerville} R.~S.,  {Dav{\'e}} R.,  2015, \mn@doi [\araa]
  {10.1146/annurev-astro-082812-140951}, \href
  {https://ui.adsabs.harvard.edu/abs/2015ARA&A..53...51S} {53, 51}

\bibitem[\protect\citeauthoryear{{Sotillo-Ramos} et~al.,}{{Sotillo-Ramos}
  et~al.}{2022}]{Sotillo-Ramos2022}
{Sotillo-Ramos} D.,  et~al., 2022, \mn@doi [\mnras] {10.1093/mnras/stac2586},
  \href {https://ui.adsabs.harvard.edu/abs/2022MNRAS.516.5404S} {516, 5404}

\bibitem[\protect\citeauthoryear{{Stierwalt}, {Besla}, {Patton}, {Johnson},
  {Kallivayalil}, {Putman}, {Privon}  \& {Ross}}{{Stierwalt}
  et~al.}{2015}]{Stierwalt2015}
{Stierwalt} S.,  {Besla} G.,  {Patton} D.,  {Johnson} K.,  {Kallivayalil} N.,
  {Putman} M.,  {Privon} G.,   {Ross} G.,  2015, \mn@doi [\apj]
  {10.1088/0004-637X/805/1/2}, \href
  {https://ui.adsabs.harvard.edu/abs/2015ApJ...805....2S} {805, 2}

\bibitem[\protect\citeauthoryear{{Stierwalt}, {Liss}, {Johnson}, {Patton},
  {Privon}, {Besla}, {Kallivayalil}  \& {Putman}}{{Stierwalt}
  et~al.}{2017}]{Stierwalt2017}
{Stierwalt} S.,  {Liss} S.~E.,  {Johnson} K.~E.,  {Patton} D.~R.,  {Privon}
  G.~C.,  {Besla} G.,  {Kallivayalil} N.,   {Putman} M.,  2017, \mn@doi [Nature
  Astronomy] {10.1038/s41550-016-0025}, \href
  {https://ui.adsabs.harvard.edu/abs/2017NatAs...1E..25S} {1, 0025}

\bibitem[\protect\citeauthoryear{{Subramanian}, {Mondal}  \&
  {Kalari}}{{Subramanian} et~al.}{2023}]{Subramanian2023}
{Subramanian} S.,  {Mondal} C.,   {Kalari} V.,  2023, \mn@doi [arXiv e-prints]
  {10.48550/arXiv.2310.02595}, \href
  {https://ui.adsabs.harvard.edu/abs/2023arXiv231002595S} {p. arXiv:2310.02595}

\bibitem[\protect\citeauthoryear{{Taverna}, {D{\'\i}az-Gim{\'e}nez},
  {Zandivarez}  \& {Mamon}}{{Taverna} et~al.}{2022}]{Taverna2022}
{Taverna} A.,  {D{\'\i}az-Gim{\'e}nez} E.,  {Zandivarez} A.,   {Mamon} G.~A.,
  2022, \mn@doi [\mnras] {10.1093/mnras/stac200}, \href
  {https://ui.adsabs.harvard.edu/abs/2022MNRAS.511.4741T} {511, 4741}

\bibitem[\protect\citeauthoryear{{Taylor}}{{Taylor}}{2005}]{Taylor:2005}
{Taylor} M.~B.,  2005, in {Shopbell} P.,  {Britton} M.,   {Ebert} R.,  eds,
  Astronomical Society of the Pacific Conference Series Vol. 347, Astronomical
  Data Analysis Software and Systems XIV. p.~29

\bibitem[\protect\citeauthoryear{{Tully}}{{Tully}}{1987}]{Tully1987}
{Tully} R.~B.,  1987, \mn@doi [\apj] {10.1086/165629}, \href
  {https://ui.adsabs.harvard.edu/abs/1987ApJ...321..280T} {321, 280}

\bibitem[\protect\citeauthoryear{{Tully}}{{Tully}}{1988}]{Tully1988}
{Tully} R.~B.,  1988, {Nearby galaxies catalog}

\bibitem[\protect\citeauthoryear{Tully}{Tully}{2015}]{Tully2015}
Tully R.~B.,  2015, \mn@doi [The Astronomical Journal]
  {10.1088/0004-6256/149/2/54}, 149, 54

\bibitem[\protect\citeauthoryear{{Tully} et~al.,}{{Tully}
  et~al.}{2006}]{Tully2006}
{Tully} R.~B.,  et~al., 2006, \mn@doi [\aj] {10.1086/505466}, \href
  {https://ui.adsabs.harvard.edu/abs/2006AJ....132..729T} {132, 729}

\bibitem[\protect\citeauthoryear{Wainer \& Thissen}{Wainer \&
  Thissen}{1976}]{Wainer1976}
Wainer H.,  Thissen D.,  1976, Psychometrika, 41, 9

\bibitem[\protect\citeauthoryear{{Weinberger} et~al.,}{{Weinberger}
  et~al.}{2017}]{Weinberger2017}
{Weinberger} R.,  et~al., 2017, \mn@doi [\mnras] {10.1093/mnras/stw2944}, \href
  {https://ui.adsabs.harvard.edu/abs/2017MNRAS.465.3291W} {465, 3291}

\bibitem[\protect\citeauthoryear{{Weinberger}, {Springel}  \&
  {Pakmor}}{{Weinberger} et~al.}{2020}]{Weinberger2020}
{Weinberger} R.,  {Springel} V.,   {Pakmor} R.,  2020, \mn@doi [\apjs]
  {10.3847/1538-4365/ab908c}, \href
  {https://ui.adsabs.harvard.edu/abs/2020ApJS..248...32W} {248, 32}

\bibitem[\protect\citeauthoryear{Welzl}{Welzl}{1991}]{Welzl91}
Welzl E.,  1991, in Maurer H.,  ed., New Results and New Trends in Computer
  Science. Springer Berlin Heidelberg, Berlin, Heidelberg, pp 359--370

\bibitem[\protect\citeauthoryear{{Wetzel}, {Deason}  \&
  {Garrison-Kimmel}}{{Wetzel} et~al.}{2015}]{Wetzel+2015}
{Wetzel} A.~R.,  {Deason} A.~J.,   {Garrison-Kimmel} S.,  2015, \mn@doi [\apj]
  {10.1088/0004-637X/807/1/49}, \href
  {https://ui.adsabs.harvard.edu/abs/2015ApJ...807...49W} {807, 49}

\bibitem[\protect\citeauthoryear{{White} \& {Frenk}}{{White} \&
  {Frenk}}{1991}]{White&Frenk1991}
{White} S. D.~M.,  {Frenk} C.~S.,  1991, \mn@doi [\apj] {10.1086/170483}, \href
  {https://ui.adsabs.harvard.edu/abs/1991ApJ...379...52W} {379, 52}

\bibitem[\protect\citeauthoryear{{White} \& {Rees}}{{White} \&
  {Rees}}{1978}]{White&Rees1978}
{White} S.~D.~M.,  {Rees} M.~J.,  1978, \mn@doi [\mnras]
  {10.1093/mnras/183.3.341}, \href
  {https://ui.adsabs.harvard.edu/abs/1978MNRAS.183..341W} {183, 341}

\bibitem[\protect\citeauthoryear{{Willmer}}{{Willmer}}{2018}]{Willmer2018}
{Willmer} C. N.~A.,  2018, \mn@doi [\apjs] {10.3847/1538-4365/aabfdf}, \href
  {https://ui.adsabs.harvard.edu/abs/2018ApJS..236...47W} {236, 47}

\bibitem[\protect\citeauthoryear{{Yang}, {Mo}  \& {van den Bosch}}{{Yang}
  et~al.}{2008}]{Yang2008}
{Yang} X.,  {Mo} H.~J.,   {van den Bosch} F.~C.,  2008, \mn@doi [\apj]
  {10.1086/528954}, \href
  {https://ui.adsabs.harvard.edu/abs/2008ApJ...676..248Y} {676, 248}

\bibitem[\protect\citeauthoryear{{Yaryura} et~al.,}{{Yaryura}
  et~al.}{2020}]{Yaryura2020}
{Yaryura} C.~Y.,  et~al., 2020, \mn@doi [\mnras] {10.1093/mnras/staa3197},
  \href {https://ui.adsabs.harvard.edu/abs/2020MNRAS.499.5932Y} {499, 5932}

\bibitem[\protect\citeauthoryear{{Yaryura}, {Abadi}, {Gottl{\"o}ber},
  {Libeskind}, {Cora}, {Ruiz}, {Vega-Mart{\'\i}nez}  \& {Yepes}}{{Yaryura}
  et~al.}{2023}]{Yaryura+2023}
{Yaryura} C.~Y.,  {Abadi} M.~G.,  {Gottl{\"o}ber} S.,  {Libeskind} N.~I.,
  {Cora} S.~A.,  {Ruiz} A.~N.,  {Vega-Mart{\'\i}nez} C.~A.,   {Yepes} G.,
  2023, \mn@doi [\mnras] {10.1093/mnras/stad2300}, \href
  {https://ui.adsabs.harvard.edu/abs/2023MNRAS.525..415Y} {525, 415}

\makeatother
\end{thebibliography}



\appendix

\section{Cumulative distributions of halo properties}
\label{app:cumul_dists}
\begin{figure*}
    \centering
    \includegraphics[width=\textwidth]{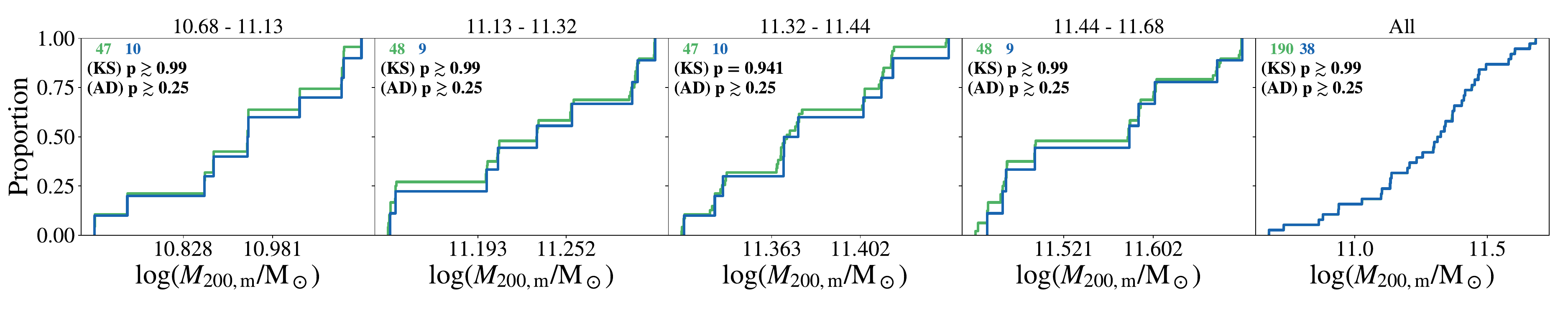}
    \includegraphics[width=\textwidth]{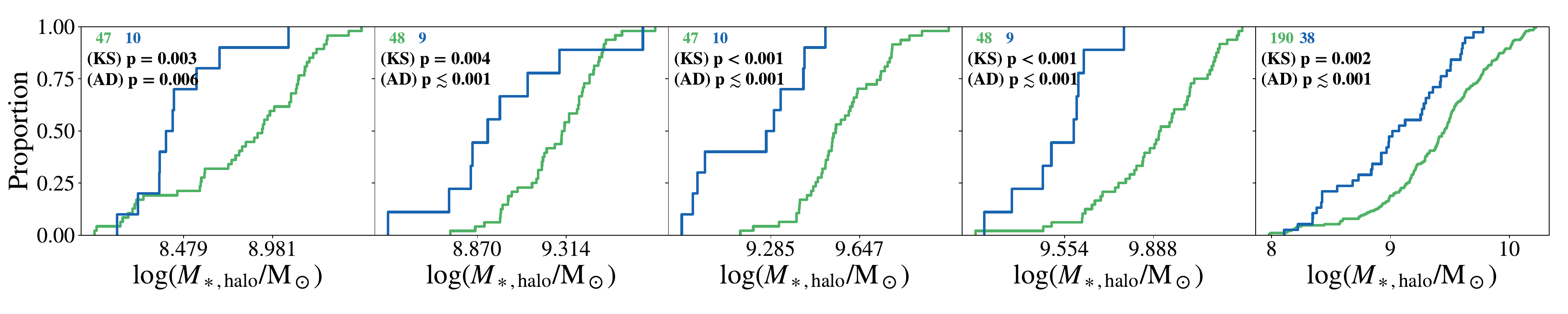}
    \includegraphics[width=\textwidth]{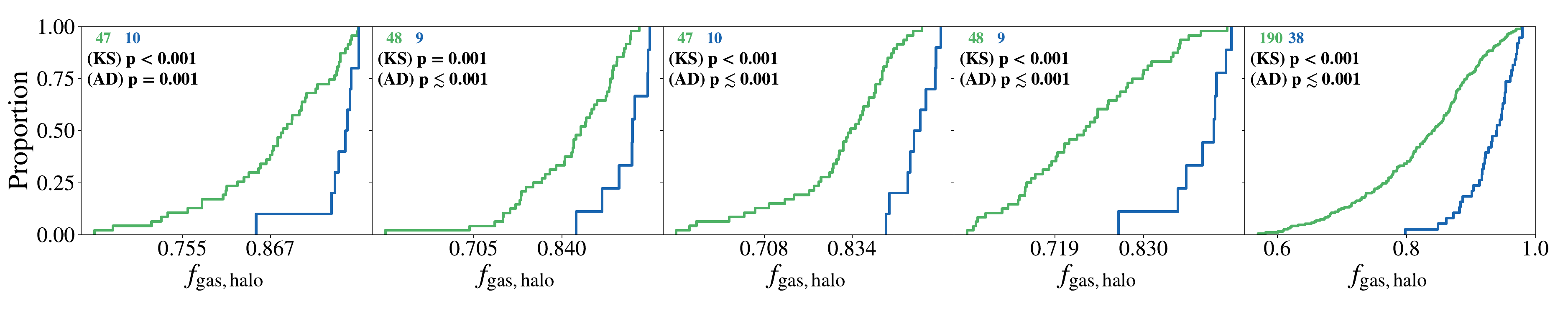}
    \includegraphics[width=\textwidth]{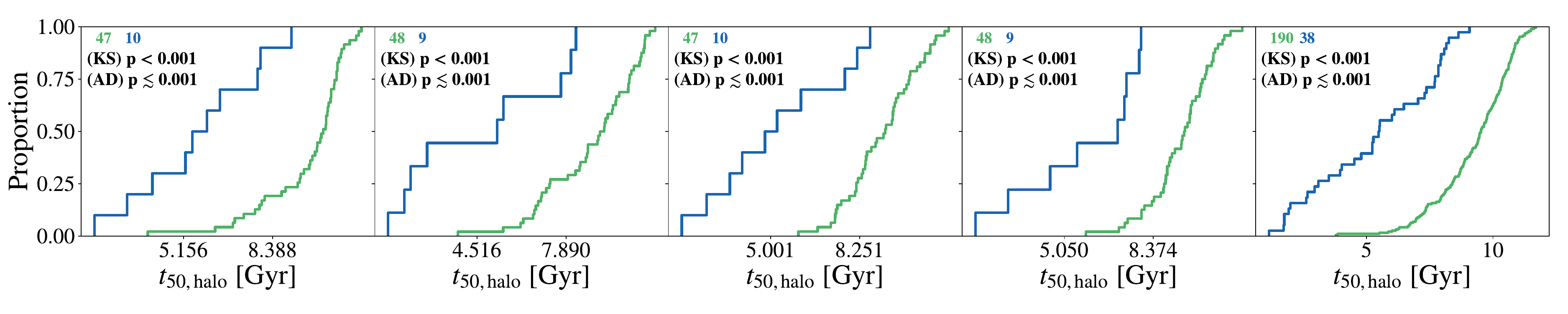}
    \includegraphics[width=\textwidth]{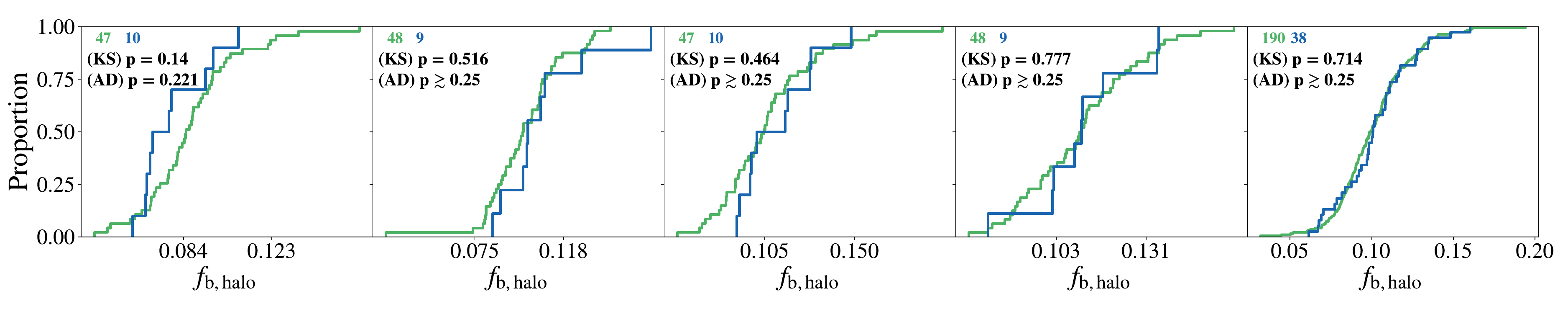}
    \caption{Empirical cumulative distributions of the quantities shown in the scatter plots of Fig. \ref{fig:host_halo_properties}. Each row shows the distribution of a halo property (top to bottom: adopted halo mass, total stellar mass, total gas fraction, 50\%-mass age, baryonic mass fraction), and each column shows the distribution of the properties for the different halo mass quartiles. Coloured numbers annotated on the top left of each panel show the number of objects in that specific mass bin. Numbers in black show the p-value of Anderson-Darling (AD) and Kolmogorov-Smirnov (KS) tests. As in the colour scheme of Fig. \ref{fig:host_halo_properties}, green refers to halos in the control sample, and blue refers to halos hosting CGDs. 
    }
    \label{fig:cum_bins_KS_halo_properties}
\end{figure*}

In Section \ref{results:halo_properties}, we presented results to show the difference between the properties of halos hosting CGDs and halos on a control sample paired by mass. For completeness, in Figure \ref{fig:cum_bins_KS_halo_properties}, we show the empirical cumulative distribution of the halo properties plotted in Figure \ref{fig:host_halo_properties}. In the first row of Figure \ref{fig:cum_bins_KS_halo_properties}, we show the distributions of halo mass, and it is clear that both samples have similar distributions of mass even in the mass bins. As seen in the second, third and fourth rows, the shape of the cumulative distributions is different for all mass bins. We quantified the difference between the distributions through KS-tests discussed in Section \ref{results:halo_properties}, and we found the same results using the Anderson-Darling $k$-sample test, which is more sensitive to the tails of distributions.

\section{Supplementary Material and Nearest Neighbour Catalogue}
\label{app:sup_material}
We provide videos and catalogues as supplementary material for this work. The material includes videos showing the evolution of gas and stellar density of CGDs. Catalogues of simulated and observed CGDs analysed in this work will also be provided as machine-readable tables. 
Here, we also present details of the nearest neighbour catalogue, which will be provided as supplementary data for the IllustrisTNG simulations. The catalogue is constructed by measuring Euclidean distances between galaxies in the simulation box. The minimum stellar mass ($M_{\ast \rm,min}^{\rm nn}$) for target galaxies and neighbours depends on the resolution of the simulation run, with $M_{\ast \rm,min}^{\rm nn} = 10^7 \ \rm M_\odot$ for TNG50-1, $M_{\ast \rm,min}^{\rm nn} = 10^{8.2} \ \rm M_\odot$ for TNG100-1 and $M_{\ast \rm,min}^{\rm nn} = 10^{9.1} \ \rm M_\odot$ for TNG300-1. For each galaxy, the catalogue stores information about its ten nearest neighbours within 15 Mpc and the total number of neighbours within a fixed spherical aperture centred on the target galaxy. Only subhalos with \texttt{subhalo\_flag=1} are used to build the catalogue.

\begin{landscape}
\begin{table}
\centering
\caption{Description of columns in nearest neighbour catalogues provided as supplementary catalogues to IllustrisTNG simulations. All stellar masses in the description refer to masses inside the aperture of two half-mass radii.}
\label{tab:landscape}
\begin{tabular}{cl}
\hline
Column name & Description of column \\
\hline
\texttt{subhalo\_ID} & Subhalo ID of the target galaxy. \\
\texttt{nn\{i\}\_ID} & Subhalo ID of the $i$-th nearest neighbour with stellar mass greater or equal to $M_{\ast \rm,min}^{\rm nn}$, where \texttt{i} spans from 1 to 10. \\
\texttt{nn\{i\}\_ID\_massive} & Subhalo ID of the $i$-th nearest neighbour with stellar mass greater or equal to $10^{10} \ \rm M_\odot$, where \texttt{i} spans from 1 to 10. \\
\texttt{nn\{i\}\_distance} & Euclidean distance to the $i$-th nearest neighbour with stellar mass greater or equal to $M_{\ast \rm,min}^{\rm nn}$, where \texttt{i} spans from 1 to 10. \\
\texttt{nn\{i\}\_distance\_massive} & Euclidean distance to the $i$-th nearest neighbour with stellar mass greater or equal to $10^{10} \ \rm M_\odot$, where \texttt{i} spans from 1 to 10. \\
\texttt{N\_aper\_\{r\}\_Mpc} & Number of galaxies with stellar mass greater or equal to $M_{\ast \rm,min}^{\rm nn}$ within a radius of $r$ Mpc from the target galaxy, with \texttt{r} being 1, 2 or 5. \\ 
\texttt{N\_aper\_\{r\}\_Mpc\_massive} & Number of galaxies with stellar mass greater or equal to $10^{10} \ \rm M_\odot$ within a radius of $r$ Mpc from the target galaxy, with \texttt{r} being 1, 2 or 5. \\ \hline
\end{tabular}
\end{table}

\begin{table}
\centering
\caption{First rows of the catalogue of simulated groups at $z=0$ in TNG50-1. Each row shows information about a dwarf galaxy in a CGD. From left to right, the columns present: group identifier, subhalo identifier, $r$-band absolute magnitude of the dwarf, stellar mass of the dwarf, specific star-formation rate of the dwarf, number of members in the group, total stellar mass of the group, total mass of the group (from projected mass estimator), group radius, total HI gas mass of the group, virial mass of host halo, 50\%-mass age of host halo, baryonic mass fraction of the host halo, total gas fraction of the host halo, maximum lookback time in which the group have $r_{\rm group} \leq 100$~kpc, maximum lookback time in which all group members belong to the same halo, inertial radius, group velocity dispersion based on Y20, group velocity dispersion based on \protect\cite{Wainer1976}, total B-band luminosity of the group.}
\label{tab:landscape}
\resizebox{\columnwidth}{!}{
    \begin{tabular}{cccccccccccccccccccc}
    \hline
    ID & SubID & $M_r$ & $\log M_\ast$ & $\log \rm sSFR$ & $N_{\rm m}$ & $\log M_{\ast \rm , group}$ & $\log M_{\rm PME,group}$ & $r_{\rm group}$ & $\log M_{\rm HI,group}$ & $\log M_{\rm 200,m}$ & $t_{\rm 50,halo}$ & $f_{\rm b,halo}$ & $f_{\rm gas,halo}$ & $t_{100}$ & $t_{\rm FoF}$ & $R_I$ & $\sigma_{\rm Y}$ & $\sigma_{\rm WT}$ & $\log L_{\rm B,group}$ \\
     &  & (AB mag) & ($\rm M_\odot$) & ($\rm yr^{-1}$) &  & ($\rm M_\odot$) & ($\rm M_\odot$) & (kpc) & ($\rm M_\odot$) & ($\rm M_\odot$) & (Gyr) &  &  & (Gyr) & (Gyr) & (kpc) & (km/s) & (km/s) & ($\rm L_\odot$)\\
    \hline
    2 & 619504 & -19.583 & 9.439 & -10.042 & 3 & 9.778 & 11.444 & 96.453 & 9.946 & 11.660 & 7.370 & 0.118 & 0.873 & 0.136 & 1.660 & 71.685 & 44.456 & 51.028 & 9.909\\
    2 & 619505 & -18.541 & 9.276 & -10.517 & 3 & 9.778 & 11.444 & 96.453 & 9.946 & 11.660 & 7.370 & 0.118 & 0.873 & 0.136 & 1.660 & 71.685 & 44.456 & 51.028 & 9.909\\
    2 & 619507 & -15.954 & 7.898 & -9.609 & 3 & 9.778 & 11.444 & 96.453 & 9.946 & 11.660 & 7.370 & 0.118 & 0.873 & 0.136 & 1.660 & 71.685 & 44.456 & 51.028 & 9.909\\
    3 & 628814 & -18.719 & 9.005 & -9.502 & 3 & 9.698 & 10.388 & 48.241 & 9.260 & 11.296 & 5.508 & 0.098 & 0.862 & 3.149 & 3.268 & 38.001 & 21.756 & 24.321 & 9.491\\
    3 & 628816 & -15.482 & 7.759 & -9.555 & 3 & 9.698 & 10.388 & 48.241 & 9.260 & 11.296 & 5.508 & 0.098 & 0.862 & 3.149 & 3.268 & 38.001 & 21.756 & 24.321 & 9.491\\
    ... & ... & ... & ... & ... & ... & ... & ... & ... & ... & ... & ... & ... & ... & ... & ... & ... & ... & ... & ... \\
    \hline
    \end{tabular}}
\end{table}

\begin{table}
\centering
\caption{First rows of the catalogue of groups found at $z \sim 0$ in SDSS DR18. Each row shows information about a dwarf galaxy in a CGD. From left to right, the columns present: group identifier, right ascension of the dwarf, declination of the dwarf, mean redshift of group members, $r$-band absolute magnitude of the dwarf, stellar mass of the dwarf, specific star-formation rate of the dwarf, number of members in the group, total stellar mass of the group, total mass of the group (from projected mass estimator), projected group radius, total HI gas mass of the group, inertial radius, group velocity dispersion based on Y20, group velocity dispersion based on \protect\cite{Wainer1976}, total B-band luminosity of the group.}
\label{tab:landscape}
\begin{tabular}{cccccccccccccccc}
\hline
ID & RA & DEC & $\bar{z}$ & $M_r$ & $\log M_\ast$ & $\log \rm sSFR$ & $N_{\rm m}$ & $\log M_{\ast \rm , group}$ & $\log M_{\rm PME,group}$ & $R_{\rm group}$ & $\log M_{\rm HI,group}$ & $R_I$ & $\sigma_{\rm Y}$ & $\sigma_{\rm WT}$ & $\log L_{\rm B,group}$ \\
 & (deg) & (deg) & & (AB mag) & ($\rm M_\odot$) & ($\rm yr^{-1}$) & & ($\rm M_\odot$) & ($\rm M_\odot$) & (kpc) & ($\rm M_\odot$) & (kpc) & (km/s) & (km/s) & ($\rm L_\odot$)\\
\hline
24 & 261.159 & 56.477 & 0.028 & -18.242 & 8.294 & -8.407 & 3 & 9.308 & 11.109 & 76.649 & - & 72.564 & 30.938 & 34.234 & 9.867\\
24 & 261.156 & 56.477 & 0.028 & -18.675 & 8.964 & -9.240 & 3 & 9.308 & 11.109 & 76.649 & - & 72.564 & 30.938 & 34.234 & 9.867\\
24 & 261.030 & 56.494 & 0.028 & -18.020 & 8.962 & -9.779 & 3 & 9.308 & 11.109 & 76.649 & - & 72.564 & 30.938 & 34.234 & 9.867\\
28 & 184.937 & -2.759 & 0.022 & -18.253 & 8.923 & -9.651 & 3 & 9.560 & 10.800 & 85.223 & - & 54.336 & 21.654 & 25.169 & 10.049\\
28 & 184.921 & -2.795 & 0.022 & -18.579 & 8.808 & -9.254 & 3 & 9.560 & 10.800 & 85.223 & - & 54.336 & 21.654 & 25.169 & 10.049\\
... & ... & ... & ... & ... & ... & ... & ... & ... & ... & ... & ... & ... & ... & ... & ... \\
\hline
\end{tabular}
\end{table}

\end{landscape}


\bsp	
\label{lastpage}
\end{document}